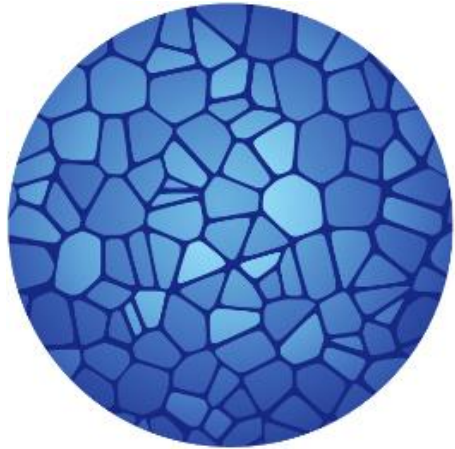

# THE HUMAN CELL ATLAS
White Paper

The HCA Consortium
October 18, 2017

# TABLE OF CONTENTS

**Acknowledgments**

White Paper Writing Committee: *Aviv Regev, Peter Campbell, Nir Hacohen, and Sarah Teichmann*

Writing Team: *Jennifer Rood, Mary Carmichael, Robert Majovski, Aviv Regev, and Orit Rozenblatt-Rosen*

We thank HCA Organizing Committee members *Dana Pe'er, Gary Bader and Henk Stunnenberg* for multiple critical readings of the entire white paper and their insightful comments.

# SUMMARY

The Human Cell Atlas (HCA) will be made up of comprehensive reference maps of all human cells — the fundamental units of life — as a basis for understanding fundamental human biological processes and diagnosing, monitoring, and treating disease. It will help scientists understand how genetic variants impact disease risk, define drug toxicities, discover better therapies, and advance regenerative medicine. A resource of such ambition and scale should be built in stages, increasing in size, breadth, and resolution as technologies develop and understanding deepens. We will therefore pursue Phase 1 as a suite of flagship projects in key tissues, systems, and organs. We will bring together experts in biology, medicine, genomics, technology development and computation (including data analysis, software engineering, and visualization). We will also need standardized experimental and computational methods that will allow us to compare diverse cell and tissue types — and samples across human communities — in consistent ways, ensuring that the resulting resource is truly global.

This document, the first version of the HCA White Paper, was written by experts in the field with feedback and suggestions from the HCA community, gathered during recent international meetings. The White Paper, released at the close of this yearlong planning process, will be a living document that evolves as the HCA community provides additional feedback, as technological and computational advances are made, and as lessons are learned during the construction of the atlas.

# 1. OVERVIEW

## VISION

For the past 150 years scientists have classified cells by their structures, functions, locations, and, more recently, molecular profiles, but the characterization of cell types and states has remained surprisingly limited. We do not yet comprehensively know our cells — how they are defined by their molecular products, how they vary across tissues, systems, and organs, and how they influence health and disease. This has limited our ability to study fundamental domains in biology – such as physiology, developmental biology, and anatomy – in health and disease, and to translate our knowledge to accelerate diagnosis and treatment of disease[1].

But an extraordinary opportunity is emerging because of transformative advances in experimental and computational methods (**Figure 1**; **Section 3** and **Section 4**). Massively parallel single-cell genomics assays can now profile hundreds of thousands of cells. Technologies to profile DNA and proteins in single cells, as well as a combination of DNA, RNA, and proteins in the same cell, provide important additional layers of information. New spatial analysis techniques, including in situ assays, imaging approaches, spatial coding, and computational inference, allow high-resolution analysis of large tissues in two (2-D) or three (3-D) dimensions. Computational algorithms have emerged to determine cell types, states, transitions, and locations from these new data, at increasing scale and resolution (**Section 4**).

Together, these advances have catalyzed a growing sense in the scientific community that we are finally capable of realizing a long-sought goal of identifying and understanding human cells and molecular states within tissues and systems in their full diversity and glory. Indeed, many researchers have already begun to apply these techniques to identify new cell types in tissues ranging from the blood to the brain — an unprecedented achievement.

We have therefore launched the Human

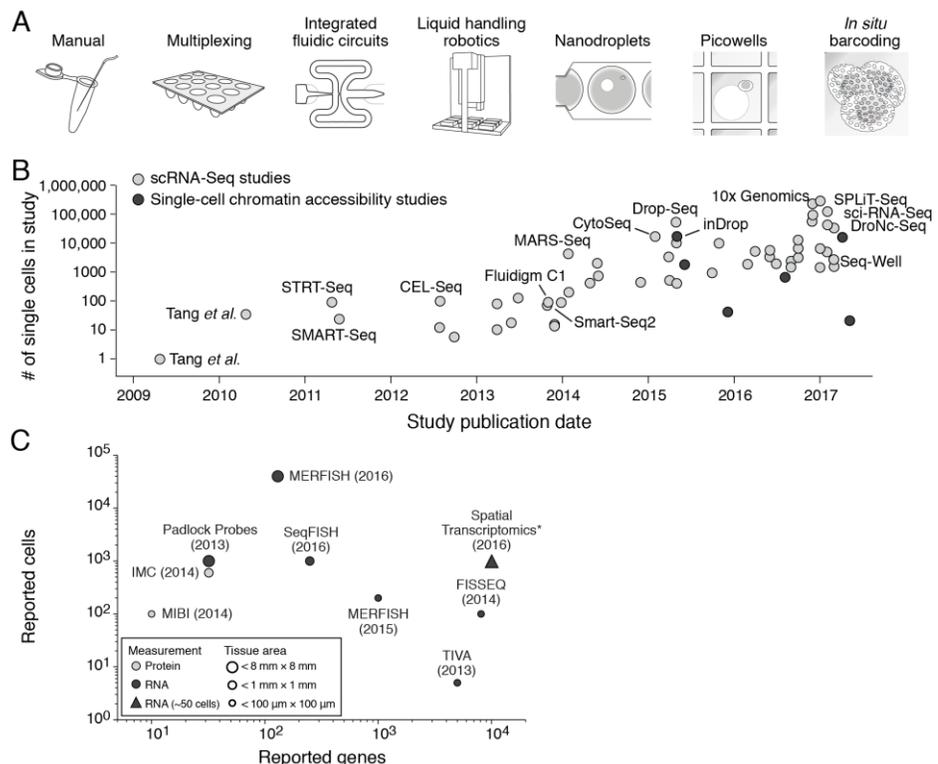

**Figure 1. Advances in experimental technologies empower the HCA.**
(A) Technologies for single-cell genomics. (B) Timeline and scale of single-cell RNA-Seq (grey circles) and single-cell ATAC-Seq (black circles). (C) Timeline and scale of methods for highly multiplexed spatial analysis of intact tissue, including the measurement type (protein, RNA) and tissue area.

Cell Atlas (HCA) Consortium — an international, collaborative effort that aims to define all human cell types in terms of their distinctive patterns of gene expression, physiological states, developmental trajectories, and location. This consortium builds on the work of previous consortia to map the human genome, starting with the Human Genome Project, and leverages and engages with recent efforts to characterize and interpret functional genomic (ENCODE), epigenetic (IHEC, BLUPRINT), transcriptomic (GTEx), and proteomic (the Human Protein Atlas) elements.

The HCA will be a foundation for biological research and medicine: a comprehensive reference map of the types and properties of all human cells and a basis for understanding and monitoring health and diagnosing and treating disease. The project will help propel translational discoveries and applications, ultimately laying a foundation for a new era of precision and regenerative medicine.

The diverse, international consortium that builds the HCA will be open and collaborative, bringing together and aligning experts to form networks focused on biological topics. Some of these networks — such as the Immune Cell Atlas (ICA), the Developmental Cell Atlas (DCA), and the Skin Cell Atlas (SCA) — have already emerged and initiated pilot efforts with scientific leadership and committed participants, including relevant clinical and biological experts. Scientists from across the globe have enthusiastically joined the HCA, in meetings, through social media, and electronically — helping to design the effort and participating in it. Indeed, each meeting thus far, and the work that preceded and followed it, has helped formulate key aspects of this White Paper (**Figure 2**).

The initiative will progress in phases to generate reference maps at increasing resolution. Google Maps serves as an analogy: instead of geographical features, such as continents, countries, cities, streets, and houses, the HCA's maps of the human body will "zoom in" on molecular and organizational features of organs, tissues, and cells.

The first draft of the HCA — a focus of this White Paper — will profile 30 million to 100 million cells, both isolated and in their tissue context, from major tissues and systems from healthy research participants of both genders (**Section 2**; **Table 1**). It will combine single-cell profiling of dissociated cells and single-nucleus profiling of frozen samples with spatial analysis of cells in the context of tissues. It will also integrate data from other projects and consortia, as appropriate. In this first draft, a set of representative organs and systems will be analyzed in depth; a broader range will be analyzed to a more limited extent. While the first draft will be assembled with geographic, age, disease, and ethnic diversity in mind, it will not yet aim to be comprehensive with respect to these features.

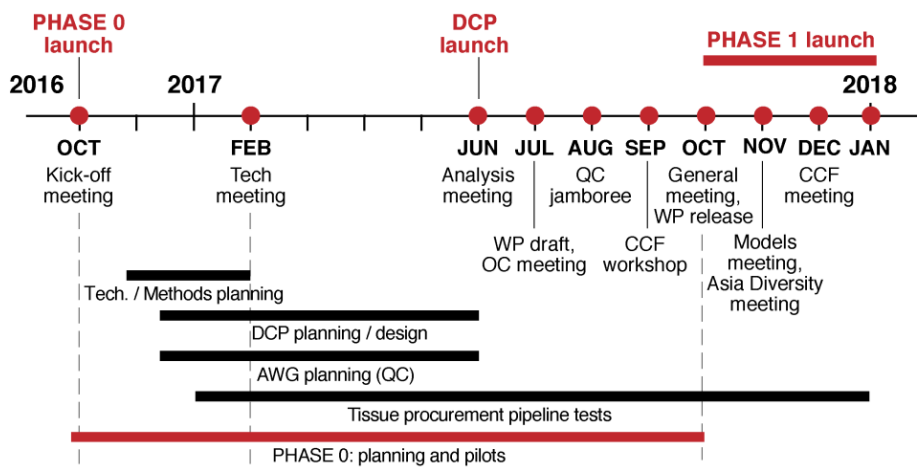

**Figure 2. Timeline of HCA activities**, October 2016 through January 2018.

This first draft and the lessons learned in building it will serve as the basis for a comprehensive atlas of at least 10 billion cells, covering all tissues, organs, and systems — the necessary reference for future comparison and biological insight across disease areas, genetic diversity, environments, and ages. The cells will come from both healthy research participants and small cohorts of patients with relevant diseases, as these are critical to reflect on cells' diversity. The cells will be studied using a broad range of techniques to capture both breadth and depth and will fully represent the world's diversity.

As with previous genomic projects, bounds for the HCA must be defined. First is the desired resolution: the rarity of cell types and states to be detected, analogous to the bounds on human-variant frequency in human genetic studies. Second is disease and diversity. The draft atlas will endeavor to characterize healthy samples that capture as much genetic, geographic, environmental, and age diversity as possible. In particular, the HCA is committed to genetic and geographic diversity and equity at every phase, even if mapping genetic variation from very large cohorts may not be feasible in the initial stages. Although disease will not be a focus of the first iteration of HCA, we expect some disease samples to be captured that may include, in addition to disease cells, additional cell types relevant for correctly referencing healthy cell types; for example immune cells from the tumor microenvironment. The HCA methods and framework, together with dedicated partnerships, will also empower disease-mapping efforts in individual tissues and across cancers. A third bound will be cellular function. The HCA will validate the *existence* of identified cells and enable their *functional assessment*, but the functional characterization itself is not included in its scope. By analogy, the functional characterization of the genes discovered in the Human Genome Project is still an ongoing endeavor, through numerous inspired studies by individual investigators as well as concerted efforts by international consortia.

The HCA should help answer questions in all areas of human biology, from the taxonomy of cells and histological tissue structure, to developmental biology and cell fate and lineage, to physiology and homeostasis and their underlying molecular mechanisms. With corresponding atlases of model organisms that facilitate functional assessment, the HCA will allow us to better understand how faithful our models are to human physiology and pathology and to validate findings through perturbation.

Because the HCA will be an open resource, it will dramatically accelerate discoveries by biological researchers, data scientists, and translational scientists and clinicians worldwide, inspiring insights in therapeutic discovery, drug development, and diagnosis. The HCA will provide crucial information about the cell types in which a given gene and its disease-associated variants are expressed; will empower us to develop better drugs and more readily predict their unintended toxicities; and can transform today's standard diagnostic practices.

In this White Paper, we detail a research and organizational strategy for the HCA as a comprehensive, open, global resource — one with the potential to transform our understanding of biology and ultimately allow us to fulfill the promise of precision medicine.

# VALUES

The HCA will be built on a set of guiding principles and values to ensure its success and maximize its utility to the research community and humanity at large. Some of these build on lessons from earlier consortia; others reflect the unique opportunities at this moment in time. These value include:

- *Transparency and open data sharing.* Data will be released as soon as possible after it has been collected so it can be used immediately.
- *Quality.* The HCA community will be committed to producing the highest-quality data and establishing rigorous standards, shared openly and broadly and updated regularly.
- *Flexibility.* The HCA community will maintain intellectual and technical flexibility, so it can revise the design of the HCA as new insights, data, and technologies emerge.
- *Community.* The HCA community will remain global, open, and collaborative, led by a scientific steering group. It will remain open to all interested participants who are committed to its values.
- *Diversity, inclusion, and equity.* The selection of tissue samples will reflect geographic, gender, age, and ethnic diversity. Similar diversity will be reflected in the distribution of participating researchers, institutions, and countries.
- *Privacy.* We are committed to ensuring privacy of research subjects, consistent with the consent of research participants.
- *Technology development.* The HCA community will develop, adopt, and share new tools to empower others.
- *Computational excellence.* The HCA community will develop new computational methods, leveraging and driving the latest algorithmic advances, and share these through scaled, open-source software.

## FRAMEWORK FOR A FIRST DRAFT ATLAS

We currently envision a Draft Atlas v1.0 that contains data from 30 million to 100 million profiled cells and their matching tissues, though the scale and scope may grow as measurement methods increase in throughput, robustness, and affordability (**Section 2**).

Building the first draft will require careful decisions about the **organs, systems, and tissues** to be analyzed; the level of **resolution**, such as the rarity of cells to be detected and spatial resolution; the **sampling and measurement approaches** applied to the selected samples; and the **data analysis and a data platform** needed to store, analyze, and visualize the data.

**Organs, systems, and tissues**

A complete Human Cell Atlas will map all tissues, systems, and organs. The first draft will incorporate a carefully chosen subset that is immediately useful and addresses key representative examples, from which general lessons can be learned for the next phase (**Section 2**).

*Choice of tissues: depth and breadth.* The first draft will not cover every organ. Instead, it should incorporate data from several major tissues and systems (**Table 1**). Some tissues will be sampled deeply; for others, only a portion will be sampled. All studies will include adults; some will also incorporate pediatric samples. This will nonetheless yield global insights into the intersection of systems and tissues and into what is needed to handle different kinds of tissues.

*Acquisition*. The core activities of the HCA project will be the profiling of normal — ideally, freshly obtained — human tissues. Multiple tissues from the same donor should be processed using the same technology in the same place, then preserved and banked consistently in appropriate form for subsequent processing (e.g., Formalin-fixed, Paraffin-embedded [FFPE] or flash frozen in optimal cutting temperature [OCT]) for single nucleus RNA-Seq, spatial analysis, and additional possible future applications. To achieve this, we will need reliable sources of suitable tissues. Normal tissue samples can be challenging for individual labs to acquire, but concerted efforts — e.g., the Genotype-Tissue Expression (GTEx) Project and the Cambridge Biorepository for Translational Medicine (CBTM) — have demonstrated how to assemble and bank excellent sample collections for use by the community. It is essential to have the ability to procure all human organs and tissues from postmortem samples, as this is the only source from which an entire human body can be studied. At the same time, samples from live research participants or transplant organ donors are likely to be closer to normal physiology and, therefore, should be obtained, banked and analyzed whenever possible.

*Reproducibility and diversity*. For each selected tissue, a minimum of 20 ethnically diverse samples will be collected and banked from at least three geographically distinct sites. The number of required samples is based on our community's current experience and on the impressive reproducibility observed in healthy cells across individuals in a few studied tissues. Possibly, a larger number will be required or a smaller number will suffice. Adaptive power analyses will be conducted as part of the collection effort to revise these choices.

Each site will collect and bank samples from individuals of both genders, all adults (with the exception of the human development atlas — see below), ranging from 20 to 55 years of age. Ethnic diversity will be ensured at each site to the maximal extent possible, given the number of individuals and that site's location.

*Additional optional elements*. For each tissue, it is expected that the relevant biological community will incorporate deeper investigation of at least one additional dimension, such as more extensive genetic diversity or disease state. For example, to properly sample normal immune cells, disease challenges must be included. Similar rationale would apply in many other tissues.

**Resolution**

A draft atlas must have predefined technical bounds, such that we can determine the completion of phase I *within those bounds*. This involves determining how tissues are sampled within organs, the rarity of cells to be recovered, the resolution of spatial coordinates, and the depth of molecular information.

*Cellular resolution*. Each biological tissue or system will have a predefined cell-rarity threshold achievable within the proposed phased sampling procedure. We note the need to detect both discrete subsets of cells (stable cell types or states), often presented as dense regions in high-dimensional space (where the dimensions are gene expression or other cellular features), as well as continuous processes (dynamic transitions) reflected as paths in the space.

*Spatial resolution*. Each analyzed tissue will have a predefined scheme for spatial resolution. We anticipate the use of rapidly evolving technologies, such as MERFISH, Seq-FISH, FISSEQ, Spatial Transcriptomics, CODEX, MIBI, and targeted in situ RNA sequencing. In the first draft, we do not expect to generally analyze complete organs, but rather cells in their histological context. The spatial scheme for a tissue will define the level of histological unit to be analyzed (macro resolution) and its internal partitioning.

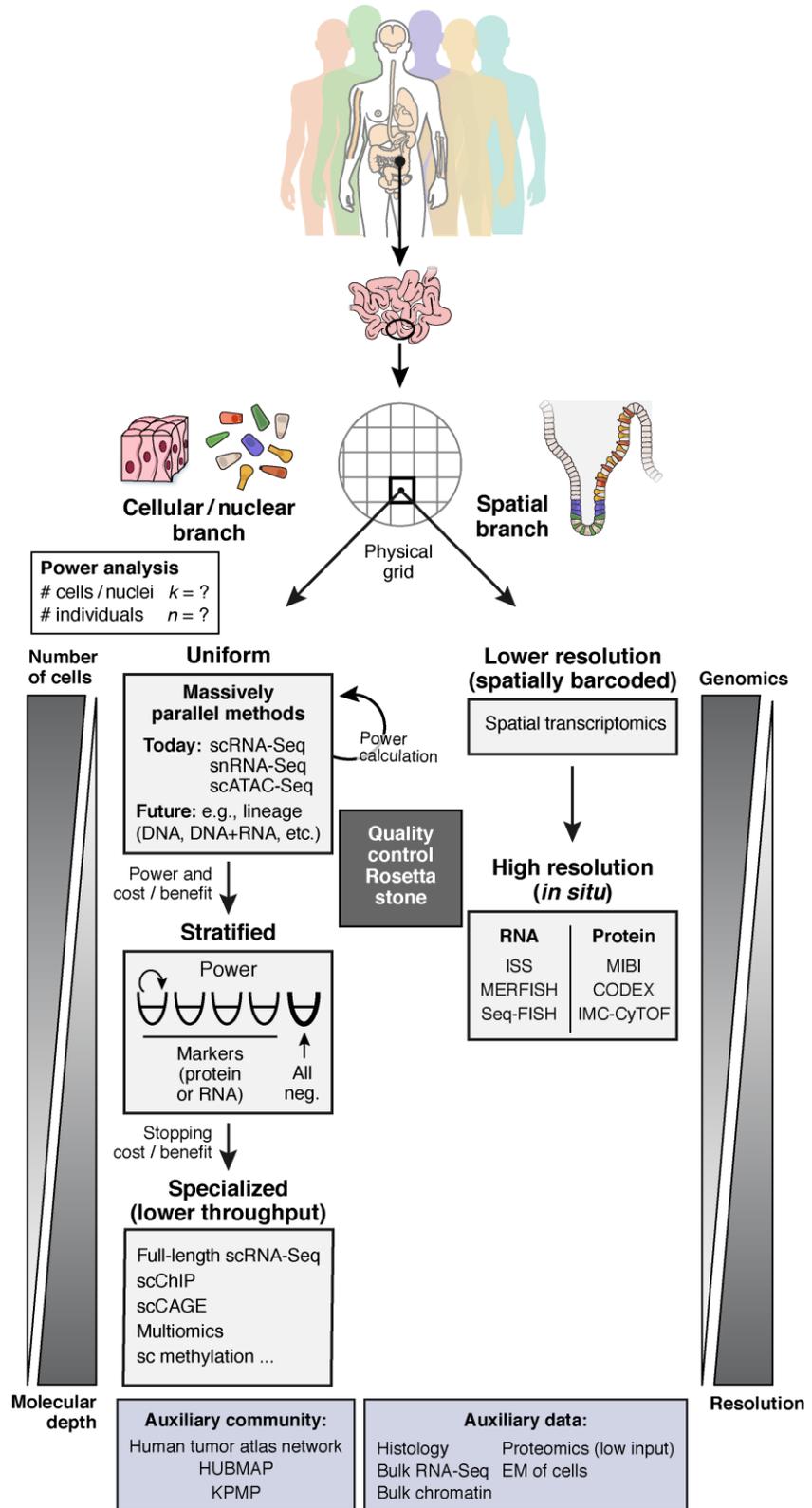

**Figure 3. Framework for sampling and measurements for an initial draft.** Any given tissue specimen will be analyzed through a two-pronged strategy combining [1] single-cell molecular profiling of dissociated cells or single-nucleus molecular profiling of nuclei from frozen tissue (cellular/nuclear branch, left) with [2] highly multiplexed spatial analysis of intact tissue (spatial branch, right). To relate the two, before tissue processing, physical specimens will be registered and imaged for their physical coordinates, and matching portions will be analyzed by cellular and spatial approaches.

*Molecular resolution*. Currently, there is a trade-off between the number of cells a technique measures and the depth (complexity of data) afforded per cell. Massively parallel techniques profile many cells, but show less of the molecular complexity of each cell, while more in-depth technologies typically handle far fewer cells at a time. We will strike a balance between these techniques.

### Sampling and measurements

An important role for the first draft is to learn key lessons on appropriate sampling, sample handling, and measurement of representative tissues. The experience of past consortia, such as the Human Protein Atlas and GTEx, may serve as a guide. Any given tissue specimen will be analyzed through a two-pronged strategy, combining single-cell molecular profiling of dissociated cells (cellular branch) and/or single-nucleus molecular profiling of nuclei from frozen tissue (nuclear branch) with highly multiplexed spatial analysis of intact tissue (spatial branch) (**Figure 3**). To relate the two, before tissue processing, physical specimens will be registered and imaged for their physical coordinates, and matching portions, such as adjacent sections, of the same specimen will be analyzed by cellular and spatial approaches. Within and across the branches, we expect to use multiple methods (each standardized with appropriate quality controls and benchmarks) to analyze the same type of tissue or actual specimens. We expect the HCA effort to catalyze further technology development and to adapt to such changes in an agile manner (**Section 3**).

*For the cellular branch*, we will address the current trade-off between scale and molecular complexity, as well as aim to efficiently sample complex cell mixtures when different cell types or states may be represented at radically different proportions. To this end, we will follow a **"Sky Dive" strategy**, with initial uniform ("agnostic") profiling by massively parallel approaches, followed by profiling of stratified samples (e.g., through sorting and/or enrichment or depletion of subpopulations). The number of cells to be profiled in each phase will be determined by statistical calculations with prespecified goals for detection sensitivity (cell rarity). These calculations will initially be informed by prior biological knowledge ("educated guess") and then adaptively re-estimated as data are collected. Both the uniform and stratified phases may be repeated until a predefined stopping threshold is met. When possible, cells from multiple individuals will be mixed in a single assay, and then distinguished based on the genetic differences, reducing batch effects and costs, and streamlining the process. Specialized, but lower-throughput, techniques, which provide deeper and/or more diverse molecular profiles, will be applied to limited numbers of cells in the stratified groups. Auxiliary data (e.g., corresponding bulk molecular profiles of RNA, chromatin, or protein for annotation) will be generated within the HCA effort or in coordination with other consortia. We will engage with related communities (e.g., Tumor Cell Atlas Network, the BRAIN Initiative) to share best practices, collection strategies, and data platforms.

*For the spatial branch*, we will address the current trade-off between the number of distinct molecules to be analyzed and spatial resolution. Ideally, each specimen will be analyzed by both genomic-profiling approaches with lower spatial resolution and by RNA- and protein signature–based assays or in situ sequencing with high spatial resolution. Because some of the key techniques are not yet widely disseminated or fully scaled, we will aim to preserve and bank specimens in appropriate form for subsequent processing. Fortunately, most methods are compatible with common preservation strategies.

*The two branches are highly complementary and intimately connected.* Currently, most profiling approaches do not preserve spatial information, whereas most spatial techniques either rely on predefined signatures of genes and proteins or do not have single-cell resolution. Thus, the cellular branch can help define signatures for spatial measurements or single-cell deconvolution; the spatial branch can validate cells defined by profiling approaches, position them in their tissue context, and help identify any compositional biases in the profiling branch introduced during tissue dissociation.

*The key results of the draft must be validated for their reproducibility, integrity, and predictive value. Reproducibility (stability)* is defined by the ability to recover, through prospective isolation or repeated analysis, cells with the same profiles and features predicted by the initial

data and analysis. *Integrity* is defined by the ability to capture all cells in the tissue in the correct proportions and appropriate profiles. *Predictive value* is defined by the ability to determine that a subset characterized with one distinctive set of features (e.g., molecular profiles) either appropriately maps to a known, previously validated biological entity or predicts a new entity with distinctive features of a different nature (e.g., cell morphology or histological or anatomical context).

**Data analysis and a Data Coordination Platform (DCP)**

As soon as the first HCA data are generated, they should be of immediate use to the research community. For this reason, data will be openly released (**Section 7**) through a data platform (**Section 5**) as soon as possible after it has been collected. The Data Coordination Platform (DCP) will take on all relevant data types, from both the cellular and spatial branches. The platform will be open source and open data and can be fully cloned.

Key analysis methods to address each of the central questions regarding a cell's identity — types, states, transitions, developmental history, and tissue organization — will be built by grassroots efforts across the HCA community. The open source DCP, governed by HCA, will allow any method to connect to the platform through dedicated APIs. In addition, HCA will designate some of these as HCA official pipelines, an integral part of the open-source data platform. These official pipelines will be applied immediately to any new data and will be part of any streamed version and formal releases. Data processed through the pipelines will be accessed through portals and apps, with functionalities for both computational users (through APIs) and biologists (through GUIs).

In addition to streaming, the HCA will have formal releases, where data has been processed, analyzed, and vetted. Such cohesive drafts should ensure that the project:

- is able to integrate data from multiple organs, techniques, and researchers;
- possesses a data infrastructure that synthesizes data in a way that maximizes the knowledge that the data provide;
- is indeed a true *atlas* by revealing relationships between the cell types that are included; and
- reflects on the lessons learned in the HCA effort up to that point to guide practices for the next phase of the project.

# SCIENTIFIC ORGANIZATION

The HCA consortium will be built on four scientific pillars: Collaborative Biological Networks, the Technical Forum, the Data Coordination Platform, and the Analysis Garden.

The **Collaborative Biological Networks** will bring together the scientific community's domain experts in specific systems and organs, together with genomics, computational, and engineering experts to construct the atlas of each tissue, system, or organ.

The **Technical Forum** will develop new technologies and run dedicated pilot projects to test, compare, and disseminate existing technologies.

The **Data Coordination Platform** will be a centralized way to "bring researchers to the data" by creating software to perform data ingestion, storage, processing, analysis, visualization, and access controls.

The **Analysis Garden** will be a rich, diverse, open, and easily accessible ecosystem where computational methods and algorithms developed by any interested group can bloom and be shared across the community.

The full and formal governance of the HCA and its membership is stated in **Section 7** and **Appendix I.** Briefly, the HCA is governed by an **Organizing Committee** (OC), currently comprising 27 scientists from 10 countries with diverse areas of expertise. It is led by two **co-chairs**, who are members of the OC, and has an **Executive Committee** (EC), which includes the two co-chairs and five additional OC members. The OC establishes **Working Groups** in specific key areas. It also governs the **Data Coordination Platform** (DCP). It establishes and appoints a **DCP Governance Group** (DCPGG), which reports to the OC, to oversee the implementation of its policies for the DCP. The HCA is coordinated by **Executive Offices** (EOs), currently in four locations (U.K., U.S., European Union, and Asia), which staff the OC in performance of its duties. The OC also convenes members of the community in meetings, workshops, and jamborees. The OC also convenes but does not monitor the Funders' Forum, an opportunity for funders and potential funders of the HCA to discuss the project.

Any individual may become an **HCA Member** by registering at the HCA Member Registry and agreeing to abide by the principles of the HCA (as stated in **Section 7**), especially its ethical standards. In addition, an HCA member who is a participant in an HCA project, or a member of an OC-designated HCA group, will be designated as an **HCA Collaborating Member**.

Any scientific project related to systematic biological characterization at single-cell resolution may become an **HCA Project** by registering in the HCA Project Registry. Projects will fall into three categories: HCA Participating Project, HCA Network Project, and HCA Flagship Project.

*Each of the pillars and the overall organization will require dedicated engagement, support, and funding*. Examples include:

- **Biological Networks:** tissue acquisition, data generation, and HCA Flagship Projects (see below and **Section 7**) for each tissue, including development and testing of the Common Coordinates Framework;
- **Technical Forum:** technology development, benchmarking, dissemination, and training;
- **Data Coordination Platform:** software development, revision, and maintenance, plus support of data portals, data storage, and computational resources;
- **Analysis Garden:** computational methods development, software dissemination (including through the DCP), method comparison, testing, and training; and
- **Overall Organization:** workshop and jamboree planning, research progress tracking, connecting with and onboarding of community members, HCA meeting planning and convening, and engagement with the broader scientific community, the public, and the media.

## IMPACT

The HCA will help answer fundamental questions in all aspects of biology as well as serve as a guide to unravel the secrets of human disease (**Figure 4**). The translational promise of the cell atlas ranges from basic biology of the human organism, to disease mechanism, diagnosis, prognosis, and treatment monitoring, to immunotherapy, drug development, and cell and organ replacement. General areas of medical impact include:

- **Genes to drugs.** The cell atlas will enable researchers to identify the cell types in which a given genetic variant acts, thus helping to pursue therapeutic targets identified by genetic studies of disease. For example, analyzing tens of thousands of neurons in the retina revealed new subtypes that eluded neuroscientists before, which can help us find in which cells the genes important in blindness actually act.
- **Regenerative medicine.** An atlas of cell types that are lost in disease will enable efforts to generate such cells faithfully. Similarly, an atlas of healthy human tissues and the matching organoids or in vitro differentiated cells will help determine if the engineered samples faithfully represent normal tissue composition and identify ways to complete any missing components. For example, efforts are under way to produce dopamine neurons in vitro or, alternatively, to reprogram cells in vivo into dopamine-producing neurons to treat Parkinson's disease; an atlas of cell types will pinpoint characteristics that must be programmed into these cells for them to succeed.
- **Disease mechanisms.** Because the cell atlas will provide detailed maps of cells and their roles in tissues, researchers will be able to understand the mechanisms underlying any disorder at both the cell and the cellular-ecosystem level. For example, an atlas of the small intestine will help map the cell of action for genes associated with Crohn's disease, food allergy, obesity, and colon cancer.
- **Drug discovery**. The cell atlas will provide guidance as to which gene signatures to pursue in drug screens to represent desired cell phenotypes. For example, it can give us a molecular map of which genes and signatures drive cell development and how it goes awry in, say, cancer and provide targets for drug discovery.
- **Toxicity**. It will be possible to determine where else in the body a particular gene is expressed, helping to identify potential off-target effects prior to drug trials. For example, a cell atlas will help CAR-T immunotherapy cell developers ensure that the cells do not inadvertently target healthy essential cells that express the same gene or that drugs will not have off-target effects in other tissues (for example, causing blindness by targeting genes expressed in the retina).
- **Drug efficacy and resistance**. The atlas will provide the tools necessary to understand why drugs work — or don't — at the level of cells and tissues, both prior to and after treatment. For example, a "cellular-ecosystem map" that identifies both target cell types and target molecules of immunotherapy will help predict and monitor tumor response and provide new leads for immune modulation in resistant patients.

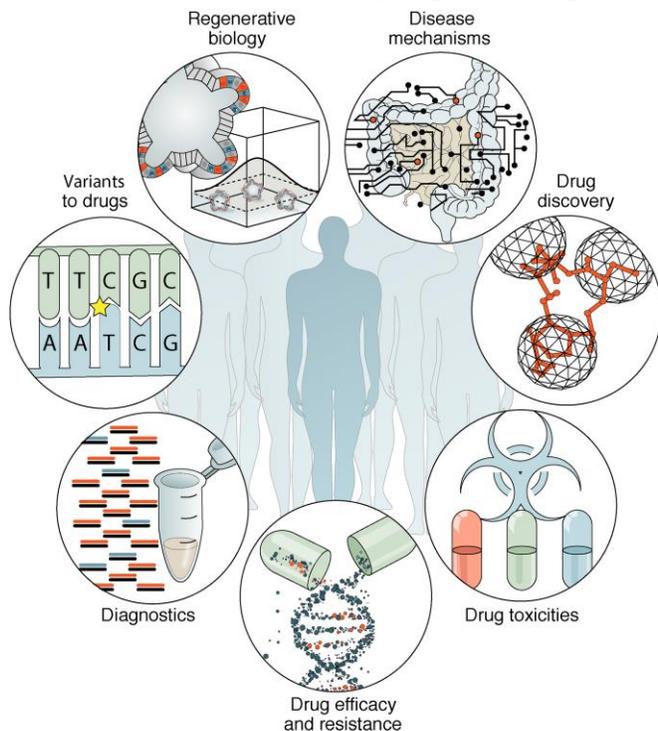

**Figure 4**. **The Human Cell Atlas** will have profound impact on biology and medicine.

- **Diagnostics**. Knowledge of all the cell types in the body and their role in disease will enable updated and much more powerful versions of common diagnostic tools, such as the Complete Blood Count (CBC) and next-generation biopsy. For example, the CBC, a census of a limited number of blood components that is used in a variety of diagnostic settings, could be supplemented by a "CBC 2.0" that would provide a high-resolution picture of the nucleated cells in, for example, blood disorders, infectious disease, autoimmune disease, and cancer. Tissue biopsies from patients could also be analyzed with unprecedented resolution.

# 2. A FRAMEWORK FOR CONSTRUCTING A PHASE I ATLAS: EARLY DRAFT

While our overall goal is to build a comprehensive Human Cell Atlas, we think it is wise to set intermediate goals for draft atlases of increasing resolution, comprehensiveness, and depth. This phased approach will define milestones along the way, provide guidance for better planning of subsequent milestones, hold the project accountable, and provide immediate utility to the scientific community.

In this section we provide a general framing for constructing Phase I of the HCA. We discuss the key design considerations and provide an overall approach to address them. Nevertheless, many details must be specified precisely for each data collection effort, and some will vary by tissues (**Section 6**).

## TISSUE ACQUISITION

Accessing the tissues needed to meet the aims and objectives of the HCA will present significant biological, logistic, and regulatory challenges. For adult samples, we will primarily acquire healthy tissue from rapid autopsies and deceased organ donors (who have already donated any organs useful for medical purposes). In some instances (**Section 6**) these will be complemented with disease tissue. For samples of human development, we will acquire tissues from available repositories. Whenever possible, we will obtain resections, biopsies, and other sources from living healthy research participants, but we recognize that this will not be possible for some tissues and organs. Comparing tissues obtained from live healthy donors to their counterparts from post-mortem sources would help determine the validity of the latter for those cases when postmortem sources are the only ones available.

**Tissue sources**

Tissue sources used for the HCA must have the following essential features:

- Access to **normal live cells and tissues** from male and female research participants or donors across a range of ages and ethnicities.
- Ability to obtain **adequate tissue mass** from all organs and tissues of interest to enable multiple comprehensive analyses of the same samples.
- Full **regulatory approval and participant informed consent** for comprehensive (including genetic) sample analysis as required by legal and ethical standards.
- Unrestricted **open access to the data** generated.

Additional features that could ideally be met, at least for some sources, include:

- Ability to perform **comprehensive sampling** from multiple organs and multiple samples from the same donor.
- Access to a range of **developmental tissues**.
- Access to **full donor demographic data**, including relevant past medical, family, and social history.
- Comprehensive **infectious disease screening** of the research participants and donors.
- Access to live cells and tissues from **disease cohorts of interest**.

To satisfy these requirements, the HCA effort will use three major complementary tissue sources: (**1**) live individuals as research participants; (**2**) deceased transplant organ donors; and (**3**) rapid autopsies. Each presents a trade-off between the breadth of available tissue and the faithfulness of the source to normal living humans.

- *Live research participants*. Access to healthy live tissue and cells is mostly limited to samples donated by living volunteers (e.g., blood, fat, or skin biopsies) or resected tissue collected as part of an invasive procedure (e.g., surgery or endoscopy) performed as part of screening or on patients with known or suspected pathologies. Viable tissues are naturally limited in quantity and range and may be challenging to obtain for some tissues without data-access restrictions. However, these tissues represent truly healthy individuals and can also be the starting material source for many organoids.

- *Deceased transplant organ donors.* This source allows controlled and planned tissue acquisition in a very rapid manner after death. There is detailed anonymized medical and social history and screening data available for each donor, along with ethical approvals, and many (albeit not all) organs and tissues can be obtained from each deceased donor. Tissues can be perfused with cold organ preservation solution, ensuring cell and tissue viability for >24 hours after donation, or with newer *ex vivo* perfusion systems that keep organs warm and operational for hours after death. Tissue can also be acquired from donors with a range of pathologies that are not absolute contraindications to organ donation (e.g., diabetes, hypertension, ischemia-related heart disease, and autoimmunity — but not cancer). Similar strategies can be applied for developmental tissue.

- *Postmortem examinations* can provide tissue from any organ, with appropriate consents for open-access data, and can be collected via rapid, or "warm," autopsy. Many rapid autopsies are pre-consented, or individuals have predeclared their interest in donating organs to scientific research after death. Tissue can be acquired from deceased donors with relevant disease histories, including cancer, or can focus on pathologically uninvolved tissue.

**Feasibility case studies**

Two use cases highlight the potential and power of deceased transplant organ donors and rapid autopsies for systematic collection of organs and tissues for HCA.

*Feasibility Case Study 1: Cambridge Biorepository for Translational Medicine* (CBTM)

CBTM ([www.CBTM.group.cam.ac.uk](www.CBTM.group.cam.ac.uk)) was founded in Cambridge, U.K., in April 2015 to support translational research projects that require access to fresh or biochemically viable tissue. It currently supports 31 major research projects, led by 24 PIs in 14 departments and institutions, and has collected tissue from >50 deceased human donors with comprehensive ethical approval and informed consent. The research projects supported by CBTM cover diverse use cases, including genomics, transcriptomics, regenerative medicine, immunology, metabolism, physiology, and pathology. CBTM has successfully enabled projects, which, like HCA, required access to tissue as detailed below.

- *Normal live tissue that is very rarely available* from living volunteers or through resections (e.g., dorsal root ganglia, heart, thymus, and spinal cord).

- *Normal tissue that is not available in sufficient quantities* from individual patients or living volunteers (e.g., large volumes of peripheral blood).

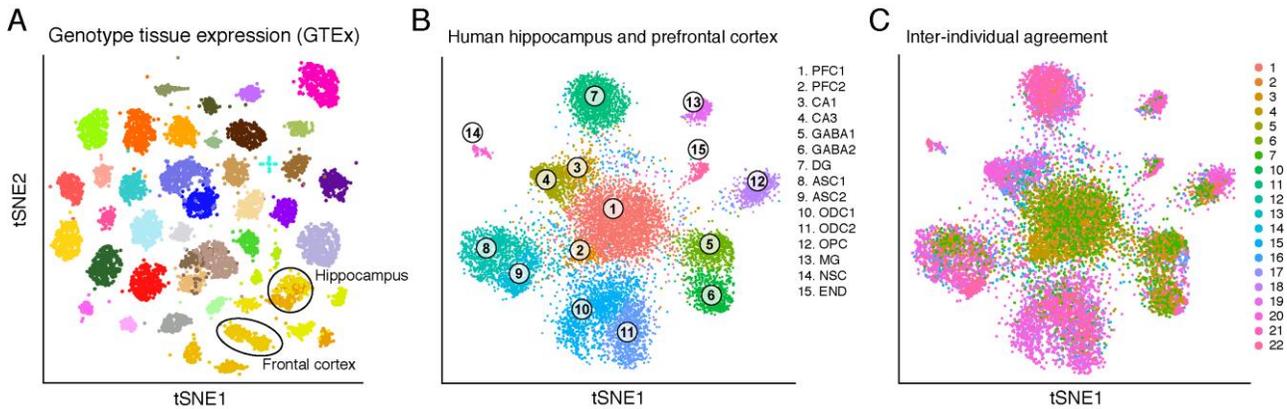

**Figure 5. Retrospective samples from GTEx can be successfully profiled using single-nucleus RNA-Seq.** (A) Bulk gene expression profiles from all GTEx tissues. Hippocampus and frontal cortex sample clusters, from which samples in (B) are obtained, are circled. From Auget and Ardlie, *Current Genetic Medicine Reports* (2016); 4:163-9. Reprinted with permission from Springer Nature. (B) Single nucleus RNA-Seq (by DroNc-Seq) of hippocampus and frontal cortex samples from the GTEx collection. tSNE plots are colored by k-NN graph clustering and labeled *post hoc* by cell type. (C) Each cluster is supported by multiple individuals (from relevant tissue). (B) and (C) from Habib *et al.*, *Science* (2016) Aug 26;353(6302):925-8. Reprinted with permission from AAAS.

- *Multiple normal live tissue samples from the same organ* (e.g., entire length of the gastrointestinal tract and its secondary lymphoid organs).
- *Multiple normal live samples from several organs* and anatomical locations from the same donor.
- *Normal live tissue that must be processed very rapidly* and within seconds of interruption of the blood supply (e.g., oxygenated myocardium).

CBTM is already providing tissue for pilot projects to perform tissue dissociation and scRNA-Seq and is planning a CBTM-HCA collaboration for up to 3,000 specimens per donor from 5 to 10 deceased donors.

*Feasibility Case Study 2: The GTEx project collection network*

GTEx is a U.S. NIH Common Fund project initiated in 2010 to determine how genetic variation affects gene expression across 44 normal human tissues. To support this effort, it developed a tissue-collection platform, spanning multiple organ procurement sites, that meets the ethical, scientific, informatic, and operational challenges of large-scale, rapid, viable postmortem biospecimen collection. To date, GTEx has collected ~30,000 tissue specimens from 960 donors, across 53 distinct tissue sites (median of ~26 tissues per donor), from both transplant organ donors and rapid-autopsies. A small portion of the archived samples have been flash frozen and can be profiled by using single-nucleus RNA-Seq[2-4]. Unfortunately, most of the samples were not archived for such purposes, and the original project does not specify consent that data would be shared and released openly, so previously collected samples cannot be used for the open-access HCA.

Nevertheless, the GTEx collection network has demonstrated that it can fulfill all of the aims and objectives of the HCA for tissue acquisition from postmortem donors, including open access.

- It conducted a successful pilot project to enroll four donors who consented to full open-access sharing of research results (performed together with the ENCODE 4 consortium).
- All of the protocols and standard operating procedures developed for the GTEx collection network have been made publicly available and remain in place.
- All of the aims and objectives outlined above were met by the GTEx collections.

- Although GTEx tissues were not specifically collected for single-cell analysis, the organ-procurement sites involved have considerable experience with the collection and stabilization of tissues for single-cell research. Indeed, it was recently demonstrated with retrospective samples, that rapidly collected, frozen samples from GTEx can be successfully used in snRNA-Seq (**Figure 5**[3]).
- The GTEx collection network also obtains an extensive collection of detailed histology images, enabling integrative computational analyses of molecular and spatial features.

**Challenges and related pilot studies**

There are three key challenges to create the necessary infrastructure and optimize and validate protocols for tissue acquisition for the HCA: (**1**) engage or establish tissue-acquisition networks beyond the U.K. and the U.S.; (**2**) determine the impact of ischemia time; and (**3**) perform tissue acquisition in the context of a Common Coordinate Framework for anatomical mapping. We discuss each in turn, along with pilot studies to address them.

*Deceased transplant organ donors for tissue acquisition beyond the U.K. and U.S.* The CBTM and GTEx pipelines provide proof-of-principle for the suitability of deceased organ and rapid-autopsy donors for HCA tissue acquisition. It is now necessary to identify a small number of transplant centers in these and other countries to maximize international access and geographical and ethnic diversity. Representatives from several European organ-procurement operations, who participated in two recent GTEx meetings, have begun exploring such involvement. A potential **pilot study** (of 3 to 6 months) could examine the feasibility of extending this model to one more European Transplant Unit. In parallel, for the HCA it will be advantageous to tap into other organ-collection efforts from ethnicities and geographic regions traditionally neglected by the research community — for example, by collaborating with scientists in Africa, Asia, and Latin America.

*Impact of ischemia time.* Understanding molecular changes during warm and cold ischemia after circulatory arrest, in a tissue and organ-specific manner, is critical for setting acceptable limits for processing time after circulatory arrest and will have significant operational implications. Moreover, pathophysiological changes after brain-stem death and premorbid medical conditions can impact the donor's cells and might require the development of criteria for inclusion in or exclusion from the HCA. To date, analyses of bulk samples in GTEx have suggested that the ischemic interval affects the transcriptome in a tissue-specific manner. A **pilot study** at the single-nucleus and spatial level could leverage the pre-existing Biospecimen Methods Study (BMS) in GTEx, in which samples from 30 donors and six tissues were collected at four sequential ischemic time intervals; alternatively, new collection can be performed to test this.

*Anatomical mapping in a Common Coordinate Framework.* A Common Coordinate Framework (CCF) is essential for mapping the anatomical location of every sample from each organ in a comparable manner. A CCF must have higher granularity and precision beyond classical anatomical descriptions, be robust to natural variations in donor and organ size and shape, be bespoke for each organ, and should flexibly enable higher levels of precision sampling, such as spatial grids. This is a general challenge for HCA and will require proper standard operation procedures, encoding of current anatomical knowledge, and computational inference. Clearly, it has a direct impact on tissue acquisition. A **pilot study**, will aim to develop an initial CCF for one or two organs (e.g., kidney, liver, heart, or colon) that will include appropriate referencing to constant anatomical landmarks, nomenclature, and the necessary data platform for linking to spatial coordinates within the organ. We discuss the CCF further below.

# COMMON COORDINATE FRAMEWORK

A workshop focused on developing, piloting, and deploying a CCF will be held in Washington, D.C., in December 2017, engaging anatomists, pathologists, clinicians, organ experts, technology experts, machine-learning experts, software engineers, and visualization experts. This section will be written after the workshop and will appear in a later version of the White Paper.

# DATA COLLECTION

We will collect data for the HCA using a two-pronged approach: a **cellular branch** for molecular profiling of dissociated cells or nuclei; and a **spatial branch** for analysis of cells in tissue (**Figure 3**).

**Cellular branch: a Sky Dive for single-cell or single-nucleus profiling**

Key operational challenges in building a cellular atlas include the absence of a priori ground truth on the number of cell types and states and their relative proportions and rates of transitions; the need to ensure that understudied cell populations — whose molecular markers or mere existence may be entirely unknown at the outset — are not ignored; and the possibility of technical biases in the recovery of different cells because of, for example, differential loss during sample handling.

To address these challenges, the first cellular draft will be built using a Sky Dive: a step-wise, iterative, and adaptive design strategy. Akin to skydiving — in which the jumper sees a wide landscape upon leaping from the plane, then continually resolves finer and finer detail on the ground as she falls — this strategy starts with a broad uniform survey of cells in a specimen, followed by stratification into specific subsets for additional profiling guided by accrued knowledge. This iterative strategy ensures that rare cells can be discovered and adequately profiled. Similar strategies have been deployed for the cost- and time-efficient sampling of large human populations for census taking. To make the atlas most useful in the long-term, we expect to use multiple technologies to examine the same tissues, both at the same and/or at different levels of the Sky Dive. This strategy has already been used to profile the cells in the mouse retina, where first the whole retina was profiled[5] and then a more focused effort to profile a specific cell type, bipolar cells, followed[6]. A similar approach was taken in colon organoids[7] and the subsequent analysis of rare enteroendocrine cells[8]. Key to the success of this Sky Dive will be an "all negative" bin that will recover even those cells that could not yet be distinguished in the initial survey. Finally, to address the possibility of differential loss, both single cells and signal nuclei will be profiled in the cellular branch. The latter can be isolated from tissue without dissociation in a manner than appears to minimize compositional biases[2-4,9].

*Initial uniform sampling* will be performed to profile a defined number of cells, $k$, from a specimen with minimal, if any, additional stratification. The number of cells to be profiled must be determined based on a statistical model. An early example (**Box 1**) accounts for the number of cell types expected in the tissue, the proportion of the rarest type that we desire to detect, the minimal number of cells desired of the rarest type, and the desired confidence level to achieve this number. The key unknown parameter is the number of cell types; it is initialized by prior knowledge or an "educated guess." This early model makes several simplifying assumptions and should be enhanced, for example, to address biases in sampling of cells (e.g., loss of specific cell types because of differential viability or capture efficiency) and to handle cases where the distinctions between cells may be subtler and require larger numbers of cells. (Similar models will be required for the case of continuous paths, such as cell differentiation.) The Analysis Working Group (AWG, **Sections 4 and 7**) is already working on improved approaches for experimental design and power calculations.

The process may be iterated. Following initial profiling and analysis, the key parameter of the number of expected cell types may be revised, in which case a recalculation will be performed and additional cells may be profiled. Based on the composition of the tissue revealed in this phase, an additional power analysis and a cost-benefit analysis will determine the empirical rate of discovery and will determine when to move to a stratified analysis. Notably, in these iterations it may be possible to select — from the Whole Transcriptome Amplification (WTA), barcoded material — a subset of cells or nuclei barcodes of interest for deeper sequencing of such transcripts, without any a priori cell enrichment. This will not recover new cells, but could deepen the characterization of rare cells in the sample without the need to similarly sequence molecules from more prevalent cells.

*Stratified sampling* requires that cells be first separated into "buckets" by sorting, enrichment, or depletion strategies. In addition to any affirmative buckets, a final "all negative bucket" is maintained to help discover additional, possibly rare subsets. The buckets aim to optimize collection when cell proportions vary substantially, but critically depend on the availability of reagents and experimental procedures. For example, we may not have many antibodies for less-known cell types and can possibly use RNA probes on "fixed cells" for sorting; this would also simplify enrichment for nuclei. Within each bucket, the same power calculations are applied as in the initial uniform approach, along with the same procedure for iteration and the same stopping criteria. In each iteration, buckets may be further partitioned. Notably, the goal will be to determine composition up to the predefined rarity *P*, but, depending on tissue composition, rarer subsets may be detected.

The first two phases of the Sky Dive — uniform and stratified sampling — will rely on massively parallel profiling. At the moment, two types of molecular profiles can be collected at such scale: RNA, using one of several strategies for scRNA-Seq or single nucleus snRNA-Seq, and chromatin, using a single cell Assay for Transposase Accessible Chromatin (scATAC-Seq). RNA and chromatin profiles are highly complementary, especially in the context of cell differentiation. For each of these methods, it may be possible to mix cells from multiple individuals prior to profiling, and using their genetic variation as natural genetic barcodes to distinguish them. This has both operational and analytic advantages, but is applicable only to

---

**Box 1: Statistical models to determine the number of cells to profile.** A simple statistical framework can help answer the question: "Given a tissue with *N* discrete cell subsets, the rarest of which is present at proportion *P*, how many cells *k* need to be sampled such that at least *n* cells are recovered in each subset with confidence level *C*?" The model assumes the conservative (worst-case) scenario in which there is one dominant cell type and many equally rare cell types. It further assumes that the Central Limit Theorem (CLT) holds, and the samples are independent: that is, the set of cells sampled is far larger than the number of cells desired, a safe assumption in nearly any case but the smallest samples. While the sampling of each cell type is not actually independent it is not a serious violation if the number is large enough. Given these assumptions, the number of cells from each type, $T_i$, will distribute as $E[T_i] = N*p$, $SD[T_i] = sqrt(N*p*(1 – p))$, and we solve for N such that all (N – 1) subtypes to have at least n cells. The last (dominant) subtype easily clears this threshold since its proportion is much higher. This model is now available at http://satijalab.org/howmanycells. We expect further refinements, as well as a similar model to be developed for continuous transitions.

those samples that can be accrued (e.g., frozen) prior to processing. Whenever single-cell profiling methods are similarly scaled, they can be considered as potential additions to this process. An example approach recently scaled to achieve high throughput is lineaging through targeted tracking of selected DNA mutations[10]. Other approaches will be added as they scale and based on the extent of independent information that they provide.

*Deep analysis with specialized but lower-throughput methods.* Massively parallel methods do not reflect all key aspects of the cell, do not currently integrate multiple profiles in the same cell, and may provide relatively lower depth (complexity). A host of sophisticated methods currently available only at lower throughput has emerged to measure DNA mutations (for lineaging), epigenomic profiles (e.g., histone modification, 5' CAGE, DNA methylation, which may reflect cell types at longer time scales) and multiomics (e.g., RNA and DNA methylation, RNA and protein; chromatin accessibility and RNA; which provide mechanistic insights). Furthermore, sensitive methods are available for full-length scRNA-Seq and snRNA-Seq, but they are currently available only in lower throughput. These methods will provide a fuller profile of the molecules present, including splice variants and epigenomic states, thus providing a more comprehensive characterization of each cell type.

As the Sky Dive reaches high resolution, with subset stratification, the massively parallel profiles will be complemented by these lower-throughput and/or more specialized and costly methods applicable with more modest cell numbers within predefined cellular subsets.

*Auxiliary bulk profiles for annotation and interpretation.* Analysis of single-cell profiles, especially those from massively parallel methods, relies on auxiliary annotations. For example, massively parallel scRNA-Seq quantifies RNA levels relative to transcript annotations: if 3' end and splice isoforms are not well-reflected in pre-existing annotations, as is likely to be the case for less frequent and less studied subsets of cells, important information may be lost. To address this challenge, "annotation grade" bulk RNA-Seq profiles (deep, long, paired-end reads) will be collected at each level of the Sky Dive — from the entire specimen to iteratively finer buckets — and used to generate high quality annotations. Other profiles — for example, of histone modifications and proteomics — should be similarly collected. This is a remarkable opportunity for partnership between the HCA consortium and other consortia, such as ENCODE and the Human Protein Atlas, to share the same, open-consented specimens.

**Spatial branch: Molecular microscopy into cells in their tissue context**

Direct- and virtual-imaging technologies can relate molecular measurement with cellular, subcellular, and structural features of tissues. Current techniques have a clear and continuous trade-off between speed, resolution, and molecular complexity (**Figure 3**): high-resolution methods mostly use predefined molecular signatures (e.g., sets of RNA, proteins), whereas profiling methods are mostly applied at lower spatial resolution. While technologies advance, this can be addressed by iteratively using the lower-resolution methods, including single-cell profiles, to define signatures and the higher-resolution methods to obtain exquisite spatial details.

Ideally, portions of every specimen will be analyzed with methods in each of three increasing levels of resolution.

- *Anatomical tiles*. At the coarsest level, tissue will be assayed with Anatomical Tiles (ATs), collected through serial 3-D sectioning that produces voxels of tissue for further processing, which are registered to imaging information. Ideally, each AT is further partitioned to provide samples both for single-cell profiles and for spatial analysis of the same specimen. The length scales of tiles should reflect the anatomy and histology of the specific tissue.

- *Spatial barcoding*. At the next level, molecules in tissue can be spatially barcoded, with techniques currently at the resolution of 100 µm (~2 to 50 cell diameters) and improving.
- *High resolution, in situ measurements*. Multiplex hybridization, in situ sequencing, and protein-capture methods can measure RNA or protein signatures at the highest resolution, from cellular to single molecule, depending on the technique. This is also an opportunity to collaborate with the Human Protein Atlas and measure the spatial distribution of the human proteome at a subcellular level[11,12].

*Iterative signature selection and spatial power analysis*. The need to choose molecular signatures for high-resolution spatial profiling introduces questions about the power to detect different spatial patterns ("spatial power analysis"). We will address this iteratively: Initial signatures will be selected by a combination of prior knowledge (known landmarks) and from signatures defined from single-cell profiles or lower-resolution spatially barcoded profiles. Once a signature is measured, entire genomic profiles can be computationally projected through the landmarks[9,13-15] to identify additional putative patterns and to define new spatial signatures for measurements. A key remaining concern exists if profiling methods suffer from systematic biases, but this can be addressed by performing multiple compositional measurements, each assessing the sum abundances of multiple molecules[16], and use decoding strategies to determine an entire profile.

*Tissue preservation.* Because some of the spatial techniques are not yet broadly disseminated and scaled, in some cases high-resolution spatial data may be collected only after substantial single-cell profiles have been collected from the same specimen. Fortunately, most spatial methods are compatible with long-term tissue preservation, allowing for this staggered approach. Thus spatially and anatomically registered tissue specimens will be routinely preserved.

**Validation: reproducibility, integrity, and predictive value**

In assessing the validity of the generated draft, we should consider *reproducibility or stability*, defined by the ability to recover (through both planned replicates and follow-up prospective isolation) cells and cell subsets with the same profiles and features; *integrity*, defined by the ability to capture — using all techniques and data in aggregate — all cells (up to the defined rarity thresholds) in the tissue in the correct proportions and appropriate profiles, without disproportionate loss of specific subsets or change in molecular patterns; and *predictive value*, defined by the ability to determine that a subset defined by a distinctive set of features (e.g., molecular profiles) either appropriately maps to a known, previously validated biological entity (cell type, state, or transition) or predicts a new entity with distinctive features of another nature (e.g., histological).

To assess *reproducibility or stability*, we will compare across multisite, multi-individual replication studies of both profiling and spatial analyses. In addition, the later phases of the Sky Dive will include prospective isolation, followed by single-cell and/or bulk profiling. In this context, *Rosetta Stone* samples will be collected early in the project, as an archived, large set of specimens, available for comparison of protocol application, at least when fresh cells are not required.

To assess *integrity*, bulk RNA-Seq or proteomics measurement can help determine the probability of residual, unascertained types residing within samples, if those are not too rare and are sufficiently distinctive. For cases with such residual signals unaccounted for in dissociated cells, additional cell profiles may need to be acquired, either randomly by comprehensive tissue dissociation or through stratification. Single-nucleus profiling should also help identify biases in cell composition or state introduced by dissociation. In the case of tissues analyzed in situ, this will require new panels of probes. It may be advisable to use multiple, randomly selected combinations of probes in such instances.

Consistency between profiles of dissociated cells and spatial measurement of intact tissue will provide another estimate of integrity. Comparison to spatial transcriptomics of intact tissue can be applied in a similar way, for more sensitive analysis. Comparison to in situ measurements can help determine changes in single-cell profiles introduced during dissociation.

To assess *predictive value*, cross-referencing between distinct molecular signatures, defined from profiles, and distinct spatial features (cell morphology, spatial localization, histological structures, cellular neighborhood) will be a primary tool. We note that functional assessment — such as studies of the physiological function of cells and their characteristic molecules — are generally not within the direct scope and focus of the HCA, but will be highly enabled by the atlas.

# 3. TECHNOLOGY

Building the HCA will require careful experimental design, including consideration of which technologies we should deploy to generate reproducible and high-quality data and which we will need to develop to enable new measurement capabilities or increased scale. Indeed, the HCA has been inspired by and will be facilitated by technological innovations that address a broad range of measurements and experimental challenges, including single-cell and single-nucleus "-omics" profiling (e.g., transcriptomics, genomics, epigenomics, proteomics, and multiomics); tissue dissociation; tissue and cell preservation; and spatial and temporal characterization. Some of these are currently more mature and broadly disseminated than others. As a result, it is reasonable to assume that, at any given moment, some data will be collected using established technologies ("production"), while next-generation, emerging methods progress through refinement and dissemination ("scaling") to production. We also envision that, similar to the Human Genome Project and other consortia, the HCA will inspire the development of new approaches by helping provide a sense of community and mission-driven focus, as well as resources and drive that foster innovation.

**The Technical Forum**

To serve such an effort, the HCA's Technical Forum (Forum) will have two goals:

- bring together groups and centers to compare and disseminate *existing* methods; and
- ensure that *new* techniques are developed and adopted through a combination of academic efforts and the activities of technology companies that are eager to help the HCA and to learn from its collective expertise and experiences.

To compare, test, and scale methods, the Standards and Technology Working Group (STWG; **Section 7**) will initiate the Forum with groups spanning different countries and areas of expertise. These groups will carry out well-designed technical pilots in a handful of biological systems, in close partnership with computationally focused researchers. Pilots will be initiated among both the cellular and the spatial branches of the HCA and span the entire life cycle from invention to proof-of-concept and finally dissemination in optimized systematic pipelines at scale.

*Development of new technologies.* To promote the development of new techniques, the STWG and the Forum will first identify key areas in which new technologies are greatly needed and promote work within them. Advancing new methodologies will require

- dedicated support to individual labs that are inventing methods;
- dissemination of nascent protocols into additional labs beyond those that developed the initial technique, including additional benchmarking;
- generation of preliminary open data sets on which computational groups can hone different analytic techniques; and
- providing ongoing, iterative feedback from computational experts to experimentalists that will expedite the improvement of techniques.

To fulfill these latter needs, the analysis working group (AWG) will identify potential pitfalls, provide early and rapid feedback that helps pinpoint artifacts, speed optimization, and generally guide the technology development efforts. For example, jamborees co-organized by AWG and STWG (**Sections 4 and 7**) can help rapidly develop initial analytics or quality-control measures for new technologies.

*Technology optimization and comparison.* Techniques for sample preservation, tissue dissociation, "-omic" preparation, and spatial profiling should be evaluated, and benchmarking or validation data sets generated. Comparisons can be made between different protocols that have the same goal (e.g., different methods of scRNA-Seq or of multiplex fluorescence *in situ* hybridization [FISH]). They can also be made between laboratories using the same protocol to assess robustness and reproducibility. Finally, they can be made by using complementary methods (scRNA-Seq and spatial analysis) on the same kind of sample for a feasible subset of samples. The AWG will help facilitate this process by developing metrics and benchmarks for the comparison that are broadly agreed, including through community-wide efforts such as jamborees, co-organized with STWG.

*Technology dissemination.* Early single-cell sequencing techniques required fairly ubiquitous skills and equipment, leading to wide and rapid adoption (for example, SMART-Seq required only 96-well plates, a method of isolating single cells, and limited infrastructure). By contrast, many spatial technologies require highly specialized tissue-handling protocols and sophisticated microscopes. Moreover, the handling and dissociating of human tissue specimens requires substantial domain know-how, which is often not broadly disseminated. We will need to ensure hands-on training and equipment to disseminate both of these techniques to a wider community, as well as system-specific and general benchmarks to ensure effective adoption. This will require investment in both infrastructure (e.g., equipment) and in training, including on-site hands-on guidance by Specialized Work Acquisition Teams (SWAT teams), which will convey expertise and know-how from one site to another.

Below, we discuss these considerations more fully in the context of technologies underlying the two key strategies for the HCA: molecular profiling of dissociated single cells or nuclei (cellular branch) and highly multiplexed spatial analysis of intact tissue (spatial branch). In each case, we present the state of the art, key areas for development and optimization, approaches for evaluation and benchmarking, and strategies to disseminate, combine, and deploy techniques in building the HCA.

## SINGLE-CELL MOLECULAR PROFILING OF DISSOCIATED CELLS

The revolution in single-cell genomics has enabled genome-wide quantification of mRNA in thousands of individual cells at once. Additionally, multiple techniques have been developed to study the genetic and epigenomic characteristics of single cells, including DNA mutations (and associated lineage information), cytosine modifications, higher-order chromosome conformation, histone modifications, and regions of accessible chromatin (**Figure 1**)[17-19]. Each of these methods will need to be further optimized and deployed to generate the HCA — some as the main workhorses and others as auxiliary methods, with agile reassessment of chosen techniques in a fast-evolving landscape.

We have also begun to integrate and couple multiple "-omics" measurements on the same individual cell. Recently, the first methods for integrating genomic DNA and mRNA sequencing from the same cell[20,21], single-cell methylome and transcriptome measurements[22], and even chromatin accessibility, methylation, and transcriptomes[23] from the same single cell were described[23]. Several methods have been developed that allow coupling of single-cell transcriptomics to protein measurements[24-27]. Collectively, these and newly developed single-cell multiomics methods would help integrate information on the genome, lineage, (hydroxy)methylome, genome accessibility and structure, transcriptome, and proteome, and would provide more comprehensive pictures of the properties of, and relationships between, each cell type in the atlas.

**Current technologies**

*mRNA*. Because of the close connection between a cell's function and its transcriptional program, as well as the relative simplicity of amplifying and detecting nucleic acids, gene-expression profiling at the single-cell level has become a major form of molecular phenotyping[28]. As the most mature to date, single-cell transcriptomic tools will form the initial foundation of a draft taxonomy of cell populations across whole tissues.

Building on molecular techniques for amplifying minute quantities of mRNA, which first made it possible to analyze gene expression from single-cell lysates, rapid advances in single-cell handling — including flow cytometry-based sorting[29,30] and microfluidic partitioning[31,32] — enabled the study of entire transcriptomes in up to hundreds of cells. Most recently, innovations in DNA-based cellular barcoding using primer-coated microparticles have been combined with droplet microfluidics (Drop-Seq[5], InDrop[33]) or nanowell arrays (Seq-Well[34], CytoSeq[35]) to scale single-cell profiling to hundreds of thousands of cells at once. Commercialization and simplification of these techniques have enabled widespread adoption of high-throughput scRNA-Seq across labs and research areas[36,37]. In situ tagging, propagated through split-pool processing cells and nuclei, also allows extensive scale[37,38].

*DNA and epigenetics*. Single-cell profiling of DNA sequence and epigenetic state — including cell type–specific differences in genetics, gross DNA folding, DNA methylation, chromatin accessibility, and histone modifications — have also seen marked technological improvement. Detection of DNA mutations in single cells allows precise cell lineage reconstruction[39-43], whereas epigenetic measurements[19] provide a unique lens on cellular state, can be more robust across time than the more dynamic transcriptome, and often convey more detailed mechanistic information.

Some of these techniques already have combinatorial potential — for example, methods that use restriction enzymes to detect epigenetic marks generate unique overhangs, meaning that several can be used together to gain information about multiple epigenetic marks in the same cell[18]. Meanwhile, combinatorial indexing[38,44,45] and microfluidics[46,47] have significantly improved the throughput of several of these approaches, allowing, for example, for measurement of chromatin accessibility (scATAC-Seq) at a large scale[4,48,49]. Epigenetic and transcriptomes measurements may often be complementary, with one emphasizing more cellular features that are invariant on longer-term scales and the other able to capture more subtle, dynamic variations in cell state.

*Proteins*. Since proteins most directly carry out the cell's functions, their levels can provide more biologically meaningful information. (Their expression is also more buffered from intrinsic sources of noise than mRNA and, sometimes, may be more reproducible[50].) While some progress has been made in single-cell proteome profiling[51], it is not currently possible to profile entire proteomes at scale in single cells because of limitations in detection reagents and lack of direct amplification. Instead, antibodies to specific proteins can be multiplexed either by coupling to a suite of heavy metals[52,53] or nucleic acid tags[54,55], permitting the profiling of dozens of epitopes in individual cells by CyTOF. In addition, antibodies barcoded with DNA can be co-detected with cellular nucleic acids through sequencing, enabling simultaneous high-throughput profiling of whole transcriptomes and targeted epitope panels[56]. The availability and consistency of antibodies and similar capture reagents is a substantial hurdle. This is an opportunity for partnership with the Human Protein Atlas efforts[11,12].

**Key areas for development and optimization**

*Incorporating lineage information in single-cell -omics measurements.* Lineage information will help determine how cell states change with cell division and differentiation. The ability to add lineage information to single-cell measurements of accessibility, the (hydroxy)methylome, and the transcriptome should be within reach. Currently, this could be achieved in model organisms by methods that change the endogenous genome sequence by editing (e.g., with CRISPR/Cas9) or by integration of barcoded viruses. In humans, meanwhile, it is possible to infer lineage relationships between cells from the accrual of genetic mutations with each cell division[39-43]. Coupling single-cell genotyping across multiple loci with epigenomic and/or transcriptomic profiles — especially if performed at scale — could lead to promising noninvasive methods that could be highly attractive for human lineaging work within the HCA[19].

*Simultaneously recording nucleosome accessibility, methylation, and the transcriptome.* Methods are emerging to combine single-cell methylation profiling (by bisulfite sequencing) and transcriptomics or for coupling nucleosome or chromatin accessibility and transcriptomics[23]. Some are still in initial development, and none has been demonstrated yet at scale; the information trade-off between scale in one metric and collecting several orthogonal ones is being actively explored. Further incorporation of nuclear RNA into these measurements could provide a more direct understanding of the relationship between epigenetic marks and transcription initiation.

*Time-resolved multiomics.* Single-cell states, particularly the epigenome and the transcriptome, are highly dynamic, but we currently lack time-resolved measurements, particularly in vivo but also in vitro because genomic protocols are destructive. Imaging approaches for readers of epigenetic and transcriptional state in live cells could add to these capabilities.

**Approaches for evaluation and benchmarking**

*Variety of methods.* Often, large-scale "-omics" projects adopt a single technology, with the aim of minimizing technical variability in sample processing and allowing cross-sample comparisons. By contrast, we imagine that a useful long-term atlas will instead foster new technologies and use multiple methods to examine the same tissues. At the same time, however, there is great value in systematically applying a smaller number of technologies with consistent standard protocols *within* each. Analyzing samples across different technology platforms will allow us to cross-validate data, while remaining nimble to adopt disruptive technologies and able to engage a wider community of scientists; applying consistent, standardized protocols within each technique would reduce batch effects, facilitate comparison, and streamline HCA operations. To gather scRNA-seq data in the cellular branch, for example, we will use a combination of techniques, since some methods (e.g., droplet and nanowell) generate "broad" information, while others (e.g., SMART-Seq, CEL-Seq) provide depth. Each method will be applied in a stringent, standardized way, and additional valuable methods will be incorporated as they become ready.

*Evaluating technologies.* To assess the suitability of existing and emerging technologies to generate the HCA, we should focus not only on technical measures, such as yield, cost-effectiveness, robustness, reproducibility, and comparability, but also on how well each method identifies relevant data-driven biological structures in tissues compared with predefined biological benchmarks, whenever possible (**Section 2**). We caution that some technical metrics, such as per-cell/per-gene sensitivity, may have less final impact on data utility[57].

*Benchmarking and Rosetta Stones.* A multiplatform, multilab atlas must rest on extensive benchmarking tools and reagents — a set of vetted standards and protocols that can be used to calibrate new experimental and computational methods, as well as Rosetta Stone sample resources that allow comparisons. Benchmarks and Rosetta Stones are critical not only to enable data reconciliation between HCA labs working on similar tissues, but also as reagents for HCA end-users in the longer term. Possible benchmarking tools and reagents include barcoded mRNA capture beads, synthetic cells, spike-in reagents, a repository of frozen cells (primary or cultured) of demonstrated uniformity, and heavy-metal isotopes.

**Strategies to standardize, disseminate, combine, and deploy techniques in building the HCA**

Optimizing each method will require the coordinated efforts of the STWG and AWG. Once convergence has been achieved, members of the STWG will work in conjunction with a given method's developers to generate publicly available standard operating procedures for performing the protocol and for quality-controlling all necessary reagents against validated, centrally coordinated benchmarks. Notably, all HCA Projects (**Section 1** and **Section 7**) are required to openly release and share all protocols. We envision that initial testing of and iterative improvements to these standard operating procedures will be driven by SWAT teams, first in centralized, controlled facilities and then in remote locations. Dedicated resources will be required to actuate this vision. For both current and future technologies, the STWG and AWG will need to determine and then disseminate best practices for ensuring maximum consistency, ease of cross-comparison, and protocol optimization, carefully integrating in feedback from those actively engaged in the HCA.

ScRNA-Seq, snRNA-Seq, and (increasingly) scATC-Seq can all be used on a massive scale and, therefore, will be employed as the main profiling methods in the top of the Sky Dive in Phase I of the cellular branch of the HCA (**Section 2**). As the HCA evolves, we will add other techniques. For example, if human-cell lineaging is possible at scale (e.g., by massively parallel sc(DNA+RNA)-Seq, with targeted DNA sequencing[10]), it could be an excellent candidate for similar deployment. Similarly, scBS-Seq at scale could be used to provide highly informative methylation profiles. As additional profiling approaches are introduced, especially at the top of the SkyDive, one should take into consideration the distinct information content about cell types or states provided by it compared to the information conveyed by other existing profiles. The STWG and AWG will partner on such guidelines, along with the Biological Networks.

By commercializing single-cell sequencing and adapting it to multiple sample types—including fixed[58] and frozen[59] tissues, which are the default options for certain organs (**Section 2**) — the scientific community will gain widespread access to reproducible mRNA analyses, allowing independent research teams to generate atlases of each tissue, organ, and system. Those atlases will, ultimately, lead to the full HCA. This wide accessibility will allow scRNA-Seq data, as well as other single-cell "-omics" data, to serve as a common "information currency" that can be readily collected for every tissue under consideration.

We envision that, initially, transcriptomic maps could serve as a scaffold for multiplexed information. To achieve this, we could couple at lower levels of the Sky Dive (**Section 2**) specialized methods that provide depth but do not yet scale — for example, linking cellular transcriptional profiling with epigenetic or genetic measurements. These capabilities already exist in low throughput[20-22] and may soon extend to more cells through combinatorial indexing or microfluidic approaches[4]. Ultimately, if and when such technologies mature, the HCA effort should support the deployment of multiomic assays at the top of the Sky Dive across many cells and tissues, while considering the sequencing needs of the entire effort and the distinctive information provided about cell type or states that each measurement provides.

## SPATIAL TECHNOLOGIES OF INTACT TISSUES

A high-definition spatial representation of cell types, cell boundaries, neighbors or interacting cells, niches, and tissue contexts is necessary to generate a complete 3-D picture of the cells in the human body. We envision this data harnessed to a Common Coordinate Framework that other cellular data can be mapped onto. For instance, with the help of computational inference, multiplexed imaging of RNA signatures can give spatial placement for scRNA-Seq gene-expression profiles. Meanwhile, proteinaceous components of tissues (such as the extracellular matrix, which makes up nearly half of brain tissue) could be mapped by staining and imaging of relevant protein targets. Together, merged 3-D protein and RNA gene-expression profiles would provide a common reference for normal and pathological tissue structure.

Imaging technologies have long provided cellular, subcellular, and structural windows into tissues. Recently, several new technologies have been developed to facilitate the direct mapping of proteins and/or RNA into tissue slices at a highly multiplexed level. Other technologies for high-resolution tissue imaging, such as optical sectioning and aberration correction methods, super-resolution imaging, tissue expansion, and associated analyses, are all converging to enable near molecular-scale mapping of cellular states in tissues and to determine the key abstract relationships within and between tissues.

**Current technologies**

There are two broad classes of technology for highly multiplexed spatial analysis: approaches that measure known proteins or nucleic acids in a targeted manner — using antibodies or nucleic acid probes, respectively — at the cellular or subcellular level; and approaches that profile RNA using sequencing.

*Targeted approaches*. Targeted approaches can be further partitioned to mass spectrometry or fluorescence-based measurements. In general, mass spectrometry can enable targeted multiplexing using from fifty to hundreds of mass labels, depending on the resolution. The fluorescence-labeling systems are deployed in a variety of approaches to reach high-throughput, spatially resolved detection of proteins (50 to 70 species) and RNAs (hundreds to thousands of species).

*Mass spectrometry-based imaging.* "Conventional" mass spectrometers have long been used to detect metabolites, lipids, proteins, and other cellular material directly from tissues on slides. These approaches generally enable an unbiased measurement of all ionizable analytes of a tissue region within a defined mass range. However, they are currently limited to high-abundance analytes because of the physics of ionization and detection, and quantification of molecules is challenging[60].

To overcome these limitations, two targeted imaging methods based on mass spectrometry[61-63] use cocktails of antibodies attached to metal isotopes of a defined atomic mass as reporters to label epitopes on a tissue of interest[64]. Given the absence of auto-fluorescence, these imaging approaches are particularly applicable to formalin-fixed, paraffin-embedded (FFPE) tissue, which is a mainstay of pathology sample handling and storage because of its superb maintenance of tissue integrity.

One system, Imaging Mass Cytometry (IMC), uses a high-energy, high-resolution ultraviolet laser to systematically ablate a region of tissue, and the resulting tissue particles are analyzed by a inductively coupled plasma time-of-flight mass spectrometer, the mass cytometer, or CyTOF[63,65]. It currently allows routine measurement of 52 antibodies simultaneously at a resolution of 500 nm to 1,000 nm with comparable sensitivity to fluorescence microscopy (~50 molecules per pixel[63]). The Z depth of the image can be controlled via the thickness of the analyzed tissue section. A square millimeter of tissue can be analyzed in ~30 minutes, and the method has already been applied to 3-D tumor models at single-cell resolution and used to simultaneously image transcripts, proteins, and protein modifications.

The second system, Multiparameter Ion Beam Imaging (MIBI)[61], employs secondary ion mass spectrometry (SIMS) to result in near-single antibody sensitivity. In this technique, beams of ions ranging in width from 50 nm to 4,000 nm ablate the tissue at a Z depth of 1 nm to 250 nm. The technique can be performed to establish 3-D maps. Although MIBI uses isotope-labeled antibodies as tags, it can also employ lighter elements to simultaneously read carbon, nitrogen, oxygen, etc., in the tissue, which is useful for metabolic labeling. The current speed of the instrument is 1 minute for a 1-mm square tissue at about 1,000 nm beam width (~500 nm resolution). It is also possible to do a rapid (15-second) survey scan at low resolution (2,000 nm) and then reimage at high resolution (200 nm to 500 nm).

*Fluorescence-based imaging.* Two kinds of fluorescence-based targeted imaging approaches are relevant to building the HCA: multiplexed fluorescence in situ hybridization (FISH) for RNA imaging and iterative fluorescence imaging of protein epitopes.

Among RNA imaging approaches, single-molecule fluorescence in situ hybridization (smFISH) is the gold-standard method to determine the copy numbers and localizations of RNA molecules in individual cells[66,67]. Various multiplexed versions of this method exist, allowing simultaneous imaging of 10 to 1,000 RNA species in individual cells[68-72]. For example, multiplexed error-robust FISH (MERFISH) allows imaging and profiling for RNA species at the transcriptome scale, with high detection efficiency, by error-robust barcoding, combinatorial labeling, and sequential imaging; simultaneous imaging of 1,000 RNA species in single cells has been demonstrated with MERFISH[69]. MERFISH can create cell atlases of sizable tissues and has demonstrated the ability to measure several tens of thousands of cells, or several tens of square millimeters of brain tissue, in a single-day experiment[73,74]. A different highly multiplex FISH method is seqFISH, which uses color-based barcodes and sequential imaging[72]. SeqFISH has also recently demonstrated the ability to image a few hundred RNA species in single cells in brain tissues[75]. These methods can be combined with other technologies, such as recent advances in tissue clearing[73,76] or expansion[77,78], to improve the detection accuracy and the density of RNA that can be probed.

Serial fluorescence imaging approaches also allow multiplex analysis, especially of proteins[79,80]. CODEX (CO Detection of EXpression) is a recently developed rapid multiplexing method for imaging cytometry, which cheaply and easily allows the imaging of 50 to 100 cellular components with nearly any fluorescence microscope. Antibody cocktails are used to stain tissues, and DNA tags on the antibodies are revealed iteratively, to successively image 2 to 5 fluorophores at a time, based on the oligonucleotide sequence conjugated to the antibody. This approach has already been used to create initial maps of human immune organs at 400 nm resolution—the limit of light, with comparable sensitivity to standard fluorescence. With a regular microscope 30 hours are required to image 40 components in a 1-cm$^2$ region of a tissue with 12 Z stacks at 400 nm resolution, or 18 minutes for 1 mm$^2$. This is reduced to ~2 minutes with a confocal microscope (or for 1 Z plane).

<u>Untargeted profiling approaches</u>. Strategies that rely on genomic profiling in tissue are emerging, either using in situ sequencing or by spatial barcoding.

*In situ sequencing*. Transcriptome-scale RNA imaging of single cells has also been achieved with fluorescence in situ sequencing methods[81,82]. In these methods, cellular RNAs are converted into cross-linked cDNA amplicons and then sequenced in situ by imaging. This can be done in a multiplex, targeted assay[81], or without targeting as demonstrated in FISSEQ[82]. The detection efficiency of fluorescence in situ sequencing is currently lower than multiplexed FISH, making the quantification of low-abundance RNAs more difficult. However, unlike multiplexed FISH, fluorescence in situ sequencing does not require preselection of genes, thus allowing untargeted transcriptome measurements in single cells.

*Spatial barcoding methods* allow visualization and quantitative analysis of the transcriptome with spatial resolution in 2-D tissue sections. In one approach, Spatial Transcriptomics (ST)[83], histological sections are placed on glass slides with arrayed oligonucleotides containing positional barcodes to impart the spatial barcodes prior to RNA-Seq. Although ST does not generate single-cell data, it captures tissue transcriptomes comprehensively within each tissue voxel, thereby providing a sensitive standard for tissue composition that is complementary to tissue dissociation methods. We expect additional methods to emerge in this area.

**Key areas for development and optimization**

*Mass spectrometry-based imaging* can improve in sensitivity, speed, resolution, and in the number of molecular species that can be assayed in multiplex. *Sensitivity* can improve through higher detection of ions to the range that will enable single-molecule detection. *Speed* should improve for both laser and ion beam ablation to reduce the analysis time. The extent of *multiplex* can improve by adapting metal nanoparticles to greatly increase the number of simultaneously measurable mass channels, up to 100 or so. *Resolution* should improve for both methods through more sensitive instrumentation and nanoparticle reagents (IMC) and smaller tip size and antivibration systems (MIBI). With such improvements, methods like MIBI can be deployed at the scale and speed required for the HCA. Additional advances may allow detailed mapping of multicomponent structures and RNA in 3-D using ion-tagged oligonucleotide probes instead of fluorophore methods.

*Fluorescence-based imaging.* Pending improvements to multiplexed FISH methods make it conceivable to simultaneously measure several thousand or more RNA species in single cells. In addition, measurement throughput — the number of cells measured per day per instrument — will likely be increased by orders of magnitude, which would be of particular use to the imaging of very large tissue volumes for the HCA. Meanwhile, the detection efficiency of fluorescence in situ sequencing methods is also being improved. Notably, measuring more RNA species at high-detection efficiency will require the resolution of RNA molecules that are in close proximity to each other — in some cases, beyond the diffraction limit. We envision that it will be beneficial to combine these RNA profiling methods with super-resolution imaging methods, either by optical approaches or by tissue expansion[77,78]. Finally, highly multiplexed RNA and protein imaging of the same samples would provide simultaneous information on these complementary aspects; the compatibility between RNA imaging and immunofluorescence-based protein imaging suggests that such combination should be feasible.

**Approaches for evaluation and benchmarking**

Several key aspects require evaluation and benchmarking in the context of spatial approaches: validation of reagents; comparison to gold-standard methods; comparison between related techniques; and relation to analyses on dissociated cells.

*Reagents*. Spatial approaches that currently rely on probes require validated reagents. In particular, antibody reagents must be validated for their performance in relevant specimens. For example, more than 150 antibodies and reagents to determine absolute copy numbers of a given epitope have been validated for FFPE, and IMC has generated reference samples for 50 markers[84,85]. Such efforts will remain ongoing and should lead to an HCA-wide resource. This is also an opportunity to collaborate with the Human Protein Atlas, an effort to measure the spatial distribution of the human proteome at a subcellular level (http://www.proteinatlas.org/).

*Benchmarking to gold standards*. It is important to validate the quantitative measurements made by multiplex or genomic spatial methods relative to well-accepted gold standards, such as staining for individual proteins or smFISH. Not every measurement requires validation, but new methods, new reagents, and their application in new tissue should be validated.

*Comparison between related techniques*. As is the case for single-cell genomics, the HCA consortium does not expect to converge on a single method for all spatial analysis, but rather to have the same tissues analyzed by multiple methods, with each method having established standard operating procedures and protocols. However, this approach relies on careful benchmarking of multiple techniques on matching samples — for example, consecutive sections of the same specimen. The compared techniques could be of the same class (e.g., multiplex in situ hybridization) or of different classes (e.g., MERFISH, FISSEQ, and Spatial Transcriptomics). The ability to preserve tissue prior to analysis will facilitate benchmarking as technologies evolve through the prospective banking of Rosetta Stone samples; maintaining such specimen collections should be an important goal of the consortium. Because these technologies are less mature than many sequencing techniques, additional steps will need to be taken to fine tune them.

*Comparison to measurements of dissociated cells*. Tissue dissociation often introduces biases in both cellular composition (as fragile cells are disproportionately lost) and intrinsic cell states (through the stress of dissociation and even RNA degradation). Conversely, targeted in situ measurements are limited by the set of probes they recover, and spatial genomics to date either does not have single-cell resolution or is limited in the number of recovered transcripts. Comparing the same specimens between the two strategies will help assess their respective strengths and limitations, help optimize each, and inform how the data they generate can be integrated between the cellular and spatial branches of the atlas. This effort would be assisted by computational integration, in collaboration with the AWG.

**Strategies to disseminate, combine, and deploy techniques in building the HCA**

Today, most spatial technologies are performed only by a specific lab or a small number of labs, chiefly because most require specialized tissue-processing procedures and some require specialized equipment. For methods that require complex and expensive instruments, we should consider setting up a few instruments in centralized locations for centralized data collection and for training HCA community members in their use. For other methods that are considerably easier to use and do not require sophisticated equipment, a SWAT team from the labs that invented and/or developed the technologies could travel and disseminate them across multiple HCA sites to ensure broad adoption. Conducting hands-on workshops could also aid in widening use of these methods, as would robust software for data preprocessing. Finally, most of these techniques are not deployable at genomic scale yet. Therefore, the developer labs could optimize and scale them to make them available as systematic pipelines.

*Well-chosen pilots will help drive this effort.* Absent vast increases in resources or technical development, the initial focus of the HCA should be a select number of target tissues important not only for their biology and clinical relevance but also for representing distinct tissue architectures and properties and thus posing diverse technical challenges. Such tissues could include kidney, pancreas, colon, liver, immune tissues (spleen, tonsil, lymph nodes), breast, prostate, and the brain.

# 4. DATA ANALYSIS

Computational methods and analysis will form the backbone of the HCA. They will be required for design of data collection, for low-level data processing and normalization, for data organization and for data interpretation, all of which will require the design of novel algorithms, statistical models, and intuitive visualization strategies. Consequently, a wide range of innovative, rigorous, and appropriately tested computational methodology will be critical to make the HCA a success.

Computational methods are needed to integrate data generated using a wide variety of (noisy) technologies to create a comprehensive reference catalogue of all human cells. This reference catalogue will not only define individual cell types and their spatial architecture but also map by inference their lineage relations, regulatory circuitry, and the interactions between them. Consequently, key tasks for the HCA's computational pillar will be to generate strategies for ascertaining the best existing methods, identify and prioritize problems that lack solutions, and develop new methods as appropriate. Additionally, each emerging technology and each biological question will likely require distinct computational solutions, from guidance through initial quality control, benchmarking, and optimization of new technologies to methods that yield new biological insights.

We distinguish three key layers of computational methods.

***Basic data processing and normalization*** of the raw output of the different technologies. For example, it will be necessary to convert reads generated from single-cell sequencing technologies into normalized cell-by-molecules count matrices. In addition, experiments should be designed to allow proper modeling of batch effects. When considering data collected using multiplex imaging technologies, it will be necessary to establish quality-control metrics and to register and segment the raw files.

***Data organization in the atlas.*** This includes discovery and taxonomy of the different cell types and their relationships, including formally addressing the fundamental question "*What is a cell type?*" The Atlas will be organized using multiple complementary maps and coordinate systems: spatial and molecular systems followed by functional, physiological, and lineage-derived maps. A key challenge will be to generate these maps and to understand how they relate to one another. To do so, we will need to develop methods to link different layers (including data generated using very different technologies) and understand common features. This will shed light on the relationship between phenotype and function.

***Queries, visualization, and analysis methods for exploiting the atlas.*** These include both simple queries and visualizations and more complex computational analyses. First, *simple queries and data visualization* will be required to make the data accessible to the biological community as quickly as possible. This will facilitate identification of markers that can be used to isolate selected cell types for further study, as well as allow interrogation of cell type-specific and spatially localized patterns of gene expression, similar to a BLAST query for cell phenotypes. Finally, there will need to be a suite of sophisticated methods that utilize the Atlas for *more complex computational analyses*. For example, single-cell data allow circuits and regulatory networks to be inferred at the cellular and tissue level. Additionally, analysis methods built upon approaches honed in Genome Wide Association Studies (GWAS) will reveal how genetic variants alter cell-type composition and spatial location. Ultimately, methods that harness the HCA to elucidate mechanisms of disease by projecting cells from new cohorts of disease samples onto the reference atlas will also be vital.

## ANALYSIS WORKING GROUP (AWG)

To achieve these goals, we are building a vibrant community of computational biologists, who will interact with one another at regular meetings, to stimulate the development and critical evaluation of appropriate methods. To provide structure, the Analysis Working Group (AWG) has been convened as a small steering committee by the HCA Organizing Committee (**Section 7**). The AWG organizes working groups, data-analysis teams, meetings, jamborees, and recommends experimental designs, analysis frameworks, and best practices for different aspects and needs of the HCA, from data collection to the Data Coordination Platform (DCP).

*Guidance to the DCP*. A key goal of the AWG is to formulate computational challenges and to establish "benchmark data sets" and metrics that facilitate comprehensive comparison of a wide variety of strategies. The AWG organizes activities to help guide the DCP and the DCP Governance Group (DCP GG) and publishes best practices in data processing and analysis. In some cases, such as quality control and basic data processing (e.g., from sequencing reads to a gene count matrix or from raw image to cell segmented), the AWG will decide on a single best protocol to be implemented in the analysis pipelines of the DCP, while in others, such as trajectory detection and network analysis, the AWG will foster a wide range of methods that are designed for different goals.

*Guidance to and participation in biological networks*. An important role of the AWG is to provide advice and guidelines for experimental design. This will ensure cost-effective and consistent collection of data that can be both integrated into the atlas and used to address key biological questions. To make this process as seamless as possible, we envision that a computational biologist who is part of the data-analysis team be part of the relevant working group for each organ or technology. This representative will disseminate the relevant computational methods to the group and liaise with those who best understand the specific biological challenges and questions faced by each community. Then, they will bring these questions back to the working group and help to define the computational methods that need to be developed. Finally, the representative will participate in the planning of data collection, thus ensuring that the experimental design is well suited for computational analysis and statistically powered to address the project goals.

*Building, engaging, and training the HCA computational community*. To facilitate the participation of a large group of computational scientists and establish a vibrant analysis community, the AWG will hold regular scientific meetings, which will allow members of the community to present updates on new methods through talks or scientific posters and to formulate and pose new computational challenges in breakout sessions and informal discussions. Analysis jamborees are focused on a specific question (e.g., the best method for removing doublets from scRNA-Seq data), with the goal of collectively reaching a conclusion by the end of the meeting. Best practices — and transparent reporting on how these were derived — will not only be disseminated to the entire HCA community but also made public and readily available to the global scientific community. Gold-standard open-source data sets and code will be provided as well. Jamborees also serve to train young scientists and foster a collaborative community across labs.

The first AWG jamboree was held in August 2017 at the European Bioinformatics Institute (EMBL-EBI) in the U.K., and involved 48 young scientists from 19 computational biology labs, who worked together intensively in multi-lab teams to solve open challenges facing the HCA. These challenges, along with problem definitions, unique benchmark data sets, and necessary infrastructure, were prepared by 10 lead PIs, nine of whom attended the event. Three additional software engineers provided computational support for the event. Enabled by the collaborative nature of the event, participants tackled six problems:

- identifying ambient RNA and cell events
- removing cell doublets in scRNA-Seq data
- identifying dead and damaged cells
- normalization (and batch effect correction) between samples
- comparing clustering strategies; and
- defining experimental design strategies.

Substantial progress has been made on most tasks, whereas some (normalization and clustering comparisons) were more challenging and will be further addressed in future jamborees. These advances will likely result in publications. The postdocs and graduate students found the jamboree valuable to their training and reported a high degree of satisfaction with the event in a post-jamboree survey. Additionally, the jamboree helped foster further interactions among labs. Future jamborees will tackle both other algorithmic tasks and "live data analysis," focused on a specific organ or biological question.

## SELECT KEY CHALLENGES

The computational challenges faced by the HCA are numerous, rapidly evolving, and encompass a diverse range of areas. We outline a few immediate challenges, but note that there are many other areas where methods development will be vital. Through the lifespan of the project, new challenges will continue to emerge as new technologies and biological questions arise. The AWG will help convene and guide the community to identify and articulate these challenges, formulate the problems into a more mathematical framework, define benchmark datasets and engage the wider computational community to address them.

***Processing and normalization of scRNA-Seq.*** Data collected from massively parallel scRNA-Seq are rapidly accumulating. However, like most experimental techniques, these powerful and cost-effective recent technologies also include noise, batch effects, and other biases. It is critical to help select the best experimental approaches and best processing pipelines to increase signal and reduce noise in the resulting data so that the wider community can best exploit it.

A key challenge to address in this context is to establish a "best-practices computational pipeline" that can support initial processing, quality assessment, standards, and normalization for the technologies selected for the HCA, as well as provide rigorous quality metrics to guide the selection of these technologies. The pipeline will need to include quality control, removal of erroneous and low-quality reads, and the removal of barcodes that represent debris, dead cells, doublets, and ambient RNA. Establishing the most appropriate strategy will require rigorous metrics and computational comparisons, exploiting both real and simulated data. Of critical importance will be the creation of benchmark data sets that can form the basis of comparison.

The primary output of this pipeline will be a matrix of cell-by-gene expression values that can be leveraged for downstream analyses. We envisage the existence of one cell-by-gene (or transcript) count matrix that represents actual observed transcripts, corrected for any errors in the raw data. Following this step, different data- normalization procedures will serve subsequent downstream analysis tasks. Normalization will be a major challenge when integrating data collected across different tissues, technologies, and labs or centers. To support data-driven experimental design, power analysis methods should be developed and incorporated for assessing coverage and completeness of cell states and/or types that have been sampled.

***Defining a cell type and organization of the atlas.*** A primary goal in creating the HCA is to define cell types and to characterize their biological roles. A data-driven approach to this challenge takes single-cell measurements and provides a grouping of cells into distinct subpopulations (i.e., clusters). Specifically, each cell is represented as a point in high-dimensional space (molecular phenotype), where clustering is tantamount to finding dense regions of similar cells in this space. Numerous computational methods exist for this task, but their relative merits have not been fully assessed. As spatial data accumulate, methods should be developed that integrate spatial and molecular data toward more accurate cataloging of cell types.

*Cell type definition*. Before an optimal clustering algorithm can be selected, the notion of a cell type must be formally defined. Currently, a cell type implies significant phenotypic and functional stability over time, achieved perhaps through epigenetic regulation. While a cell type suggests a discrete entity, recent data sets are highlighting that many presumed cell types actually form part of a continuum. For example, a T cell can exist in different states (mitotic, activated) while maintaining its identity as a T cell, and T cell activation is considered as a continuum, rather than as two discrete modes. Stable cell types encompass many states and, therefore, exist in a diverse but restricted set of the high-dimensional phenotype space. Epigenetic data modalities such as ATAC-Seq will help distinguish cell types from cell states (**Section 2** and **Section 3**).

*Similarity metrics*. A key computational challenge when identifying cell types is determining appropriate similarity metrics between cells. When are two cells "the same"? How do you determine a similarity metric that best serves biological questions, such as similarity in function or lineage? Feature selection will play an important role in any similarity metric and will likely depend on the envisioned tool, such as clustering, a BLAST-like query and search tool for cells, or comparing healthy tissues to diseased tissues. Benchmark data sets that include morphology and spatial context from imaging data from the spatial branch will be pivotal in guiding the development of good similarity metrics based on molecular profiles from the cellular branch.

*Taxonomies*. Consequently, clustering algorithms need to find salient features of more stable cell types and distinguish cell-type variations from those associated with physiological states (such as cell cycle), as well as to model continuous trajectories where these exist. Thus, new computational methods will be required both to discover types and to better classify cells. As these develop, they will ultimately refine the very concepts of cell type and state. Additionally, understanding the relationship between cell types and establishing new cell-type taxonomies will be critical. To achieve this, the AWG will coordinate with HCA subgroups to establish a series of benchmark data sets that can be used to determine effective similarity metrics and assess the performance of different methods.

*Curation.* Although these tools are useful in refining our understanding of cell types, a computational definition alone does not automatically convey biological meaning. However, given the vast amount of data that will be generated for the HCA, manual curation of cell-type identity may become extremely challenging. To this end, the AWG will work with relevant HCA subgroups to generate an automated way of assigning putative biological identity to groups of cells, first by using literature text mining approaches based on the genes expressed in each cell type, and later based on new technologies for functional interrogation of cells. This will facilitate comparison of cell types across tissues and enable assembly of a meaningful taxonomy of cell types.

*Spatial dimension: Imaging tissues and modeling spatial architecture.* Multiplex spatial technologies, such as MIBI and MERFISH (**Section 3**), are advancing at a rapid pace, enabling spatial mapping of fine-grained cell-population structure at unprecedented resolution. With new technologies, however, come new computational challenges, both at the level of preprocessing (e.g., faithfully calling detection of a specific protein or RNA at a specific location, based on the raw signal specific to each technology), image processing (e.g., cell segmentation), alignment of images on a Common Coordinate Framework (CCF), and analysis (e.g., identifying cell-cell interactions and other meaningful expression patterns in tissue architecture).

Several challenges must be tackled, including selecting the optimal set of spatial technologies that will contribute to the atlas; developing preprocessing pipelines for each; defining a CCF on which each image can be aligned and meaningful comparisons across samples from different individuals can be made; segmenting cells within each image; determining for each cell its appropriate cell type based on observed marker expression; and imputing unobserved genes and proteins based on integration with genome-wide data sets, especially from the cellular branch.

Methods must be developed to select the most informative markers (genes or proteins) that can best resolve cell types and their architecture and derive a reliable mapping when combining the spatial branch with the cellular branch (e.g., scRNA-Seq, snRNA-Seq and scATAC-Seq). This integration would allow the genome-wide exploration of spatial architecture of molecules in their histological context, refine the characterization of cell identity by context or neighborhood, and help characterize interactions within and between cell types. Moreover, with a map of cells, their expressed molecules, and spatial boundaries, network-learning approaches can be developed to detect signaling networks within and across cell boundaries.

Because of the enormity of this challenge and the relative lack of existing methods for spatial analysis, the AWG will work to accelerate the development of the necessary formalisms and data-analysis algorithms through publication of benchmark data sets, collaborative workshops, jamborees, and other forms of community engagement.

*Temporal dimension: cell-state transitions, pseudo-time, and lineage tracing.* In addition to a spatial map, the atlas should describe how cell types are derived developmentally and outline possible transitions between cell types. We aspire to obtain for humans what John Sulston and colleagues developed for *C. elegans* — a complete and detailed map of all cell types in the body and the developmental relationships between them. This includes both cell-lineage relations (how extant cells are related through cell divisions) and cell-fate maps (describing the types of cells from which an extant cell is derived and its potential to yield other types).

Single-cell data has highlighted the continuous nature of cellular phenotypes. Consequently, there has been a proliferation of algorithms that attempt to order cells collected in a single snapshot in so-called pseudo-time. However, these methods are still in their infancy, and substantial work remains to improve these approaches and understand when they correctly represent temporal dynamics. Success would enable the mapping of developmental processes at high resolution, thereby helping to infer the molecular logic underlying cellular decision-making.

Each cell phenotype has some probability of transitioning to another state. Because trajectories should provide dynamic information to elucidate cell fate decision-making, we need to work within a probabilistic framework, with the goal of inferring probabilities and possible paths of transition and intermediate states among all cell types. To assess the performance of these inference algorithms, it is important to develop benchmark data sets that enable the comparison and evaluation of different computational approaches.

One strategy for generating such benchmark data sets is to use emerging genome editing approaches for large-scale lineage tracing (e.g., MEMOIR[86] or GESTALT[87]) in either organoids or mammalian model organisms (e.g., mouse). Because of the current limitations of these techniques, new algorithms are required to correctly interpret the output. In humans, we could also use endogenous lineage information, such as somatic mutations in DNA, SNVs, and methylation epialleles for lineage tracing, although these may be challenging when mutation rates are very high. Combining lineage-tracing (in organoids, model organisms, or endogenously), together with validated pseudo-time trajectory algorithms in primary human tissue, would ultimately derive a temporal cell lineage and cell-fate map for humans.

*Making the HCA accessible: portals for visualization and queries.* The atlas is only as good as it is generally usable by a broad group of biomedical researchers. Therefore, it is essential to design early portals for querying and visualizing the data that are intuitive for the noncomputational user and update them frequently. Queries and search tools must be developed for cell types and tissue architectures, analogous to the way in which BLAST enabled queries of databases to find similar sequences. In addition to highlighting similarities, methods are needed that highlight differences (between cells from different healthy individuals and between cells from healthy and disease individuals). Because of the high dimensionality and complexity of the data, it is important to choose dimensionality-reduction and feature-selection techniques that capture the most salient aspects of the data. These techniques are likely to differ between biological domains and questions. Finally, since visualization is key, new methods must be developed to interactively explore the many layers and dimensions of the atlas in a way that facilitates interpretation and biomedical discovery.

# 5. DATA COORDINATION PLATFORM (DCP)

The Human Cell Atlas will contain petabytes of data on billions of cells and tissue sections across multiple modalities used by hundreds of labs around the world. A project of this scale and complexity demands an open, modular, and extensible approach to coordinating, standardizing, and sharing data.

A suitable data coordination platform will need to

- provide as many researchers as possible with simple, open, and immediate access to all of the data and some standardized, derived results;
- be built out of modular, open-source components that interact via standards and protocols, rather than monolithic systems, allowing different components to evolve independently and to be reused by other projects; and
- be extensible to integrate and support a diversity of analyses, visualizations, and other components developed by the scientific community.

Together, these three properties will maximize the opportunity for downstream innovation in how the data is analyzed and used.

Over the past year, a team of engineers has worked to formulate an architecture for HCA that aims to satisfy these three properties and could support the scientific objectives of the project. Software engineers and computational biologists, working at several research institutes, including the European Bioinformatics Institute (EMBL-EBI), the Broad Institute, the University of California Santa Cruz (UCSC), and the Chan Zuckerberg Initiative, have now started to implement it. (We stress that this is just one potential solution that satisfies the principles described above.) The following summary of this plan is derived from a more detailed document (**Appendix II**).

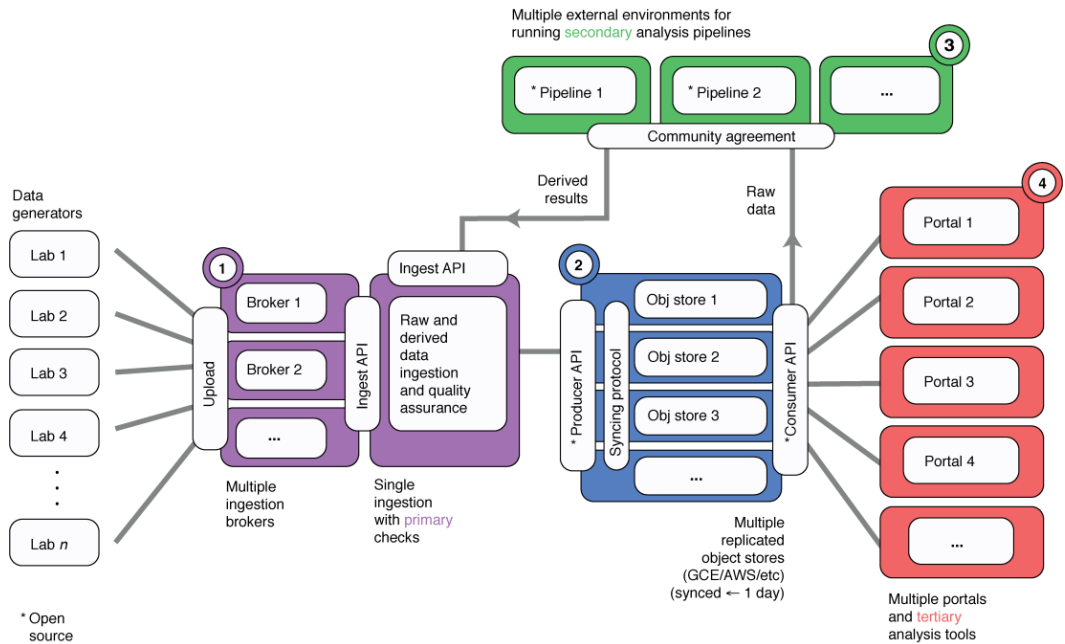

**Figure 6. The Data Coordination Platform.** A proposed open-source Data Coordination Platform, including (1) a data ingestion service (*purple*), (2) a synchronized data store replicated across multiple clouds (*blue*), (3) a collection of secondary analysis pipelines for basic data processing (*green*), and (4) a collection of tertiary portals for analyses, visualizations, and forms of data access (*red*).

The proposed architecture has four key components, each acting as a well-defined module interacting with the other modules through APIs (**Figure 6**). We refer to the four components as (**1**) the ingestion service, (**2**) the synchronized data store, (**3**) the secondary analysis pipelines, and (**4**) the tertiary portals. The proposed sequence of data flow through the system begins with ingestion of raw data and metadata by a broker into the ingestion service, followed by primary quality assurance, including basic file integrity and metadata syntax and schemas. Data and metadata are then deposited into the data store and replicated across multiple clouds. Data are processed via the secondary analysis pipelines to generate intermediate derived results and quality-control scores, which are then submitted back into the ingestion service and deposited into the data store. Basic access to both raw data and derived results are provided by the data store consumer API, on top of which a diversity of tertiary portals can be built.

Consistent with the values of the HCA, the platform will be open-source and the data will be open-access, with the ability for any third party to provide tertiary portals through APIs and to readily clone the software and data platforms for use.

Here are details about the four components.

*Ingestion service.* The primary role of the ingestion service is to form a single point of entry for all HCA data, including both raw assay data and metadata for projects, experiments, and samples (e.g., sequencing reads, microscopy images), as well as standardized analyses and quality-control metrics, which result from running vetted analysis pipelines, and appropriate provenance describing those analyses. The ingestion service will both ingest data and perform basic quality-assurance checks, including file integrity, metadata syntax and schemas, and consistency with ontologies and controlled languages. Metadata and user interaction will be developed, vetted, and agreed upon by the HCA community, with guidance from the HCA MDWG, as represented by a governance structure from the DCP GG under the HCA Organizing Committee (**Section 7**). Researchers will submit data through one of several "data brokers" that act as links between labs and a single ingestion service API. Brokers could include user-facing Websites or other Web or programmatic services and may target geographical regions or provide domain-specific data handling.

*Synchronous data store.* After ingestion and quality assurance, data will be deposited into the synchronized data store. This multisite, replicated storage system will contain all raw data, metadata, and derived results from agreed-upon standardized analyses, materialized and stored as flat files. It will ensure simple and direct data access for downstream consumers and allow both download of all the data and computing on the data directly in the cloud. Replicating the data across multiple cloud-storage solutions allows researchers to use different cloud platforms to analyze the data. It will be wrapped as a producer API (to put data into the store) and a consumer API (to consume data from the store). It will include protocols for synchronizing data across providers and a system for notifying users of content changes.

*Secondary analysis pipelines.* Although researchers will have access to all raw data in multiple cloud platforms, the majority of researchers will want to work with intermediate derived results that require at least some initial preprocessing, such as alignment and de-multiplexing of sequencing data to generate a gene-cell table; or spot detection, cell segmentation, and barcode decoding of imaging data to generate a spatial map of gene expression. The platform will include a secondary service that can automatically execute high-throughput analysis pipelines, using best-in-class algorithms for each data type, as developed, vetted, and agreed upon by the computational community, with guidance from the AWG, as represented by a governance structure from the DCP GG under the Organizing Committee (**Section 7**). It is critical that these analysis workflows are portable and reproducible, built out of open-source components, and publicly available so that the execution service can automatically run them across multiple cloud environments while researchers can also run the same analyses locally on their own data. The notification systems provided by the data store will allow the execution service to automatically rerun as new data arrive. The results of the secondary analyses, including derived analysis outputs and quality scores, are ingested and deposited back into the data store so consumers can access them alongside the raw data and metadata. This portion of the system in particular will depend on algorithms contributed by the computational community and rely on HCA governance to ensure that all selected algorithms satisfy best practices and standards.

*Tertiary portals*. To ensure good separation of concerns, the data store will provide only simple forms of data access through its consumer API. But the biological and computational communities will surely require a wide and ever-growing variety of forms of access, including rich indexing, slicing, and searching based on metadata, a diverse set of algorithms and methods for analysis, interactive data analysis and visualization, and ways to query a new sample against the complete data set. These will be provided through a collection of tertiary portals, each of which provides different functionality, analyses, and visualizations. To make sure creativity and innovation are fostered and encourage as diverse a collection of methods and portals as possible, the tertiary portals will not be generally assigned a governance committee, although HCA, through its governance structure, may choose to "bless" a particular tertiary portal for its official data release or for the community of a specific sub-atlas. Results from the portal community are expected to be less standardized because they will address differing community needs and approaches, so they will not be deposited back to the DCP. They will, however, be accessible directly from portals. The concept of portals also permits a wide variety of services, ranging from simple (a set of links to download slices of the data) to complex (an interactive visual Web-based browsing experience).

Together, these four components form a complete functioning data coordination platform. At the same time, they are sufficiently independent to allow parallel development and to allow each component on its own, or in combination, to be reused in other projects.

*Governance*. The Organizing Committee governs the Data Coordination Platform (DCP), including making all policy decisions concerning the DCP, approving the overall plan for the DCP, and ensuring the plan's successful execution by the major developers of the DCP. The OC will establish and appoint a **DCP Governance Group (DCPGG)**, which will report to the OC, to oversee the implementation of these policies, by providing guidance and making decisions concerning certain key topics, including definition of data manifest; official analysis pipelines; required metadata to reflect data collection standards; common coordinates framework; and any formal "release portal." The DCPGG will be led by two OC members; will include at least one member from each of the AWG, MDWG and CCFWG; and will include at least three additional experts from the community.

The OC will convene, on a quarterly basis, a DCP Coordination Meeting (involving the DCPGG, the major developers of the DCP, and others, as appropriate) to review progress and assist the OC in developing policy.

This is part of the governance for the HCA, as stated in **Section 7** and available in **Appendix I**.

# 6. BIOLOGICAL SYSTEMS

The HCA will be composed of atlases of individual biological tissues, systems, and organs, constructed by teams that include experts in each tissue or system.

The human body comprises the following systems (**Table 1**): central nervous, peripheral nervous, lymphoreticular, immune, urinary, respiratory, reproductive (male and female), hepatopancreatico-biliary, gastrointestinal, endocrine, skin, musculoskeletal, cardiovascular, and breast.

**Sub-atlases.** We will pursue individual tissues, organs and systems as sub-atlases, following the prioritized ranking in **Table 1**. In each sub-atlas, to ensure rapid early progress toward the first draft, we will first initiate **Pilot Networks, each representing one tissue or system of interest.** The networks aim to demonstrate the feasibility of profiling human tissues, while also generating useful data to explore the biology of those tissues. They will test and help to optimize all aspects of the HCA pipeline, from tissue acquisition through data processing and interpretation, and will grow into HCA Flagship Projects.

There is an overlap between sub-atlases, because of the natural intersection between systems, tissues, and organs. For example, immune cells in the skin will be profiled in both the Immune Cell Atlas and the Skin Cell Atlas. This is a remarkable opportunity for data validation and replication and will strengthen the overall quality and utility of the atlas.

The efforts will also inform each other technically: For instance, when multiple tissues can be taken from the same donor, the sub-atlases ideally will coordinate to ensure that the tissues are processed with the same technology in the same place. This will address inter-individual and technical batch effects and provide a framework to integrate across atlases.

While each sub-atlas focuses on a different tissue or system, they all share key principles:

- Adherence to general HCA principles and to the overall scheme for the draft atlas (**Figure 3**; **Section 2**).
- Sharing of experimental protocols for tissue procurement and processing for the sub-atlas.
- Collaborative, multicenter analysis

| Organ system | Organ / tissue | Rank |
|---|---|---|
| Central nervous | Cerebrum | 1 |
| | Cerebellum | 1 |
| | Brainstem, pituitary gland, pineal gland | 1 |
| | Spinal cord | 1 |
| | Eye | 3 |
| | Ear | 3 |
| Peripheral nervous | Nerves | 3 |
| Lymphoreticular | Spleen | 1 |
| | Bone marrow | 1 |
| | Lymph nodes | 1 |
| | Thymus | 3 |
| Immune system | Distributed / resident in other tissues | 1 |
| Urinary | Kidney | 1 |
| | Ureter | 3 |
| | Urinary bladder | 2 |
| | Urethra | 3 |
| Respiratory | Nasopharynx | 2 |
| | Trachea and bronchi | 1 |
| | Lung | 1 |
| | Pleura | 2 |
| | Diaphragm | 3 |
| Female reproductive | Ovary | 2 |
| | Fallopian tube | 2 |
| | Uterus | 2 |
| | Cervix | 3 |
| | Vagina | 3 |
| Male reproductive | Prostate | 3 |
| | Testis, epididymis, vas deferens, ejaculatory ducts | 2 |
| | Glans penis | 3 |
| Hepato-pancreatico-biliary | Liver | 1 |
| | Billiary tree | 3 |
| | Gallbladder | 3 |
| | Pancreas, exocrine | 1 |
| | Pancreas, endocrine | 1 |
| Gastrointestinal | Tongue and oropharynx | 2 |
| | Salivary glands | 2 |
| | Oesophagus | 1 |
| | Stomach | 2 |
| | Small bowel | 1 |
| | Large bowel (caecum, appendix, ascending, transverse, descending, sigmoid, rectum, anus) | 1 |
| Endocrine | Thyroid gland | 3 |
| | Parathyroid gland | 3 |
| | Adrenal gland | 3 |
| Skin | Skin, andexa, subcutanoues fat | 1 |
| Musculoskeletal | Muscle | 2 |
| | Bone | 2 |
| | Cartilage | 3 |
| Cardiovascular | Heart | 1 |
| | Aorta and other major vessels | 3 |
| Breast | Breast | 2 |
| Organoids | e.g., gut, kidney, lung, brain and retina | 1 |

**Table 1. Organs of the Human Cell Atlas.**

of all data collected in a sub-atlas.
- Engagement of computational biologists in all stages, from inception and design to analysis

Nevertheless, each sub-atlas may require modification of the overall shared scheme, with specific choices of parameters for age, gender, diversity, sites of collection, or tissue sampling from large organs as appropriate for its biological context. For example, in one setting a smaller number of individuals is analyzed across a large number of regions in one organ, whereas another system requires more individuals, each studied at lesser depth.

Below, we provide more detail on several representative biological areas and their efforts. Additional biological networks are working to provide similar synopses for their organs, systems, and tissues in the living White Paper.

## NERVOUS SYSTEM

The brain is the most structurally complex organ in the body. In particular, the human brain is distinguished by its expanded neocortex[88,89], prolonged developmental plasticity[90], and the unusually complex mental functions it serves. We still have much to learn about the brain's fundamental structure and function, but we are hampered by its size and complexity. Despite this, a comprehensive understanding of the cellular composition of the brain is within reach. The recent advent of transformative technology and analytical methods will make this possible — but we will be significantly challenged by the brain's unparalleled diversity of cell types, spread over hundreds of discrete anatomical regions.

**Why a nervous system atlas?**

Understanding how the brain works requires knowing its parts, and surprisingly little cellular-resolution information is available for the human brain. There is, therefore, enormous potential to dramatically accelerate basic and translational research through an atlas that systematically characterizes, classifies, and models neuronal and non-neuronal brain cell types and makes these data and tools available as catalytic open-access community resources. These cell types could then be mapped onto a 3-D Common Coordinates Framework (**Section 2**) of each brain region, creating a standardized nomenclature for neuroscience and building a bridge to functional imaging studies.

Prior efforts to create human and model organism whole-brain transcriptional atlases using microdissected brain regions[91-93] provide a good template for experimental strategies. For example, the Allen Human Brain Atlas (AHBA) profiled six neurotypical adult human brains using DNA microarray analysis of macro- to laser-microdissected regions spanning the entire brain and mapping the sample locations to MNI coordinate space[91]. More recently, the BRAIN Initiative Cell Census Network (BICCN) pilot has shown the promise of single-cell profiling for neural-cell classification[94]. Indeed, existing initiatives, such as the BICCN, are committed to a cell atlas of the mouse and human brain, and key members of the initiative are participants in the HCA.

**What are key considerations for a nervous system atlas?**

An ultimate atlas of the brain should include diverse molecular profiles (transcriptomic, epigenetic, and proteomic), anatomical properties (morphology, cell ultrastructure), connectivity (local, long-range), and functional properties (electrophysiology).

Several key design principles are essential for the success of the atlas and need to be carefully established in the pilot phase.

*Anatomical sampling*. Because the brain is divided into functionally distinct regions, each with its own unique cellular variations, fine spatial dissections are essential to capture cellular diversity. Thus, it will be critical to agree on a sampling strategy that will both reflect previously determined distinctions and be able to uncover novel ones. Within each brain region, we will rank regions by the extent of cellular diversity and selectively enrich rare cell types. We will combine uniform sampling, with existing knowledge on cell type proportions to guide a sampling strategy that optimizes the cost/benefit ratio for each region, and rely on stratified analysis to isolate specific cell types routinely and robustly. This would enable an iterative approach of broad, shallow sampling followed by enrichment of under-sampled subsets based on computationally identified markers to determine when the majority of cell types have been sampled (**Figure 7**).

*Analysis of nuclei and cells*. The cellular branch will aim to deliver a molecular classification of brain cell types, ideally in the form of a hierarchical taxonomy of cell types that mirrors biological principles such as developmental origin. To generate this atlas, some specimens will require distinct preparations. Adult brain tissue cannot typically be dissociated to single cells, requiring analysis of single nuclei. In the adult, spinal cord and peripheral nervous system require unique handling. Conversely, in prenatal specimens, live cells can be dissociated, offering an opportunity to use the same samples for both single-cell and single-nucleus analysis, as a Rosetta Stone for comparison.

*Spatial analysis.* The second key deliverable is a spatial census of molecular cell types based on their morphology and relative and absolute position. This will rely primarily on highly multiplexed analysis of RNA in situ, for example with single-molecule fluorescent in situ hybridization (smFISH). Applied systematically to tissue sections, this approach will quantitatively map spatial distributions, proportions, and organizational features such as topography. To determine the signatures for in situ analysis, we will rely on markers reflecting both prior knowledge, the cellular branch, and possibly random composite approaches[16] (**Section 2**).

*Sample and cell type mapping to a Common Coordinate Framework (CCF)*. To make the brain atlas as useful as possible, source tissue samples from many laboratories should be mapped to a CCF. Such a framework enables data integration and the discovery of associations between brain regions, cell types, and cell function. Existing efforts along these lines, such as the whole brain Allen Human Reference Atlas[95] and Big Brain Atlas[96], provide a 2-D (and coarse 3-D) CCF with a comprehensive hierarchical structure-level ontology.

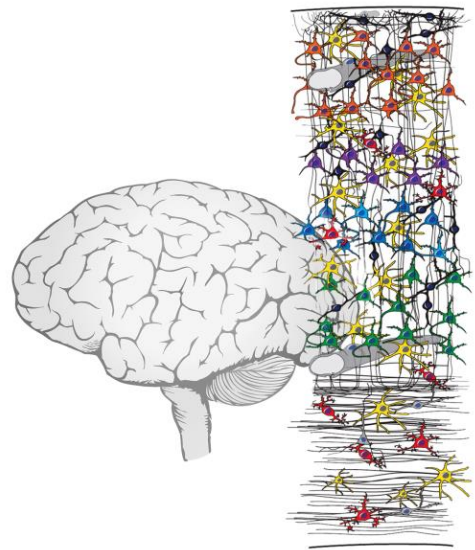

**Figure 7. The human brain.** The brain likely consists of more than a thousand distinct cell types, including a large variety of neurons, glia (astrocytes, oligodendrocytes, ependymal cells), vascular (endothelial, pericytes) and immune cells. These are arranged in an intricate three-dimensional organization that is essential for brain function. Brain regions vary enormously in cellular composition and organization across spatial domains, thus requiring sampling from a large number of defined regions.

Alternative approaches could learn some of this mapping from collected spatial data and images. In principle these structure-level atlases can be extended to the level of substructural cellular spatial positions and cell type nomenclatures, although a great deal of work remains to be done in this area. By integrating molecular cell type classifications with spatial mapping data in a CCF, we can also exploit computational approaches to assign cell types to detailed anatomical structures (e.g., relative to vessels, surfaces, ventricles), and to discover cell type communities that are organized in local niches (e.g., stem cell niches in the subventricular zone).

Ideally, the human atlas would be generated in parallel with similar mouse and non-human primate atlases, both to understand what is unique about the human brain and to understand which features are well modeled in genetically and experimentally tractable model organisms.

**Draft Atlas v1.0**

The principal goal of the first draft atlas should be to capture as much of the cellular diversity within and between brain, spinal cord, and peripheral nervous system compartments as possible. Because the brain is divided into functionally distinct regions of cells, each with its own unique variations, and because brain regions have shown a high degree of stereotypy across individuals, we propose that the draft atlas focus on a **very small number (2 to 6) of high-quality individual specimens**, where we analyze about 100 pre-defined regions of interest covering the brain, spinal cord, and peripheral nervous system. Future efforts can explore systematic variation by gender, ethnicity, and other variables. These high-quality brain specimens should be obtained through rapid autopsies from neurotypical individuals, with consent from next-of-kin through medical examiners or brain-banking operations. Because gene expression also varies significantly as a function of developmental age, we propose to focus on the stable adult period (18 to 65 years), but to include pilot projects focusing on brain development.

Each specimen will be analyzed in three ways: for the cellular branch by single cell/nucleus RNA-Seq at scale across a set of predefined brain regions; for the spatial branch by generating a spatial census of types in those regions using smFISH or similar multiplex profiling methods for RNA (and protein if possible); and integrating them by mapping specimen locations to a 2-D, 3-D, and ontological CCF, following approaches developed with the CCFWG.

# IMMUNE SYSTEM

The immune system consists of several cell lineages with distinct and synergistic functions that aim to eliminate environmental threats and damaged cells. Many immune cells circulate in the blood, awaiting injury signals that prompt them to home to the site of damage and eradicate potential threats. Other immune cells reside in healthy organs to survey for injuries, while also contributing to maintaining organ homeostasis (**Figure 8**). Immune cells thus form a complex network of cells — with distinct origin, life cycle, and function — which is only fully revealed in response to challenges.

**Why an immune cell atlas?**

Immunologists have worked for decades to discover, classify, and study the cells of the immune system and their specific functions. They have created a taxonomy of cell types based on molecular markers, cellular functions, developmental origins, and differentiation potential. However, because the human immune system is highly complex — consisting of dozens of cell subtypes whose activation states are highly dependent on tissue location, environmental triggers, and individual genetics — it has not been feasible to measure these states with existing methods.

And since most immune cell types take part in the cascade of events that define an immune response, there is a great need for methods that more comprehensively monitor immune cell types over time. Finally, there is a paucity of information about how immune cells function differently across the many tissues and organs of the human body.

Pioneering efforts by the Immunological Genome Project (ImmGen, https://www.immgen.org/) generated extensive and widely used gene expression profiles of mouse immune cells from diverse tissues at baseline and in response to challenges[97]. However, to date, such a coordinated and concentered effort has not been carried out for the human immune system. The Immune Cell Atlas (ICA) is a community-wide effort that aims to identify the molecular and phenotypic signatures and the spatial organization of all immune cell types that populate human lymphoid and non-lymphoid tissues at baseline or after many different challenges. The ICA will generate the largest unified, standardized database of the human immune cell network and make it available to the community through public data portals. Combined data on this scale should help shape the future of immunology and immunopathology knowledge and foster the development of novel immunotherapies for the treatment of human diseases including infections, autoimmunity, asthma, allergy, and cancer.

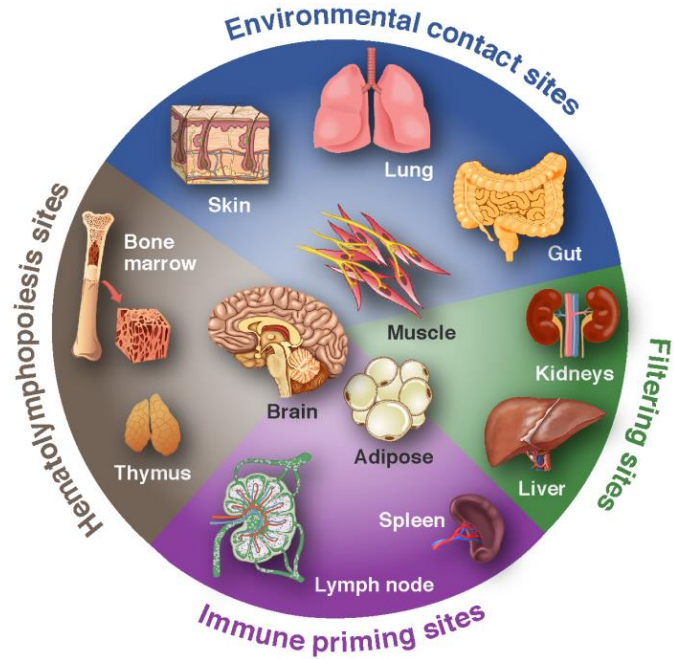

**Figure 8. An Immune Cell Atlas.** An Immune Cell Atlas will profile immune cells from primary and secondary lymphoid organs, and peripheral non-lymphoid tissues and organs, allowing for identification and characterization of all immune cells in any tissue. Example representative tissues are shown for illustrative purposes. Permission to re-use granted for skin (© Kellie Holoski); liver (© Joanna Culley); muscle cells (© John Wiley & Sons, Inc.); and bone marrow (HealthTap).

**What are key considerations for an immune cell atlas?**

The ICA will reveal the distribution and composition of the immune cellular network that resides in the different organs of the human body. It should illuminate the molecular wiring and the specific phenotype of each immune cell type, the spatial distribution and interactions between distinct immune cell types and with other cell types (e.g., epithelial cells, endothelial cells, fibroblasts) of each tissue compartment. Because immune cell function is revealed mainly in response to tissue threats, the atlas must also include immune cell populations and immune cell states that accumulate in injured tissues.

Key principles for an optimal ICA include:

*Diversity of immune cell types.* A comprehensive ICA will lead to an unprecedented global view of the immune cell populations that reside in human tissues, as was recently shown in the blood[98].

*Local impact of tissues on immune cell functions*. Immune cells of the same lineage often serve different functions depending on the tissue environment in which they reside[99]. To discover the range of immune cell functions, we must profile immune cells from different tissue sites.

*Patient-to-patient variation*. Infectious and immune diseases vary with geography and population. Therefore, to gain snapshots into the immune profiles of patients at frontline infectious-disease outbreaks, we will sample patients across gender, age, ethnicity, and geography — and this information will be anonymously linked to the profiling data according to HIPAA standards. Phase I will encompass some human diversity, though not at the full scale expected for the complete HCA.

*Spectra of cell activation states in challenged conditions*. In contrast to most cell types, immune cell fates, functions, and frequencies are determined dynamically in response to challenges such as pathogenic infection. An individual macrophage may serve a role in tissue repair or homeostasis in the steady state, but become an inflammatory, aggressive, antibacterial immune defender under challenge[99]. Thus, in addition to healthy tissue, we will profile immune cells from tissues in various states of challenges and disease, including cancers and infections, although individual diseases will not be the primary focus of the atlas.

*Transcriptional and protein profiles in cells and space*. With these considerations in mind, we will profile all immune cell populations in blood and tissues (e.g., lung, skin, gut, lymph nodes, kidney, spleen, fat, muscle) from healthy individuals and from disease states. We will harvest tissues, dissociate cells from tissues, and establish the transcriptional program of dissociated, tissue-resident immune cells via scRNA-Seq and bulk profiling, as shown recently[98,100-103]. We will extend this work by validating transcriptionally defined cell markers at the protein level, using flow/mass cytometry, and by establishing the spatial distribution of cells in tissues, using both RNA- and protein-based in situ profiling methods.

**The pilot stage** will be dedicated to optimizing standard operating procedures for tissue collection, tissue digestion, cell encapsulation, shipping, sequencing, and analysis among different collection sites. To gain information about spatial organization, we will also perform high-dimensional imaging in tissues[61,104]. We will start with pilot analyses of barrier tissues and lymphoid organs (as they are the most enriched in immune cells) from healthy research participants. Following the steps of ImmGen, we will develop strict quality-control guidelines to standardize sample quality and normalize analysis of tissue to minimize batch-dependent variation.

**Draft Atlas v1.0**

The first draft atlas aims to capture the cellular make-up of the immune system by profiling all immune cells that reside in most tissues of the human body (**Figure 8**) under baseline or challenged conditions. Each tissue will be analyzed by single-cell and tissue profiling to capture the transcriptional and phenotypic programs of each immune cell type, as well as their spatial distribution in the different tissue sties.

In the first draft atlas, we will profile "healthy" tissue specimens from 20 research participants and sample each tissue in at least three geographically distinct sites. Tissue will be collected from 20- to 55-year-old men and women from different ethnic backgrounds. In addition, we will collect "challenged" or "disease" specimens (such as cancer and inflammatory lesions) from ~20 research participants for each "challenged tissue." Each specimen will be analyzed by the HCA Sky Dive approach (**Section 2**; **Figure 3**), at first for scRNA-Seq. The transcriptional signatures will be used to develop protein markers that define immune cell populations and enable prospective cell isolation by FACS, for deeper RNA profiling and other molecular profiling (e.g., ATAC-Seq), as well as for in situ spatial characterization (at RNA and protein level).

# URINARY: KIDNEY

The beauty — and difficulty — of the human kidney lies in its complex structure and the diverse interactions among its cells during homeostasis and in pathological states[105]. Progressive chronic kidney diseases affect more than 500 million people worldwide, and yet therapies to halt or prevent these diseases remain scarce. Most individuals have two kidneys that lie at the back of the abdominal cavity and perform a number of life-maintaining functions, such as filtering and excreting waste and acid, maintaining electrolyte and water balance, and producing the hormone erythropoietin, which stimulates the production of red blood cells to prevent anemia, and the active form of vitamin D, which preserves calcium and phosphate levels and bone health. Anatomically, each kidney comprises an outer cortex containing the glomeruli, through which blood is filtered, and the medulla, where urine is concentrated. Urine flows within tubules that coalesce in the renal pelvis, draining into the ureter and on to the bladder[105]. In addition to epithelial, mesangial, and endothelial cells, the kidneys also contain a network of immune cells that contributes to organ defense and injury repair.

**Why a kidney cell atlas?**

Understanding how the kidney works in health and disease is critically important for the development of future therapies. However, information on human kidney cell diversity has been hampered by limited tissue accessibility. The kidney is difficult to sample, and percutaneous biopsies contain mostly cortical tissue, and no comprehensive map of intercellular networks in the human kidney exists. The extreme anatomical variation in tissue environment can profoundly influence local epithelial and immune cell function[106], and, as a highly vascular organ, the kidney is in constant contact with the blood and cells in the circulation. Finally, as human genetics is informing our understanding of kidney pathophysiology[107], a detailed map of gene expression and disease circuits in the kidney will be instrumental in developing precision therapies.

The Human Kidney Cell Atlas will aim to reveal with an unprecedented granularity the cell populations within the kidney. It will become the largest unified, standardized database characterizing human kidney cells, reveal previously unknown cellular diversity, and increase the resolution at which we comprehend cell function and phenotype in all anatomical regions of the human kidney. Since many diseases of the kidney are often cited as a fast recapitulation of normal aging mechanisms, these analyses will provide unprecedented insight into kidney diseases and enable further analysis and cross-validation in kidney tissue from individuals with specific kidney diseases (lupus nephritis, nephrotic syndrome, immune complex glomerulonephritis, diabetic kidney disease, acute kidney injury, etc.) or from kidney transplants with rejection or declining function to define the specific cell populations and kidney segments primarily affected in each case. Thus, the Human Kidney Cell Atlas will transform our knowledge of regional cell identity and function, providing an invaluable platform on which to base efforts to understand and treat kidney diseases (**Figure 9**).

**What are key considerations for a kidney cell atlas?**

To capture the data necessary to classify kidney cells, we will first focus on gene expression profiling by scRNA-Seq profiling to determine the combination of markers that allow the identification and distinction of major cell types in the kidney. These will be followed by validation at the protein level, using flow and mass cytometry, and establishing the spatial distribution of cells in tissues, using both RNA- and protein-based methods. Two completed feasibility studies have demonstrated that both kidney cells and kidney immune cells (CD45+) can be profiled from mechanically and enzymatically dissociated cells from nephrectomy specimens.

**In a pilot study**, key technical remain to be addressed.

*Sample source*. Discarded transplant kidneys and nephrectomy specimens can be obtained, allowing analysis of fresh tissue. The minimum amount of tissue that can be used to isolate all cell types must be assessed. It may also be important to explore how donor's co-morbidities (hypertension, diabetes, hyperlipidemia), age, sex, and ethnicity impact sample quality. Although this exploration may be limited in the pilot, it should be possible to collect relevant clinical data and begin addressing these questions, which play a key role in understanding disease.

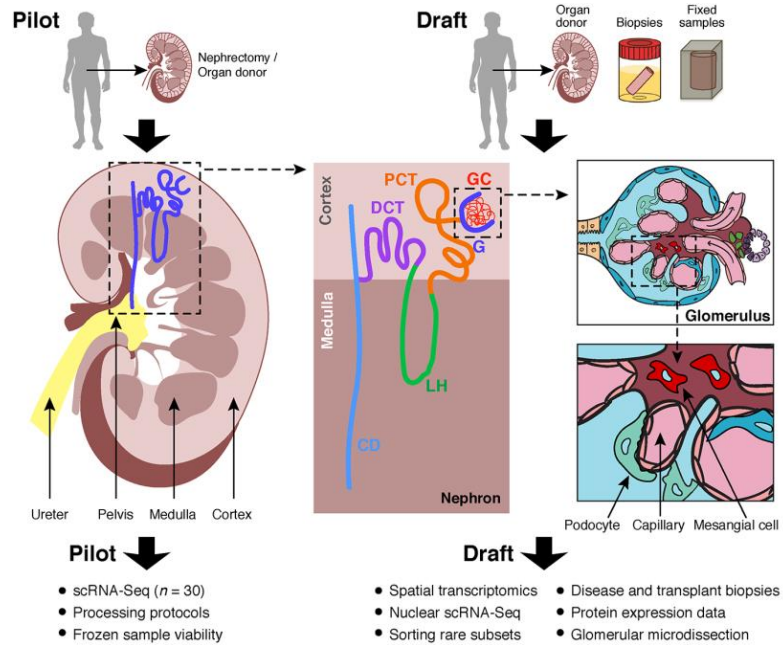

**Figure 9. Kidney atlas workflow.** In the pilot phase, samples will be obtained from *n*=30 discarded kidney transplants and nephrectomy specimens, allowing the transcriptional profile of cells from different anatomical areas of the kidney to be assessed, including cortex, medulla, pelvis and ureter. We will establish the optimal processing protocols to allow good quality scRNA-Seq data to be generated for all major cell types in the kidney, including tubular epithelial cells, endothelial cells and immune cells. The kidney is comprised of ~1 million subunits called nephrons. The tubular epithelial cells in different regions of the nephron have different functional characteristics, and the scRN-Seq data generated will allow us to compare transcriptional profiles obtained from cells in the proximal convoluted tubule (PCT), loop of Henle (LH), distal convoluted tubule (DCT) and collecting duct (CD). In the draft atlas, we will extend our sampling to include kidney biopsies and fixed samples from patients with a variety of kidney diseases. We will also use flow sorting and microdissection to enrich for rare cell subsets and cells from specific subanatomical areas, such as the glomerulus. This structure forms the filtering unit of the kidney, plays a critical role in homeostasis, and abnormalities in its cellular constituents (podocytes, glomerular capillaries (GC), and mesangial cells) contribute to disease. Spatial transcriptional and proteomic data will be generated to support and validate scRN-Seq data obtained from homogenized samples.

*Sample handling*. While freezing facilitates tissue collection, and transplanted organs are currently transported on ice, these factors, as well as delayed processing, may impact cell viability, transcriptomes, and proteomes. To this end, fresh, cold-stored, and frozen tissues should be compared.

*Single cells and nuclei*. Feasibility studies to date suggest that different protocols will be required to obtain different cell types; for example, kidney epithelial cells are more easily damaged by disaggregation protocols than kidney immune cells. Both different dissociation protocols from fresh tissue and nuclei prep from snap-frozen specimens must be tested.

*Spatial analysis*. In parallel, intact, nondisaggregated samples of all kidneys studied will be stored for future spatial analysis using both RNA and protein-detection techniques. This will include snap-frozen and formaldehyde-fixed specimens.

*Standard operating procedures*. Strict quality-control guidelines will be developed to standardize sample collection and analysis to minimize batch-dependent variation. All optimized protocols will be shared online.

By the end of the pilot, we will have generated a standard operating procedure for processing kidney epithelial, endothelial, and immune cells and ascertained the impact of cold storage and freezing on kidney cell viability. The generated data will also be a basic atlas of common cell types in different anatomical regions of the kidney.

**Draft Atlas v1.0**

The draft atlas will chart the kidney cortex, medulla, pelvis, and ureter in tissue samples obtained from discarded transplant organs, nephrectomy specimens, and kidney biopsies, as well as a moderate number of samples from diseased kidneys. Kidney tissue will be micro-dissected to anatomical areas of interest (e.g., glomerulus).

For the first draft, 3 to 4 participating geographical sites will collect and process human tissue samples locally using strict standard operating procedures. We will sample kidneys from 20 individuals from each site (20 to 55 years of age; diverse ethnic backgrounds, including African Americans, Asians, and Caucasians). In addition, we will sample 10 additional individuals, ages 55 to 80, at each site, given the high frequency of chronic kidney disease in the elderly.

Each specimen will be analyzed for both the cellular level (with uniform approaches and with stratification for rarer cell subsets using protein markers identified from the uniform analysis), measuring both RNA profiles and protein panels[108], and the spatial level (with fresh, frozen, and fixed sections).

# RESPIRATORY: LUNG

The human respiratory system is critical for gas exchange. Lung anatomy enables this through two main structures: the airways that lead the air to and from the respiratory unit and provide mucociliary clearance of inhaled particles and pathogens; and the alveoli, distal saccular structures where gas exchange occurs. The airways, beginning in the nasal cavity and connecting to the trachea, iteratively bifurcate into the branching bronchial tree, ending in terminal bronchioles, which branch into the respiratory unit encompassing the respiratory bronchioles, the alveolar duct, and the alveolar sac[109] (**Figure 10**). The entire respiratory tract is also an important part of the mucosal immune system. Consequently, a large diversity of cell types contained within a tailored extracellular matrix is required to maintain homeostasis of the lung. The recent discovery that the lung is a site of platelet biogenesis and a reservoir for hematopoietic progenitors[110] highlights the fact that the full cellular diversity of this organ is not yet known.

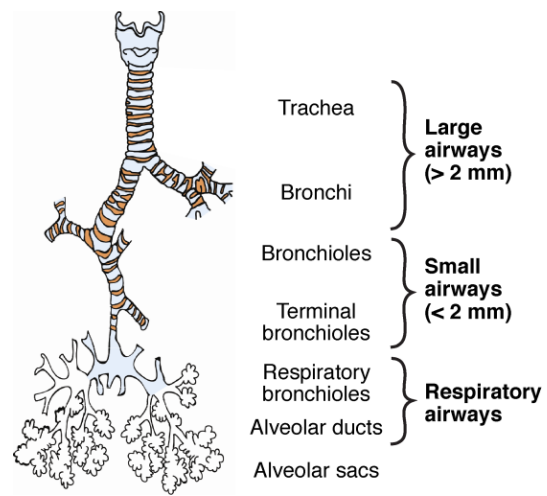

**Figure 10. Lower airway structure of the lung.** The structure of the lower airways is defined by the cartilaginous trachea and main bronchi, branching into progressively smaller bronchioles. Small airways are defined as the conductive airways with a diameter of less than 2 mm, and end with the terminal bronchioles, which connect to the respiratory unit. Adapted from *eLife* 2017;6:e30194 DOI: 10.7554/eLife.30194, licensed under CC BY 4.0.

**Why a lung cell atlas?**

The lung is a highly complex organ with at least 40 discrete cell types known to date, but there is limited knowledge about the functional relationships of these cells in disease. Lung disease is a leading cause of mortality in the U.S. and worldwide, with more than 7 million deaths attributed to lung disease annually. The Lung Cell Atlas will reveal novel insights about the identities, activities, and lineage relationships of all cells in the healthy human lung. This baseline will then serve as a starting point for profiling cohorts of chronic lung disease patients, inspiring translational research into the therapeutic strategies of the future, as well as potentially uncovering new diagnostic opportunities for lung disease. Thus, a Lung Cell Atlas is of paramount biological and clinical relevance.

**What are key considerations for a lung cell atlas?**

To build a Lung Cell Atlas, we will need to establish standardized procedures for comprehensively sampling the lung along the proximal-distal axis of the airways, aiming to develop a coordinate framework based on the 16 generations of the conductive airways in the bronchial tree up until the respiratory unit and their relative location within the five lung lobes (**Figure 10**).

**In a pilot phase**, we will establish key procedures for this process.

*Samples*. Lung and airway tissue can be obtained in several ways: direct sampling of fresh lung airway wall tissue through bronchoscopy; central or parenchymal tissue from transplant lungs; and lung parenchymal tissue from uninvolved resection material adjacent to primary or metastatic tumors. Single cell RNA-Seq data sets from freshly obtained bronchoscopy biopsies are under way and can be used as a reference for those from resection material or transplant lungs. The abundance of lung parenchymal tissue from tumor resection programs will also allow efficient optimization and standardization across centers.

*Anatomical sampling*. It is critical to standardize a coordinate framework to define location within the lung, based on the proximal-distal axis along the generations of the conducting airways up to the terminal bronchioles and the respiratory unit, and their relative position within each of the lobes. This will include harmonizing dissection protocols for larger airways and peripheral lung tissue.

*Cell and nucleus profiles*. The cellular branch of the Lung Cell Atlas (LCA) will initially be based on scRNA-Seq data, but aims to include proteomic and epigenetic profiles as single-cell methodologies evolve. Early studies have established scRNA-Seq profiles identifying the major cell types in freshly obtained airway wall biopsies and brushed airway epithelial cells, as well as in central and peripheral lung tissue from resection material and transplant donors[111]. More generally, single-cell profiling will rely on tissue disaggregation into suspensions of viable single cells, from specimens collected from multiple representative anatomical/histological sites. In addition, snRNA-Seq will be tested on cryopreserved tissue, which would allow profiling of cells across the full spatial coordinate framework of several representative whole donor lungs and avoid possible biases introduced by tissue dissociation.

*Spatial profiles*. Large-scale characterization of cellular heterogeneity from cell suspension or isolated nuclei should be integrated with corresponding in situ data for multiple sites along these standardized geographical coordinates. IHC and highly multiplexed FISH analyses on matched lung-tissue specimens will be used to validate the initial results from scRNA-Seq and snRNA-Seq data sets from both freshly obtained biopsies and resection materials and to allow spatial mapping (by inference[9,13-15] of entire transcriptomes). In addition, the recently developed nondestructive cryo-micro-CT approach[112], which allows imaging of small airway structure up to the respiratory unit, can be combined with FISH analyses to identify novel cell types and activity states onto in a tissue context. These spatially resolving methods will dramatically advance our understanding of lung tissue architecture and generate new hypothesis on functional relationships between cell types.

**Draft Atlas v1.0**

A first draft of the Lung Cell Atlas will sample tissues from the nasal cavity, trachea, primary and tertiary bronchi, small airways (diameter <2 mM) and the respiratory unit (including transitional bronchioles and alveoli). For each defined position along the respiratory tract, the "healthy" atlas will be based on 20 to 30 healthy, nonsmoking individuals aged 25 to 55 years from each of at least three geographically distinct sites, both genders, and at least two ethnicities. For every specific location along the airways, the bronchial tree and respiratory unit, we will profile at least 50,000 to 100,000 cells or nuclei and perform spatial analysis of cell type specific mRNA signatures as well as protein signatures. We will then correlate tissue architecture with spatial organization of the identified cell types together with their extracellular matrix to define distinct cellular niches along the airway tree. To derive a deeper mechanistic understanding of lung development and disease, we will additionally include airway- and lung-tissue samples of small cohorts of clinically well-phenotyped patients with specific conducting airway diseases — COPD and asthma — and interstitial lung diseases, such as idiopathic pulmonary fibrosis (IPF). We anticipate integration of our data on the respiratory mucosa with the Immune Cell Atlas in the respiratory mucosa, with data sets of the lung microbiome and proteome in our sampled tissue specimens, and with those data sets and platforms currently available in the wider lung-research community[113,114].

# HEPATOPANCREATIC-BILIARY: LIVER

The human liver is the central coordinator of the body's metabolism and performs a wide range of functions critical for body health and maintenance[115]. Unlike almost every other human organ, the liver also has an incredible regenerative capacity: it can regrow even after 80% has been removed. However, once the liver fails, the only treatment is organ transplantation. Development of alternative treatments is currently limited by our lack of understanding of the cellular landscape of the human liver. Because liver diseases — resulting from obesity, alcohol abuse, drug toxicity, and chronic infections — are a major and increasing burden worldwide, novel investigations are urgently needed to better understand how liver cells grow and work together.

**Why a liver atlas?**

A human liver cell atlas will serve as a foundation for studies of normal and diseased liver function, and will support the development of novel therapies, such as methods to reduce transplant-rejection rates and regenerative-medicine approaches to repair damaged tissue. First, the cellular complexity of the liver has been greatly underestimated, and much remains to be discovered about how the cells in the liver ecosystem fulfill hundreds of liver functions, possibly through sub-specialization of cell types, such as hepatocytes (**Figure 11**). Second, while we know liver diseases are highly localized, and their progression is likely associated with specific cell subtypes, we do not know what these cell types are. A liver atlas of the healthy baseline will serve as a reference to compare with disease samples. Third, the liver is a regenerative organ, and better understanding of liver regeneration will help the design of liver tissue repair technologies, including stem cell therapies. The human liver cell atlas will help define the molecular characteristics of cells that should be replaced.

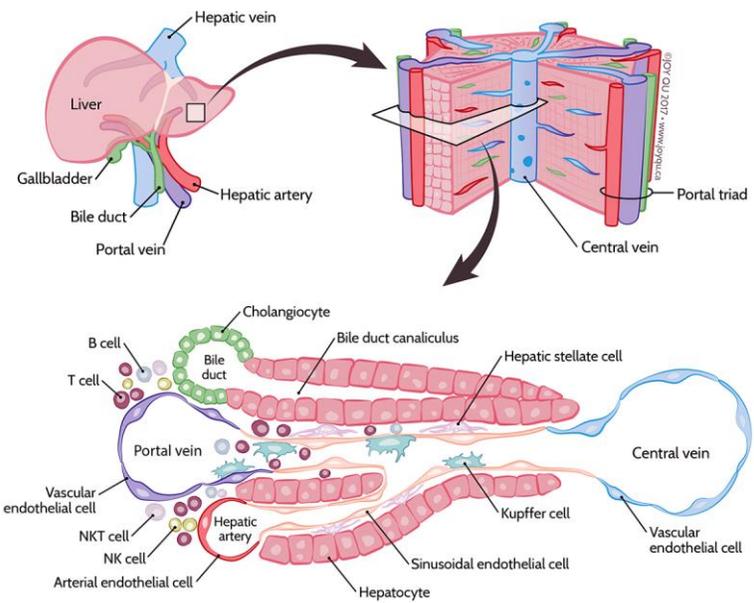

**Figure 11. Liver.** The main "building block" of the liver is the hepatic lobule, which includes a portal triad, hepatocytes arranged in linear cords between a capillary network, and a central vein. The functional unit of the liver is the sinusoid. Found within the sinusoid are parenchymal cells (hepatocytes and cholangiocytes) and non-parenchymal cells (endothelial cells, Kupffer cells, hepatic stellate cells and liver infiltrating lymphocytes- including B cells, T cells and NK cells).

**What are key considerations for a liver cell atlas?**

An ideal hepatic cell atlas will involve analyzing a large number of cells, both parenchymal [PC] and nonparenchymal [NPC], and their spatial organization, from all anatomic liver regions: two major lobes, subdivided in eight segments reflecting the major divisions of the portal vein and the bile duct, and for each segment, each of the three metabolic zones of the lobules (the segments' functional units); the gallbladder; and the common bile duct of the extra-hepatic biliary system.

However, collecting 20 samples per liver (8 segments × (PC + NPCs) + (gallbladder + common bile duct) could be challenging considering the difficulty to access entire and healthy organs and the wide variability in liver composition and activity between individuals.

Key principles for an optimal liver cell atlas include:

***Characterization of liver cells at many levels.*** The liver is structurally and functionally heterogeneous[115]. Thus, a liver atlas should include a range of molecular profiles (e.g., transcriptome to identify cell type signatures, proteome to characterize the many liver protein products, metabolome because the liver is a metabolic tissue) from the different anatomical regions, as they differ in cellular composition.

*A diversity of liver functions, human populations and environments.* Important liver metabolic activities vary tremendously among individuals and depend on age, gender and other factors. For instance, some liver diseases are so common that they should be sampled as part of the atlas. The global prevalence of non-alcoholic fatty liver disease [NAFLD] is 25%[116] and more than 70 million people are chronically infected with hepatitis C, with some geographical areas affected more than others and some human populations more prone to disease progression than others. Also, liver varies greatly over the circadian clock cycle[117]. Thus, samples must be derived from a broad number of genetically diverse individuals in different environments; information about sleep and wake cycle should be captured if available.

Disease susceptibility and progression is related to liver immune cell function. The liver cell atlas intersects with the Immune Cell Atlas, and comparing them will help characterize liver-immune cells for their similarity to and distinction from peripheral counterparts and how they are affected by an individual's environment.

*Spatial analysis.* The function of hepatocyte and accessory cells, including immune cells, is affected by their anatomical position in the liver. For example, macrophages function differently based on where they are in the sinusoid, and hepatic immune cell activity may be influenced by metabolic zonation. However, how these cells' phenotype, frequency, and functional diversity are impacted by anatomical location and histological neighborhoods remain to be fully detailed. Spatial analysis of liver cell 3-D distribution in multiple anatomical regions will help determine how location affects cell function.

*Development*. We must understand liver cell development better to comprehend disease and liver regeneration. This provides a natural connection point with the human Developmental Cell Atlas, to monitor both cell development and regeneration.

**In a pilot study, key technical questions remain to be addressed.**

*Optimizing sample preparation.* Single-cell isolation from liver samples is notoriously challenging. Liver cells such as hepatocytes and cholangiocytes are delicate and frequently don't survive standard tissue extraction. Furthermore, many liver cell types are influenced by tissue-level metabolic gradients, which may change rapidly during tissue acquisition. Standard and optimal single-cell sample preparation protocols, compatible with processing of delicate liver cells, should be developed, as well as single-nucleus methods applied on snap-frozen samples[2-4,9,118]. Fresh and frozen samples should be profiled to evaluate the impact of each preparation, as well as comparing scRNA-Seq to snRNA-Seq, including differences in cell composition.

*Sample source.* Healthy human liver tissue can be collected from several sources: organ transplants (liver segments or wedge biopsies from deceased donor prior to transplantation or whole organs that are declined for transplantation); resections from healthy regions of liver cancer; and biopsies from standard liver diagnosis. These sources will also be valuable for collecting liver tissue affected by a variety of diseases. We should define the best source of cells (or nuclei) to perform single-cell profiling (biopsy versus organ donor versus resection); spatial methods can likely accommodate all these sources. Each source has distinct advantages and drawbacks, with respect to blood supply (ischemia), cell contamination, or anatomical region diversity. Protocols to isolate high-quality cells from all studied fractions must also be developed and validated, since most existing protocols are specialized for one particular cell type while discarding the others. Single-nucleus RNA-Seq and spatial profiling can help assess compositional biases in single-cell analysis of fresh tissue.

*Data analysis.* State-of-the-art single-cell genomics data-analysis workflows will be applied in collaboration with the broader HCA effort and should be modified as needed to address liver-specific challenges, such as analysis of metabolomes, or studying zonation patterns that affect liver-cell function[15].

**Draft Atlas v1.0**

A first draft will collect samples from eight anatomic segments of the liver for at least 20 individuals collected from each of four geographically (and ideally ethnically) distinct sites (e.g., Cambridge, U.K.; Toronto, Canada; Boston, U.S.; Beijing, China); individuals from both genders; all adults 20 to 65 years of age. Three different tissue sources will be used whenever possible (biopsy, whole donated liver, and normal adjacent tissue from cancer resection) and PC, NPC, gallbladder, and common bile duct will be isolated (at least from cadaveric donors). Each collection center will use a different tissue sampling approach. Larger numbers of individuals will be necessary to encompass individual variability while larger numbers of cells for the top of the Sky Dive may be collected only with samples from some of the individuals. All samples will include preservation of matching sections for spatial analysis. A subset of the samples will be analysed by spatial profiling of RNA and proteins.

# HEPATOPANCREATIC-BILIARY: PANCREAS

The pancreas is a vital organ that is composed of cells with either exocrine or endocrine functions. It is the locus of key pathologies, including diabetes and pancreatic cancer (**Figure 12**).

The *exocrine pancreas* is comprised of acinar cells, which produce zymogens (an inactive precursor of enzymes) that hydrolyze macromolecules to aid digestion; and ductal cells that form branched tubules, whose principal products include bicarbonate, which helps neutralize gastric acid, thereby allowing effective enzymatic action. The full diversity of cell types in these two categories is still unknown. The *endocrine pancreas* is comprised of hormone-secreting epithelial cells organized in structures called Islets of Langerhans. Each endocrine cell type synthesizes and secretes a hallmark hormone product: $\alpha$-cells make glucagon, $\beta$-cells produce insulin, and $\delta$-cells produce somatostatin. The functions of islets, ducts and acinar cells are coordinated by important autonomic, paracrine, and endocrine inputs orchestrated by neuronal and specialized vascular structures[119].

Rodent studies over the past two decades have identified cellular, genetic, signaling and molecular pathways that lead to formation of the pancreatic exocrine and endocrine cells. In addition, recent progress has been made in identifying the transcriptomes and epigenetic features, including histone modifications, in human pancreatic cells[120-125]. However, much remains to be discovered about the cellular composition of the pancreas and the mechanisms regulating the developmental process in humans[119,126,127].

**Why a pancreas atlas?**

Human pancreatic diseases, including type 1 and type 2 diabetes mellitus (T1D, T2D), pancreatitis, cystic fibrosis and adenocarcinoma, affect over 10 percent of the world's population[128-130]. Thus, there is a clear need to understand the mechanisms governing the function and the development of these cells and how they work together in healthy and diseased organs.

In addition, diseases such as pancreatitis and adenocarcinoma involve interactions of pancreatic cells with "stromal" elements — (including fibroblasts, stellate cells, vascular endothelial and smooth muscle cells — as well as immune cells — including those localized to pancreatic lymph nodes. It has been widely speculated that the interaction between the stroma, immune cells, and tumor cells give rise to resistance to treatment of pancreatic cancer, which has been a major challenge in tackling this devastating disease. More detailed studies of these cell subsets are needed to understand their roles and function in physiological and pathological settings.

**What are key considerations for a pancreas atlas?**

There are good prospects for developing a detailed map of common pancreatic exocrine and endocrine tissue. All would require close collaborations with physicians, whether in general surgery, oncology, gastroenterology pathology, or transplantation surgery. Major challenges in leveraging these resources are methodological, especially standardization of tissue isolation, cell purification, and molecular studies. In particular, the high enzymatic activity within the pancreas renders extraction of high quality RNA from fresh material particularly challenging.

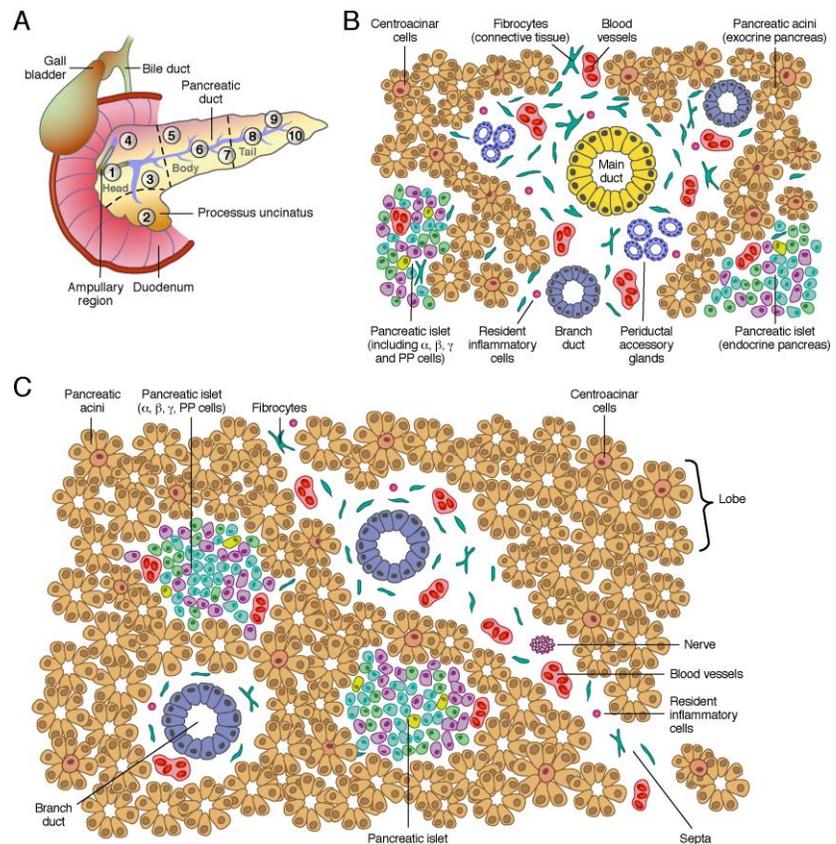

**Figure 12. Morphology of the healthy pancreas and sampling strategy**. (A) Overview of the pancreas depicting the three regions: head (with the special subsites Processus uncinatus and ampullary region), body, and tail. From the head, four sample regions are selected representing the periphery of the head (2X), the central ampulla and the Processus (periphery). From the latter two anatomical subsites (body, tail), three sample regions each will be selected, with one central location (including the main pancreatic duct) and the periphery (2X).
(B) Principle tissue composition of ampullary and central pancreatic duct regions. (C) Principle tissue composition of peripheral pancreatic tissue.

**In a pilot study,** key technical questions remain to be addressed.

*Sample source.* For endocrine tissue, cellular and molecular mapping studies of human pancreas in adults are currently feasible[131], thanks to a variety of extant robust organ procurement efforts, such as the NIH-supported Integrated Islet Procurement Program in the U.S. (IIDP)[132] and related efforts in other countries[124,133]. Similarly, there are good prospects for exocrine diseases like pancreatitis or adenocarcinoma[134-136], with specimens obtained from surgical extraction.

*Tissue handling.* Practical challenges, like rapid autolysis of cadaveric human pancreas within minutes to hours after extraction, altered gene regulation and function of cells after in vitro culture for even short periods[120], and a need for large numbers of primary cells in prior investigations of chromatin and histone modifications, have limited donor sampling and use of single-cell resolution methods[137-140].

A human pancreas atlas *requires coordinated focused efforts to procure human cadaveric organs in reliable, efficient, ethical and cost-effective ways*. This could involve multicenter teams (e.g., Arda et al.[120]) organized within interdisciplinary consortia, like the NIH-funded Human Islet Resource Network (HIRN; https://www.niddk.nih.gov/about-niddk/research-areas/diabetes/type-1-diabetes-special-statutory-funding-program/Consortia-Networks-Centers/human-islet-research-network/Pages/default.aspx), Human Pancreas and Analysis Program (HPAP; https://hirnetwork.org/consortium/hpap) and Chronic Pancreatitis, Diabetes and Pancreatic Cancer (CPDPC; http://cpdpc.mdanderson.org) consortia, or the privately-funded efforts like the JDRF Network for Pancreatic Organ Donors with Diabetes (nPOD; https://www.jdrfnpod.org). Cryogenic banking of specific cell types or structures like islets has been reported to maintain viability and function for years[133] and may provide a useful adjunct. Likewise, coordinated collection of relevant non-pancreatic cell types, such as circulating or pancreas-infiltrating inflammatory cells) — a focus of nPOD and the T1D-focused TrialNet — would ideally be coordinated with pancreas procurement.

***Comparison of cellular, nuclear and spatial measurements.*** Because of the particular challenge of obtaining high-quality profiles in a rapidly lysing tissue, a pilot effort should compare single-cell, single-nucleus and spatial methods. Single-nucleus and spatial methods can be applied to quickly preserved specimens, such as frozen (nuclei) or fixed (spatial) tissue. These can then be compared to determine the viability of each strategy in this challenging organ.

**Draft Atlas v1.0**

The principal goal of the first draft atlas should be to capture as much of the cellular diversity in the pancreas and understand the cells' histological organization. Although the pancreas has a challenging lack of stereotypical tissue architecture, powerful approaches like the tissue transparency method CLARITY[141] have begun to elucidate and quantify dynamic and previously undetected cell–cell interactions, like islet innervation, in the human pancreas[142]. A combination of cellular and spatial approaches will be able to tackle this challenge.

The draft atlas will include a sampling of at least 10 regions of the pancreas, covering both endocrine and exocrine tissue from known macro-anatomical pancreatic regions including head, body, tail, processus uncinatus, and ampullary. However, on the microstructure level, many more regions might exist that are not yet recognized by morphology. In a draft atlas, we will collect 20 samples for each of these regions from each of at least three geographically (and ideally ethnically) distinct sites; individuals from both genders; 20 to 55 years of age. To characterize cell types, we will apply both single-cell and single-nucleus RNA sequencing — the latter performed on snap-frozen tissue — and spatial techniques, including 3-D reconstruction, to identify the cell types present and understand how they are located and communicate with each other.

## GASTROINTESTINAL: SMALL INTESTINE AND COLON

The colon and small intestine serve as a critical barrier separating ingested materials and microbes from the rest of the body. Both organs are critical for absorption, with nutrient uptake occurring in the small intestine and absorption of salts and water in the colon controlling electrolyte levels throughout the body. Simultaneously, the colon excludes dietary waste and microbes from systemic exposure. To maintain these selective barriers, the epithelial lining of the small intestine is renewed every three to five days and that of the colon every 72 hours. This remarkable rate of self-renewal occurs in microanatomic structures — crypts — where intestinal stem cells divide and mature into more than a dozen specialized intestinal epithelial cell types[143]. The intestinal epithelium surface is primarily composed of epithelial cells of which there are three major types of cell types: absorptive columnar cells; mucus-secreting goblet cells; and hormone-producing enteroendocrine cells. These epithelial cells are continuously replaced by cells that are produced from stem cells that reside in the base of the crypt. The interaction of these cells with gut-resident immune cells, stroma, and enteric neurons is critical for maintaining the gut's barrier function. The gut immune system — particularly T cells, B cells, and more than a dozen types of innate immune cells — is influenced by the presence of the gut microbiome, which is responsible for "educating" the immune system. These microbes are now increasingly appreciated to affect many aspects of human health, including metabolic disease, inflammatory disease, autoimmune disease, allergic disease, asthma, anxiety and depression, and even predisposition to neurodegenerative diseases like Parkinson's[144].

**Why a gut cell atlas?**

There are more than 40 cell types identified to date in the colon and small intestine (hereafter, "the gut"), including rare cell types such as enteroendocrine cells, mast cells, innate lymphoid cells, and gut "pacemaker" cells[143,145] (**Figure 13**).

A complete atlas of the cells in the gut would provide a critical resource to not only reveal the basic biology of this essential organ but also provide the information necessary to improve clinical care and diagnostics for a wide range of diseases. These include diseases such as inflammatory bowel disease — ulcerative colitis and Crohn's disease — microscopic and collagenous colitis, colon cancer, food allergy, and neurodevelopmental or neurodegenerative conditions.

While genetic risk factors are known for many of these diseases, the initiating cells and processes that trigger disease remain unknown. Providing a baseline for the normal composition of cells across age, sex, ethnicity, and exposure to antibiotics and other drugs will serve as a foundation for more detailed atlases of disease progression, such as the evolution of colon polyps into full-blown cancer. In addition, a full catalog of cell types in the gut will advance our understanding of therapeutics, such as immune checkpoint blockade (e.g., anti-PD1) for cancer, in which tumor regression is often accompanied by a toxicity condition known as checkpoint colitis [146,147].

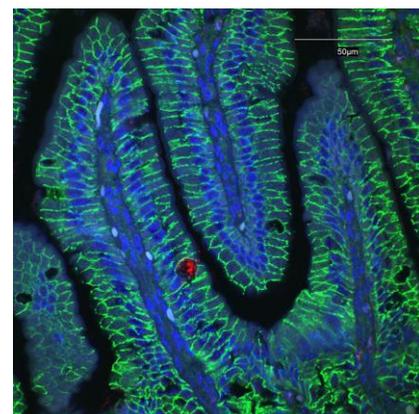

Figure 13. **Normal small intestine.** e-cadherin, an epithelial marker (green); RELMbeta, an anti-parasitic agent (red); DAPI stains the nucleus.

**What are key considerations for a gut cell atlas?**

**A pilot phase** will determine the approaches for each key aspect necessary for a first draft.

*Sample sources*. The colon (but less so the small intestine) is also an easily accessible tissue, since patients receive colonoscopies at all ages and regular screening from age 50. This means that there are opportunities to collect some biopsies from healthy volunteers, although entire organs require transplant organ donors (**Section 2**). Additional sources are resection and biopsy specimens for colon pathology, which often include healthy adjacent colon tissue. In particular, full thickness tissue will be essential to capture all non-epithelial elements. These cannot be obtained by biopsy, but will rely on the availability of surgical specimens.

*Cell sampling*. To comprehensively profile the human colon and small intestine, where many cells are present in radically different frequencies, we will need to develop appropriate protocols for tissue dissociation to capture even rare or fragile cell types from the biopsy, such as stroma and submucosal neurons. In particular, neurons may only be efficiently recovered by single-nucleus analysis, and very rare cell types may require either very large cell numbers at the top of the Sky Dive or enrichment thereafter.

*Spatial organization*. An early cell census of the mouse small intestine not only captured all known epithelial cell types and several previously unknown subsets, but also highlighted that cell proportions and subsets vary between fine structures and between anatomical regions. A sampling of both aspects, with both cellular and spatial methods, would help determine the appropriate mapping strategy across this large organ. In particular, the small intestine and colon are massive in size. From proximal to distal, the gastrointestinal tract is divided into specialized regions within the small intestine (duodenum, jejunum, ileum) and colon (right, transverse, left colon, and rectum). At the histological level, the small intestine and colon have stereotypical repeating structures. In the small intestine, villi — finger-like projections of the internal intestinal surface — are covered by a layer of epithelial cells. Of note, the surface of the colon carries no villi, yet its architecture is otherwise comparable. The crypt structures supporting the epithelial stem cell niche reside at the base of adjoining villi. Beneath the epithelial layer, is the lamina propria, which is comprised of connective tissue and stromal cells that are penetrated by blood vessels, lymphatic vessels, and nerves. Beneath the lamina propria, is a layer of innervated smooth muscle that is critical for digestive motility. Importantly, cells of the immune system can traffic to and reside in any region of the intestinal substructure. For example, intraepithelial lymphocytes reside between epithelial cells at the mucosal surface, and many additional immune cell types populate the lamina propria. The mucosa of the intestine also accommodates organized lymphoid structures such as Peyer's patches (small intestine) and isolated lymphoid follicles (colon) where immune responses are mobilized and tailored to different infectious agents. In addition, specialized cells in the follicle associated epithelia (FAE) residing above the Peyer's patches (small intestine) and lymphoid follicles (colon) facilitate antigen entry and processing by these adjacent immune residents. Thus, an atlas should include sampling from each macro-anatomical region, as well as partition the distinct micro-anatomical structures in each.

*Organoids*. The recent revolution in organoid biology has built on the pioneering work[148,149] with intestinal organoids. Because of the ease of production and the close resemblance to human organs in health and disease, organoids hold great appeal for translational research and invite an almost immediate application into the clinic. The current versions of organoids should still be considered an abstraction of their in vivo counterpart: nerves, blood vessels, supporting tissue, and immune cells are absent, and as a consequence, disease processes are only partially recapitulated. Thus, a pilot would assess the composition of matching organoids to test the extent to which it mirrors that of in vivo tissue. This will be in partnership with the Human Organoid Atlas.

**Draft Atlas 1.0**

A first draft atlas will expand the pilot to comprehensively characterize gut cellular subsets[7,8,150] and integrate these spatially into a comprehensive picture of the human gut as a reference for understanding and treating gut disease.

A principal goal of the first draft atlas should be to capture as much of the cellular diversity in the gut as possible. In a draft colon atlas, we will sample four sites: the right colon, transverse colon, left colon, and the rectum. For the small intestine, mostly the ileum will be sampled from healthy living research participants (the limit of which routine colonoscopy can reach), whereas all three regions of the small intestine (duodenum, jejunum, and ileum) will be studied from resections or organ donors. Twenty samples for each of these regions will be collected from each of at least three geographically (and ideally ethnically) distinct sites; individuals from both genders; all adults representing a variety of ages (between 20 and 70). Each of these tissue sites will be represented by two punch biopsies, 3 to 5 mm in diameter, from neighboring sites. To characterize cell types, we will apply both single-cell and single-nucleus RNA sequencing and spatial techniques, including 3-D reconstruction, to identify the cell types present and understand how they are located and communicate with each other.

## SKIN

The skin, our most visible tissue, is intimately tied to our identity, forms the outer covering of the body, and serves as a protective interface with the environment [151]. It is not only a physical barrier but also a defensive one, providing the first immunological protection against pathogen and external challenge.

**Why a skin cell atlas?**

There are several reasons to pursue a pilot-scale atlas of the skin. First, the wealth of knowledge gleaned from such an atlas will be of broad scientific, clinical, and commercial benefit. Second, the tissue's unique accessibility can help it serve as a prototype for organ-based HCA programs and inform how to sample a tissue that is distributed across the entire body. Encouragingly, initial scRNA-Seq studies of the skin of mouse models and cultured human keratinocytes, similar to those that might be involved in constructing the atlas, have already been published[152,153], and more are underway.

**What are key considerations for a skin cell atlas?**

**A pilot phase** will determine the approaches for each key aspect necessary for a first draft.

*Anatomical locations*. The skin is distributed across the body, and prior studies have shown that at least some component cells are substantially impacted by location[154], as are features such as the skin microbiome[155]. Early efforts will pilot the systematic analysis of single-cell transcriptomes and spatial profiling of skin from two anatomical sites: abdomen and breast, which are readily available from cosmetic surgery operations.

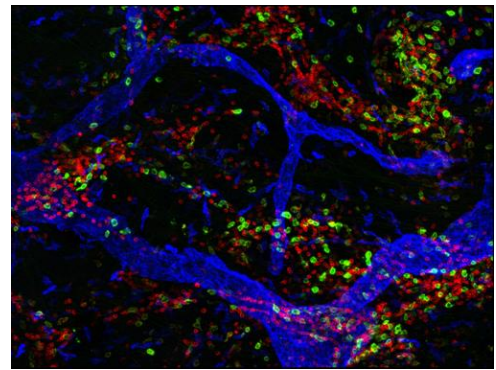

**Figure 14. Skin.** En face view of human dermis showing lymphatic vessels (blue), dendritic cells (green), T cells (red) and macrophages (blue). X10 magnification.

*Tissue disaggregation*. Skin is a challenging tissue to handle and disaggregate, while recovering all cell types faithfully (**Figure 14**). The pilot will seek to address several technical questions, including: (**1**) What is the minimum size of biopsy that can be used to isolate all cells? (**2**) Should tissues be separated into different components prior to single-cell analysis (for example, separating epidermis from dermis), or should all cells be isolated simultaneously? (**3**) What is the impact of the donor's health status, age, sex, ethnicity, and sun exposure on sample quality? (**4**) Which protocols, including single-cell and single-nucleus methods are most suitable to obtain good cellular representation? When the answers to these questions are determined, optimized protocols for cell disaggregation will be shared online. Already, three feasibility studies have been completed: an unbiased droplet-based single-cell transcriptome of enzymatically dissociated whole skin; plate-based adaptive sampling of skin mononuclear phagocytes; and spatial transcriptomic analysis of frozen OCT-embedded skin, which also helps serve as a control for biases in disaggregation methods.

*Optimizing additional methods and archiving samples*. To assay protein levels in cells, the single-cell profiling data of the pilot will be combined with legacy knowledge of human skin to devise mass cytometry or flow analysis panels for protein-level information. Other auxiliary collections will include the surface microbiome of skin samples[156] and blood archiving for subsequent correlation with the skin's immune cell populations[157].

*Spatial methods*. We will optimize the methodology for visualizing all cell types within intact skin, including, for example, rendering the tissue transparent and performing flat-mount labelling[158,159].

*Assessing relevance of additional data*. The pilot data will be related to all existing scRNA-Seq datasets in skin, including from mouse[152] and cultured human epidermis[160], previous bulk data of skin components[161], and other types of single-cell and population-based data, such as ATAC-Seq and proteomics, to assess the distinct contribution of each.

**Draft Atlas v1.0**

Once the pilot has developed the necessary methodology, a draft atlas will broaden its scope by collecting samples of different body sites (including from the same individual) and hair-follicle types. For each of the anatomical collection sites, at least 20 samples will be collected from age, site, ethnicity, and sex-matched individuals from each of three geographical sites. Detailed analysis of 10 to 20 research participants should capture all of the different cell types and states in adult skin, capturing diversity in terms of sex, ethnicity, and body site with the remaining samples. The cellular branch, will rely on both single-cell and single-nucleus profiling, the latter being important given the challenge of dissociating skin tissue. Some cell types require specific collection efforts: for example, capturing hair-follicle cell types would require a specific focus on hair follicles with at least three major ethnicities and five types represented, because these cells are difficult to disaggregate and will not be well represented in scRNA-Seq of abdominal waste skin. For the spatial branch, we will also perform spatial analysis on a subset of samples. Finally, some disease indications may be included, either within the draft or as a follow-up. Priorities include: atopic dermatitis, melanoma, basal- and squamous-cell carcinoma, scleroderma, and rare genetic skin diseases caused by known mutations but lacking a deeper mechanistic understanding.

# CARDIOVASCULAR: HEART

The heart is the earliest organ to form during embryogenesis, and its continued function is essential for sustaining a person's life, until the last heartbeat. Heart disease is the number one killer in the Western world, surpassing all cancers combined; the burden on society in terms of mortality and morbidity and the costs of health care are astronomical and rising[162]. Moreover, congenital heart defects affect one of every 100 live births and are collectively the most frequent noninfectious cause of death in the first year of life[163,164]. The heart and its supporting blood vessels have been studied for centuries, since the anatomical drawing of Leonardo de Vinci, and discoveries relating to its physiology in the 20th century led to the development of several drugs that have helped combat heart attack and heart failure. Despite a deep knowledge of cardiac cellular and organ physiology, we currently lack the detailed molecular information that would fully capture the breadth of cell types found in the complex, densely populated cardiac tissue.

**Why a heart atlas?**

Defining the cell types that comprise the heart and the coronary arteries — its supporting blood vessels — and their developmental origins will be critical to refine our understanding of heart function in health and in disease. While we understand the broad anatomical compartments that comprise the heart — left and right atrium, left and right ventricles, outflow tract, interventricular septum — these are further subdivided into sub-compartments with distinct functional and cellular characteristics. For example, the right atrium has a set of specialized cells, the sinoatrial node, which is the main pacemaker of the heart. The ventricular walls have at least three types of cardiac myocytes that promote gradients of electrical properties based on differential expression of ion channels. Furthermore, heart tissue contains, in addition to myocytes, a multitude of cell types essential for the structure and function of the heart — for example, fibroblasts, endothelial cells, cardiac nerves, macrophages, and many other cell types. The developmental biology of the heart has been well studied in model organisms[165,166], but despite current advances, a large-scale cellular atlas is lacking. A Heart Cell Atlas will be a rich resource to better define the cellular composition of the heart, which will lead to a greatly improved understanding of physiology and disease (**Figure 15**). Inclusion of iPS cell-derived cardiovascular cell types will enhance the heart atlas by helping devise both more accurate models of disease and more precise strategies for cardiac regeneration, thus facilitating rational development of regenerative strategies.

**What are key considerations for a heart cell atlas?**

A Heart Cell Atlas should include a broad diversity of anatomical sampling locations across a diverse population of healthy adults, supplemented with similar sampling from various disease types. A heart atlas should also include human developmental samples. An additional consideration for current efforts in disease modeling would be to include in an atlas a comprehensive set of cells obtained by in vitro directed differentiation of human induced pluripotent stem (iPS) cells. These hold much promise for in vitro modeling of human disease and for high-throughput screens aimed at finding new therapies, but it is not known if the identity of the cells obtained from iPS cells reflect the cell types found in the human heart.

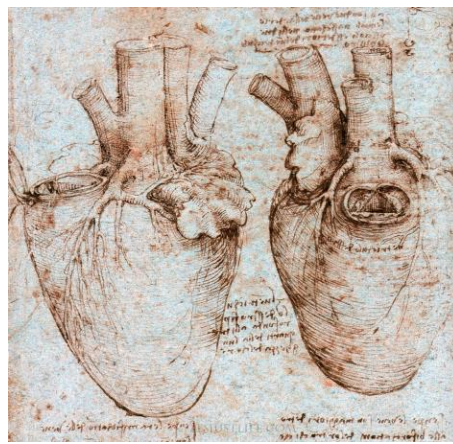

**Figure 15. Anatomical drawing of the human heart by Leonardo da Vinci.** Image: https://www2.warwick.ac.uk/fac/med/study/ugr/mbchb/societies/surgical/events/cardiac_leonardo_da_vinci/

The following considerations should be carefully weighed in **a pilot study**, ahead of a comprehensive draft

Heart Cell Atlas.

*Anatomical sampling.* Cardiac anatomy is well defined, and its complexities and associated physiology are well understood. This makes it possible to sample discrete anatomical compartments with precision. Ideally, this sampling should be coordinated with cardiac anatomists and the Common Coordinate Framework Working Group, who could develop a consistent and broad sampling strategy.

The key challenge is that it is uncommon to have access to healthy, intact hearts for rapid processing, and it will require considerable coordination between clinicians, anatomists, and HCA scientists to achieve this. A simpler approach would be to define a minimal set of regions that can be consistently and rapidly sampled either as live biopsies or as easily obtained post-mortem samples with limited scope. In addition, human developmental samples would ideally be collected from diverse anatomical locations based initially on known developmental landmark structures, but aiming for eventually diverse and comprehensive sampling.

*Sample preparation.* An important challenge in developing a pilot heart atlas is the ease of dissociation of adult heart tissue into single cells. The dense fibrous nature of the adult heart is a particular challenge that will need determined effort to overcome. One solution is to usesingle-nucleus RNA-Seq, which does not require dissociation and will allow profiling of archived material. Dissociation is less challenging for human development tissue.

*Cardiac cells derived from iPS cells.* A handful of effective approaches have made it possible to derive enriched populations of human cardiac and vascular cells from iPS cell lines, including recent refinements to produce cells that have characteristics of relatively specialized cell types, such as atrial versus ventricular myocytes, pacemaker cells, and endothelial and smooth muscle cells. However, the definition of these cell types is based on limited examination of functional and molecular parameters. For example, it is not clear if ventricular-like cells represent left ventricle, right ventricle, epicardial or endocardial myocardium, or a "hybrid" cell type that does not correspond to any endogenous cell type. Mapping of in vitro–derived cell types to a cell atlas reference will be invaluable. To this end, a pilot effort should include a limited number of well-characterized iPS cell lines with robust existing differentiation protocols used to derive broadly defined cardiac myocytes, endothelial cells, and smooth muscle cells for study and comparison.

**Draft Atlas v1.0**

In the first draft of the atlas, it should be sufficient to collect a relatively simple diversity of anatomical samples (similar regions of each atrium, each ventricle, and a coronary artery) across 20 individuals from each of at least three geographically (and ideally ethnically) distinct sites; individuals from both genders, and ideally of defined and varied stages in human development. A rich data set can be provided by scRNA-Seq, snRNA-Seq and scATAC-Seq, and if possible, samples suitable for spatial visualization of gene expression should be obtained. Spatial visualization will be accomplished using multiplexed fluorescent in situ hybridization on histological sections. Human developmental samples would initially be limited to developmental stages largely determined by availability. Knowledge of cellular diversity and identity from this pilot atlas will guide the development of strategies to more comprehensively map the heart and to include more variety of anatomical sampling sites as well as more diverse populations, including patients with defined diseases.

# DEVELOPMENTAL BIOLOGY AND STEM CELLS

Cells can only arise from other cells. It is therefore critical to trace where cells come from, and to identify their more immature predecessors, to fully understand their nature. Furthermore, the lineage and fate relations of cells provide a key aspect of their identity, helping in the elusive definition and validation of cell types and transitions. Developmental biology and stem-cell biology concern themselves with these questions—how a fertilized egg develops into all the various cell types that make up the fully formed organism, and how tissue-specific stem cells ensure the maintenance and repair of adult tissues.

Cell fate decisions during development and stem-cell differentiation are executed at the level of individual cells and impacted by signals from their microenvironment to the organismal level. Conventional, population-average gene expression measurement techniques therefore have only limited utility to identify the underlying molecular processes, and they cannot identify continuous trajectories that are found in such data sets. As a consequence, the developmental biology and stem-cell research communities have rapidly developed and adopted new single-cell profiling techniques[167-170]. These have the power to help create an atlas of progenitor cells that will be transformative for the developmental biology and stem-cell research communities.

**Why a stem-cell/developmental biology atlas?**

A comprehensive human single-cell gene expression atlas, complemented with corresponding mouse data, will serve as an important reference point for developmental biology and stem-cell research. These two research communities are uniquely placed to connect single-cell molecular profiles with single-cell biological function. This, in turn, will transform a broad spectrum of biomedical research with direct implications for human health, including tissue repair and regenerative medicine, developmental syndromes mostly affecting children, aging, degenerative diseases, and cancer. In particular, adult tissue stem cells, and the subversion of their function, play critical roles in major human pathologies ranging from aging to degenerative disease and cancer. Linking tissue stem-cell biologists with organ-specific pathologists will therefore also have significant long-term benefits when comparing healthy and diseased tissues.

**What are key considerations for a stem cell/developmental biology atlas?**

A unique aspect in a developmental atlas is the temporal aspect. This promises to transform static single-cell measurements into an appreciation of a cell's current state, past history, and likely future, with wide-ranging impacts for both basic and translational research.

*Tissues and stem cells*. The atlas should encompass both adult stem-cell processes in tissues such as brain, heart, liver, kidney, and pancreas (in partnership with the effort around the relevant tissue), in vitro systems, and developmental time courses (including the placenta). For example, a pilot project studying mouse gastrulation using scRNA-Seq is already producing exciting results[171]. This key stage of mammalian development provides an excellent test case for new single-cell technologies, since all major cell types are being formed and the whole embryo (up to 50,000 cells) can be profiled at single-cell resolution. Application of emerging technologies for spatial single-cell profiling, lineage reconstruction, and real-time analysis should therefore all be applied to this system. Adult human tissues with adult stem cells, such as the skin or mammary gland, are readily accessible and/or have powerful clonal stem-cell assays and, therefore, also represent ideal test cases, to be done in partnership with the effort around the relevant tissue. In particular, stem-cell biologists should become fully engaged in any planned organ-specific HCA pilots to ensure that cell-sampling strategies capture adult stem cells, which can be exceedingly rare. Finally, placental samples from diverse patients should be collected for study.

*Spatiotemporal analysis.* The ability to perform complex single-cell analysis in 3-D could also be applied to in vitro–cultured developmental samples, organ slices, or cells grown on tissue-engineered scaffolds. This, in turn, could be exploited to perform time-course analysis of developmental or stem-cell differentiation processes at single-cell resolution (4D analysis). It could also provide novel avenues to investigate disease phenotypes from patient tissue or investigate the potential therapeutic effects of small molecules at both the molecular and cellular level.

**Draft Atlas v1.0**

Analysis of samples taken along developmental time courses will be important to reveal insights into human development in addition to those generated from adult tissues. Only some time points of human development can be obtained. Therefore, analysis in model organisms would be particularly complementary and could be potentially pursued. An initial Developmental Cell Atlas HCA project will focus on later stages of development and specific tissues (e.g., liver, kidney, and skin during human development are in a current pilot) and will process 20 samples for each of these tissues across two geographical sites, with scRNA-Seq or snRNA-Seq, as well as spatially resolved methods. It can also include [a Placental Cell Atlas of](#) samples from research participants of diverse genetic backgrounds, ages, and disease states.

# PEDIATRIC

It is important to include pediatric tissues in the HCA as one-quarter of the world's population is 14 years of age or younger. Physiologically, children are not just small adults: their developing bodies behave differently in both health and disease. As they grow and develop, their bodies change at the cellular level in ways that are poorly understood. For the children unfortunately suffering from disease, these changes impact not only how the disease manifests but also the efficacy of treatment. A better understanding of the bodies of children at the cellular level and the changes they undergo will be critical to developing better diagnostics and treatments for the diseases of childhood, which afflict the 2% of school-age children in the U.S. in fair or poor health.

To support this vital area, a Pediatric Cell Atlas (PCA) will generate an age- and cell-based cellular-level atlas of pediatric tissues. The PCA will create comprehensive reference maps of cells from all pediatric tissues and systems, sourced across several childhood ages ranging from birth through adolescence. These references will provide the basis for both understanding the cellular basis of child health and development and for future studies in diagnosing, monitoring, and treating pediatric disease. While a full Pediatric Cell Atlas will have to mirror the effort of the entire HCA, a carefully planned draft can be generated to maximize early impact, leveraging the work in the main branch of HCA and partnering in each system.

**Why a Pediatric Cell Atlas?**

Disease and medical interventions manifest differently at the molecular level in children and adults. There are well-known age-related differences in response to anesthesia and various medications, and similar pediatric diseases differ when compared by developmental age. However, the physiological basis of many of these age-related differences is not well understood.

Any high-resolution study of pediatric disease would require comparison with similar studies of age-matched normal pediatric tissue — a resource that does not currently exist.

There is currently little to no understanding of normal cellular function within pediatric tissues, how these associated cellular processes relate to the course of normal child development and maturation, or how the cellular composition, state, and function in pediatric tissue compares with that in adults. Contrasting children's cells and tissues across ages through to adulthood will provide a lens into pediatric treatments and diseases.

Thus, a PCA will provide a normal reference for studies on the origins of diseases affecting children; reveal physiological differences in tissues between adults and children; provide insight into growth and development of human tissues across ages that may also provide a perspective into organ regeneration; and provide insight into the basic biology of growth, development and maturation.

**What are key considerations for a Pediatric Cell Atlas?**

*Sample procurement*. The scale of the PCA endeavor will require contributions from large teams spanning clinical and basic research across multiple institutions and across a range of ages in all pediatric tissue systems. This can also be enhanced through partnerships with the organ-specific teams working on adult tissues. Most tissue will be sourced as incidental from surgeries. Although acquiring tissue from live donor sources is preferred, it is also possible to rapidly harvest samples at the time of organ donation and at autopsy from appropriate donors with the generosity of consenting families. It is reasonable to expect several dozen donors a year across the current PCA network.

*Systems*. In principle, the PCA, mirroring the general HCA, would study all pediatric tissues, systems, and organs. However, that scope is excessive for an initial draft, and it does not leverage the knowledge and technical effort of the HCA in the corresponding adult tissue. Thus, a PCA pilot, followed by a first draft, will focus on four key areas: brain, skin, the immune system, and the colon.

**Draft atlas v1.0**

The first draft of the Pediatric Cell Atlas would focus on four key systems or tissues. In each system, 20 samples from each of at least three geographically (and ideally ethnically) distinct sites will be obtained; individuals from both genders; all pediatric — from birth to 18 years of age. For each sample, both scRNA-Seq and/or snRNA-Seq profiling will be performed, as well as spatial analysis for a subset of samples and auxiliary profiling as needed.

A *Pediatric Brain Cell Atlas* will map variability in brain cells from unaffected brain tissue directly after the resection from 100 resective surgeries.

A *Pediatric Immune Cell Atlas* will study children of different ethnic backgrounds, aged ~1 through 18 years, as well as newborns and preterm infants, covering primary (bone marrow, thymus) and secondary lymphoid tissues (lymphatic fluid, blood), as well as other immune relevant tissues (lung, liver, intestine, skin).

A *Pediatric Epithelial Colon Cell Atlas* will focus on colon mucosal biopsies as well as surgically resected tissue from normal and disease patients (e.g., IBD, Hirschsprung's disease, immunodeficiency).

A *Pediatric Skin Cell Atlas* will study skin variability from surgical samples sourced from diverse pediatric patients aged 0 to 18. A series of skin disorders will also be available for profiling from surgical resections, including congenital melanocytic lesions, spitz nevi, epidermal nevi, vascular tumors, vascular malformations, hair follicles and scalp, and lipomas.

## ORGANOIDS

Recent advances in 3-D culture technology allow pluripotent — ES cells, iPS cells — and adult human stem cells to exhibit their remarkable self-organizing properties[172]. When sequentially exposed to defined signaling molecules, both types of stem cells can be converted into organoids: small structures that reflect key properties of organs such as kidney, lung, gut, brain, or retina[172] (**Figure 16**). Both healthy and disease tissues can be converted into organoids, and thus organoid technology can be applied to model human organ development, physiology and various human pathologies "in a dish," including cancer "tumoroids" that complement animal experimentation as an increasingly faithful human model.

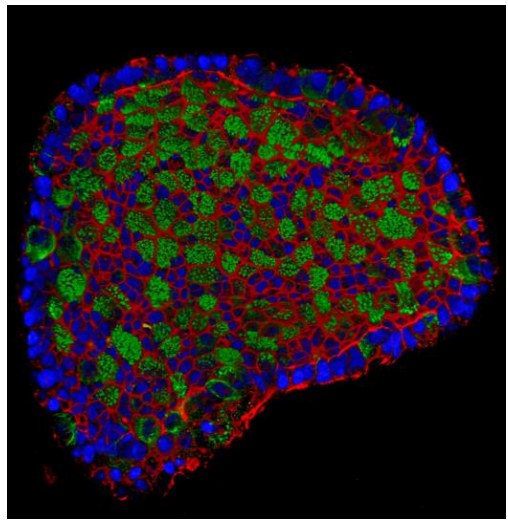

**Figure 16. The inside surface of an airway organoid.** Airway organoids can be grown from biopsies but also from airway lavages of individuals of any age. They contain the various cell types of the original tissue: ciliated cells (cilia are stained in green), basal cells, club cells etc. Blue: nuclei (DAPI). Red: actin. Image: Norman Sachs.

**Why an Organoid Cell Atlas?**

Organoids provide a useful tool for unraveling aspects of biology that have been difficult to pursue in model organisms or in living humans. For example, human brain organoids have recently been applied to document how Zika virus causes brain-development abnormalities[173]. In addition, organoids may hold the key to precision medicine: patient-derived organoids can be used as "avatars" to predict drug response in a personalized fashion — and are already being used for this purpose in cystic fibrosis[174,175]. The technology also holds great promise for regenerative medicine as an alternative to whole-organ transplantation.

However, the current versions of organoids are still an abstraction of their in vivo counterparts. Thus, it is important to have a detailed knowledge of the composition of both the original tissue and the organoid to determine whether core cell types are missing or whether variations in organoid tissue across patients result because of biological or technical factors. In addition, accessory cells such as nerves, blood vessels, fibroblasts, and immune cells are absent in the organoid, and as a consequence, disease processes are only partially recapitulated. Similarly, the process of derivation may not be optimal because some of the factors secreted by the niche are missing. Yet it is anticipated that the potent self-organizing properties of organoids may extend beyond their current boundaries and allow the proper incorporation of additional cellular (or microbial) elements — which can be determined by developing organoid and matching-tissue atlases.

An open-access Organoid Cell Atlas will be a road map for researchers taking on new organoid challenges, allow us to assessing the changes that occur in disease and use them for better disease modeling and drug screening. Because of the ease of production of organoids, and their close resemblance to human organs in health and disease, an Organoids Cell Atlas holds great appeal for translational research and invites an almost immediate application into the clinic.

**What are key considerations for an organoid cell atlas?**

An Organoid Cell Atlas, collected in conjunction with the originating endogenous tissue, aims to yield an unprecedented understanding of the human body, provide faithful models of exceptional utility for research, and transform precision and regenerative medicine. In particular, it should allow us to definitively assess how faithful and reproducible current organoid technology is and to improve organoid-generating techniques iteratively in an informed, data- and model-driven, way — through better representation of the true human tissue composition and organization in the dish.

*Organoids and matching tissue.* To create a Rosetta Stone of the cellular composition of a key set of organoids, along with matching originating tissue, orgnaoids will be generated from tissue collected in coordination with the relevant HCA effort in that tissue. Each initiating human tissue specimen will be partitioned in two, and one portion will be analyzed immediately, using both single-cell and spatial genomics methods to determine its full endogenous composition. In parallel, the other portion will be used to generate matching organoids, and those organoids will be subsequently analyzed with the same techniques.

*Comparison*. Computational algorithms will be used to compare the source tissue to its organoid model, for cellular composition, cell states, and spatial organization, thus determining in a precise and comprehensive way the extent to which the model faithfully recapitulates the human. (In cases initiated from iPS cells, such as brain organoids, the comparison will be to the closest possible specimen source.)

*Optimization*. Through the comparison, opportunities for optimization will be identified: the process will then be iterated, to generate better organoids with more comprehensive composition of cell populations (e.g., sub-types of epithelial cell in the gut; sub-types of neurons in brain), the effective co-culture of accessory cell type (e.g., addition of immune cells), or the introduction of new secreted niche factors for more efficient and reproducible organoid generation (e.g., based on secreted factors expressed by stromal cells in the endogenous tissue).

**Draft Atlas v1.0**

A draft atlas should sample each of the currently used experimental organoids and their corresponding tissues (which should be coordinated with the appropriate organ cell atlas). These include central nervous system, stomach, lung and thyroid, small intestine, liver, and kidney organoids grown from pluripotent stem cells, as well as small intestine and colon, stomach, liver, pancreas, and prostate organoids from adult stem cells. A key focus will be on organoid variability — across starting specimens, across replicates from the same starting specimens, and across varying protocols for derivation. For each organoid category, 20 healthy specimens for deriving organoids should be sampled for each tissue from each of at least three geographically (and ideally ethnically) distinct sites; individuals from both genders; all adults (20 to 55 years of age under baseline), and each should be derived into organoids at least in triplicate. Both single-cell profiling and spatial methods will be applied to verify that these organoid-generating protocols create organoids that are similar in cellular and spatial composition to each other and to the tissues they represent.

# MODEL ORGANISMS

> This section will be added in a later draft. First, however, a workshop is being convened to discuss the goals and best approaches for building atlases of model organisms.

# 7. THE HCA CONSORTIUM

The HCA builds on a historical legacy of large-scale scientific projects involving many international participants, and consequently can draw inspiration and knowledge from these projects while developing an organizational and governance model unique to its challenge (**Box 2**). In particular, the consortium must engage a broad range of experts across its four pillars — the collaborative Biological Networks, Technology Forum, Data Coordination Platform, and Analysis Garden — and each pillar must engage in deeply collaborative, hands-on work to build the atlas. The consortium is also responsible for governance of the Data Coordination Platform and for key decisions and policies, including those related to the data-collection scheme, common coordinate framework (CCF), and data release; for convening the community in meetings and other forums; and for representation and negotiation around the atlas effort. To enable these activities, we have established a robust organizational and governance structure, led by an Organizing Committee, that constitutes Working Groups and delegates organization, advisory, and decision-making responsibilities to them where appropriate **(Figure 17)**.

---

**Box 2: Key features of past transformative projects to generate biological resources.**

Upon reflection on past projects to generate biological resources that have proven transformative in biology in the test of time, the following features emerged:
- A comprehensive approach to a fundamental biological unit that can propel progress in thousands of laboratories on diverse problems
- An audacious, but potentially tractable, scale
- A technology landscape with rapidly decreasing costs and rapidly advancing capabilities
- Intellectual flexibility in the community that allows goals to evolve
- Commitment to quality control, with rigorous focus on quality and full transparency
- International collaboration
- A strong leadership group that is chosen and led by scientists
- Both larger centers and smaller groups contribute according to strengths and capacity
- Development of a data-sharing infrastructure
- Regular scientific meetings to bring the community together
- Strong commitment to data sharing, with associated technical solutions
- Clear, inspiring communications for scientists, funders, and the public
- Attention to ethical issues, such as global equity and privacy
- Supportive funders

---

# GOVERNANCE AND ORGANIZATION

## HCA GOVERNANCE[1]

### Organizing Committee

The HCA is steered and governed by an **Organizing Committee (OC)**, which is the decision-making body of the HCA.

**OC responsibilities.** The responsibilities of the OC include convening the community through regular meetings, workshops and jamborees; coordinating and authoring key documents; defining scientific values and ethical principles; defining and upholding processes including QC standards and analytic standards; governing the Data Coordination Platform (DCP) and Common Coordinate Framework (CCF); coordinating HCA work products; communicating on behalf of HCA; representing and negotiating on behalf of the HCA with other entities and organizations; and polling the HCA community at regular intervals for input on issues, including performance of the OC. The OC does not generate data and is not a direct grantee or grantor for such purposes.

**OC membership.** The OC will consist of up to ~35 scientists. Considerations for new OC members include expertise, geographical representation, and diversity. Additional members are added to the OC by majority vote of the OC. The OC will periodically seek input from the HCA community on the scientific scope of its members, performance, and potential new members. The current (founding) OC consists of 27 scientists from 10 countries and diverse areas of expertise (**Table 2**).

**Terms**. All members will have five-year terms, which can be renewed once by a majority vote of the OC.

**Co-chairs.** The OC is led by two co-chairs, who are members of the OC. The co-chairs have five-year terms, which can be renewed once by a majority vote of the OC.

**Executive Committee**. The OC has an Executive Committee that is responsible for performing routine tasks between OC meetings, preparing meeting agendas, and providing guidance to the executive offices. The EC includes the two co-chairs and five additional OC members, with two-year terms, which can be renewed once by a majority vote of the OC.

**Executive Offices**. The HCA is coordinated by Executive Offices (EOs), which staff the OC in performance of its duties. The OC established four Eos, which are located in the U.K. (Sanger), U.S. (Broad Institute), European Union (Karolinska Institute), and Asia (RIKEN). The EOs' responsibilities include meeting organization; community coordination; coordination of writing of community outputs (e.g., reviews, commentaries, white papers); coordination of interactions with companies; supporting interaction with funders; triage of press inquiries; triage of community inquiries; registry and tracking of projects; and registry and tracking of members. Each EO will also take the lead on some general duties, as well as regional activities.

**Quorum**. A quorum for decision making by the OC will constitute (**1**) a majority of OC members at an in-person meeting that has been announced to the OC at least one month in advance, or (**2**) at least 75% of OC members responding by email to a proposed action that has been circulated to the OC.

---

[1] The text is the formal governance of the HCA, as developed by its Organizing Committee, and ratified by it on September 19, 2017.

| OC Member | Affiliation |
|---|---|
| Ido Amit | Weizmann Institute of Science, Israel |
| Gary Bader | University of Toronto, Canada |
| Peter Campbell | Wellcome Trust Sanger Institute, U.K. |
| Piero Carninci | Riken, Japan |
| Hans Clevers | Hubrecht Institute, Netherlands |
| Roland Eils | German Cancer Research Center; University of Heidelberg, Germany |
| Nir Hacohen | Broad Institute, MGH, USA |
| Arnold Kriegstein | University of California, San Francisco, USA |
| Eric Lander | Broad Institute, USA |
| Sten Linnarsson | Karolinska Institutet, Sweden |
| Partha Majumdar | National Institute of Biomedical Genomics, India |
| Miriam Merad | Mount Sinai, USA |
| Shalin Naik | Walter + Eliza Hall Institute of Medical Research, Australia |
| Garry Nolan | Stanford University, USA |
| Dana Pe'er | Memorial Sloan Kettering Cancer Institute, USA |
| Chris Ponting | Edinburgh University, U.K. |
| Steve Quake | Stanford University / Chan Zuckerberg Biohub, USA |
| Nikolaus Rajewsky | Helmholtz Association, Germany |
| Aviv Regev, Co-Chair | Broad Institute, MIT, HHMI, USA |
| Ehud Shapiro | Weizmann Institute of Science, Israel |
| Jay Shin | Riken, Japan |
| Michael Stratton | Wellcome Trust Sanger Institute, U.K. |
| Henk Stunnenberg | Radboud University, Netherlands |
| Sarah Teichmann, Co-Chair | Wellcome Trust Sanger Institute, U.K. |
| Alexander Van Oudenaarden | Hubrecht Institute, Netherlands |
| Jonathan Weissman | University of California, San Francisco, USA |
| Barbara Wold | California Institute of Technology, USA |

**Table 2. HCA Organizing Committee members.** Current co-chairs are noted. (EC members have not yet been chosen by the OC.)

**Working Groups**

The OC establishes **Working Groups** and mandates them to take on specific key areas. At the moment, these include the Analysis Working Group (AWG), the Meta Data Working Group (MDWG), the Common Coordinate Framework Working Group (CCFWG), the Standards and Technology Working Group (STWG), and the Ethics Working Group (EWG). Each Working Group has two co-chairs; one co-chair is a member of the OC and the other co-chair is external to the OC. The Working Group co-chairs together select the other members of the Working Group. Working Group members will have three–year terms, which can be renewed once by a majority vote of the OC.

The OC governs the Data Coordination Platform (DCP), which includes making all policy decisions concerning the DCP; approving the overall plan for the DCP; and ensuring the plan's successful execution by the major developers of the DCP. The OC will establish and appoint a **DCP Governance Group (DCPGG)**, which will report to the OC, to oversee the implementation of these policies, by providing guidance and making decisions concerning certain key topics, including definition of data manifest; official analysis pipelines; required metadata to reflect data collection standards; common coordinates framework; and any formal "release portal." The DCPGG will be led by two OC members; will include at least one member from each of the AWG, MDWG, and CCFWG; and will include at least three additional experts from the community.

The OC will convene, on a quarterly basis, a DCP Coordination Meeting — involving the DCPGG, the major developers of the DCP, and others as appropriate — to review progress and assist the OC in developing policy.

**HCA Meetings**

The HCA OC convenes and advertises meetings, workshops, and jamborees.

**HCA Members**

Any individual may become an HCA Member by registering at the HCA Member Registry and agreeing to abide by the principles of HCA, as stated in the HCA White Paper, especially including its ethical standards. HCA members are invited to attend HCA community-wide meetings and discussions; join the HCA Slack channel; be on mailing lists; and so on.

In addition, an HCA member who is a participant in an HCA project (defined below) or a member of an OC-designated HCA group will be designated as an HCA Collaborating Member. Certain types of scientific meetings, opportunities, and activities will specifically engage HCA Collaborating Members.

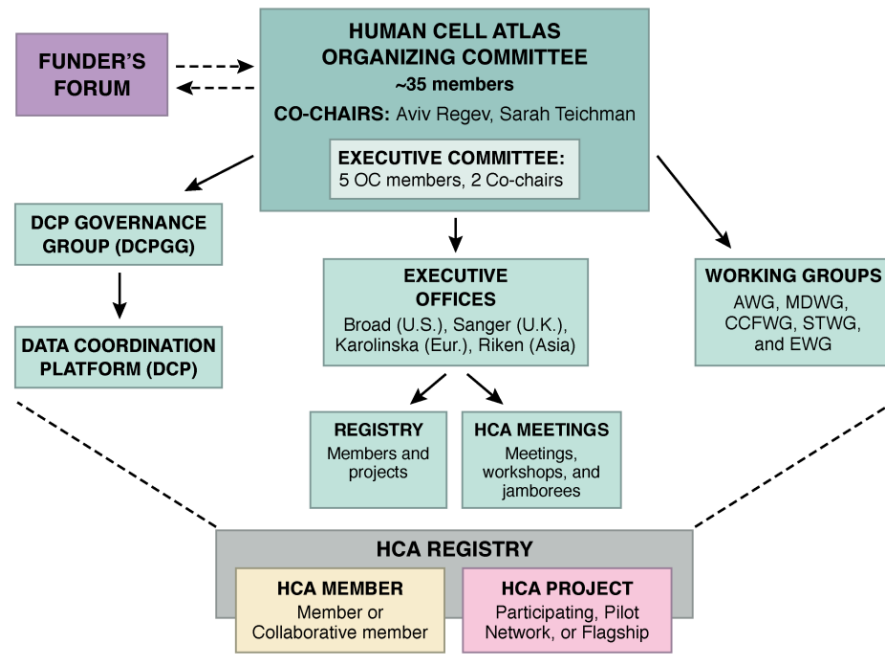

**Figure 17. Organization of the Human Cell Atlas.**

Any scientific project related to systematic biological characterization at single-cell resolution may become an HCA Project by registering in the HCA Project Registry. The registry will include a description of the project, its strategy, and its investigators. By registering a project, its investigators affirm their commitment to abide by HCA standards, including the Information Release Policy (below). Projects will fall into three categories:

1. **HCA Participating Projects.** Any project (including those focused on data generation, experimental method development, or computational methods development) may be an HCA Participating Project, simply by registering.

2. **HCA Network Projects**. An HCA Network Project is an HCA Participating Project that commits to liaise with other network participants.
3. **HCA Flagship Projects.** An HCA Flagship Project is an HCA Network Project that is aimed a delivering a component of the HCA Draft Atlas Plan 1.0; adheres to the overall framework of Plan (e.g., being comprehensive, adhering to technology standards, engaging domain expertise, having available preliminary data, and having substantial impact); and involves funding of at least 20 M Euros over its duration.

**Information Release Policy**

All HCA Projects will commit to ensuring that data developed by or for the project will be (i) deposited in the DCP through regular data release, and (ii) made available in an open access manner to the maximal extent allowed by ethics (e.g., some metadata may be restricted). The DCP will tag the data to make it citable. Users will be free to use the data as they wish, apart from the obligation to cite the generators of the data and to not attempt to identify or contact individual participants contributing samples. The OC will work with journals to define opportunities for key data collection and analysis papers.

In addition, all HCA Projects will commit to publicly releasing all experimental and computational methods used for data generation and/or analysis and source code for software developed by or for the project.

**FUNDERS' FORUM**

The HCA works in close partnership with committed and visionary science funders across the globe. To provide funders with a forum for discussion without blurring boundaries with potential future grantees, the HCA has arranged for a Funders' Forum. The forum's goal is to create an organization for interested funders. It is reasonable to expect that at some point in the future, an investment in HCA will be required to join the forum, but we believe such restrictions are premature at this time. Ultimately, we expect the funders to define their roles and the forum independently. The HCA OC has set up and administers an email list to facilitate discussion among funders and introduces potential funders to this community. Critically, however, the HCA Executive Offices cannot access any of these emails.

The Funders' Forum consists only of funders, is distinct from the OC, and meets separately. Nevertheless, the forum coordinates with the HCA leadership and, ideally, meets with the leadership regularly.

**INTERACTIONS WITH OTHER CONSORTIA**

The HCA is one in a series of major consortium efforts going back to the Human Genome Project, and it builds on the scientific and organizational foundations laid by those previous consortia. We are committed to learn from past consortia and interact with active ones to gain inspiration and knowledge about the best ways to achieve success and to find new opportunities for collaboration.

Several members of the OC have experience as members or advisors of other large consortia, including the International Cancer Genome Consortium (Bader, Campbell, Majumdar, Stratton), FANTOM (Carninci, Clevers, Shin), 1000 Genomes (Lander), ENCODE/ModENCODE (Wold, Pe'er), ImmGen (Merad and Regev), BLUEPRINT (Stunnenberg), and the Human Genome Project (Lander).

We have taken inspiration and used practical experiences from existing consortium projects in specific areas, such as the Data Coordination Platform, sample procurement and donor consent, and building networks of biological collaborators. For example, planning the Data Coordination Platform has benefitted from experiences in building the platform and portals for the Human Genome Project, the 1000 Genomes Project, the Cancer Genome Atlas, the International Cancer Genome Consortium , and the NCI Cloud Pilot, among others; procedures for tissue acquisition are based on the experience and pipelines of the GTEx project and ENCODE and the lessons learned in human genetics efforts; the experiences and participants of the Immgen consortium and the BRAIN Initiative have helped us shape and execute plans for the engagement of biological communities; and lessons from consortia spanning from the Human Genome Project to the Global Alliance for Genomics and Health (GA4GH) are helping shape legal and ethical considerations around data sharing.

More generally, past experience has informed our commitment to technology development, intellectual flexibility, quality control, international collaboration, strong governance, data sharing, clear communication, high ethical standards, and the intention to work collaboratively with supportive funding organizations. Mechanisms developed to ensure multiple opportunities to join the project and to ensure transparency are key as well (**Box 2**).

Finally, we have already identified several consortia that are complementary to HCA and where there is clear mutual benefit to coordination. For example, for the auxiliary profiles in the cellular branch, we hope to join forces with consortia like ENCODE, the 4D Nucleome or the Human Protein Atlas. Both 4D Nucleome and the Human Protein Atlas consortia have expressed similar enthusiasm for complementary partnership. We will also interface with organ- or disease-specific programs, such as the Tumor Cell Atlas from the Cancer Moonshot and initiatives in the profiling of tissues such as kidney and brain. In particular, the HCA has put together a proposal to be a GA4GH driver project (through the HCA DCP and DCP GG) and is exploring opportunities to leverage and extend the HCA DCP in disease-specific projects (e.g., Kidney Precision Medicine Program). This gives us a wonderful opportunity to engage a wider community of scientists and expert communities and to integrate preexisting data into the construction of the atlas.

## PARTICIPATION AND REACH

**Establishing equity in the HCA**. Throughout history, geographical atlases were largely developed to serve global power centers rather than the general good. Conversely, in an effort to maximize impact, the HCA should strive for equity. This means distributing power, ensuring comprehensive coverage of samples relevant to all of humanity, and empowering the global research community, while respecting local laws, mores, beliefs, and traditions.

A fundamental question then becomes: how can the HCA best achieve these goals? As a start, it can learn from previous efforts that have faced similar challenges. For example, genome-wide association studies (GWAS) have been used to identify thousands of variants associated with hundreds of phenotypes, improving our biological understanding of a multitude of traits and diseases. Yet many GWAS have focused largely on individuals of European ancestry, raising questions about how generalizable their findings are. Notably, genetic diversity is only one set of factors we need to consider; experiential and environmental diversity could also have substantial effects on cells. Broadening ethnic and experiential representation in the HCA (and generally in biomedical studies) must therefore be a priority to ensure that the biological insights gained will be broadly applicable to diverse human populations.

One way to accomplish this is to sample from different locations. For example, the 1000 Genomes Project and the Simons Genome Diversity Project used 140 different populations from diverse geographic settings as a proxy for genetic diversity. This revealed that human genetic diversity could be grouped into five major continental clusters — Sub-Saharan Africans, Oceanians, East Asians, Native Americans, and West Eurasians — with substantial sub-structure within each, suggesting that representative sampling may be sufficient to enable the generation of a relatively thorough draft atlas.

When identifying appropriate samples and deciding how and where to profile them, the HCA should aim to foster equity in selection and community engagement (scientific and otherwise), yet strive to maintain uniformity and rigor. With somewhat similar goals in mind, the International Genome Sample Resource (IGSR) was established at the EMBL in 2015 to ensure the ongoing utility and relevance of the 1000 Genomes Project, while further extending it in existing populations and expanding it to new ones.

To meet similar goals, the HCA's design should be egalitarian and comprehensive on multiple levels.

- *Compositionally*, it should include variety along axes of gender, age, ethnicity, environment, and cultural tradition, as well as, in some cases, disease or disease susceptibility. This would enable characterization of each factor that may affect the molecular profiles of cells and be representative of humanity.

- *Organizationally*, it should be inclusive of all countries and researchers across educational backgrounds and stages of training. This will ensure continuity and comprehensive, informed coverage, and allow for careful consideration of potential regional barriers to success.

- *Educationally*, it should strive to touch all sectors of society through a combination of training programs, workshops, conferences, outreach, and on-site dissemination initiatives.

**Compositional considerations**. To provide the greatest utility to humanity, the HCA should itself be representative of it. Genders must be considered because sex impacts cells in both sex-specific tissues and across the body. Diverse ethnicities must be profiled to understand how underlying genetic diversity informs the phenotypic heterogeneity of different cellular subsets. Multiple environments must be surveyed to understand how factors external to the body — such as temperature, altitude, pollution, allergens, and microorganisms — influence cells. A range of cultures must be sampled to understand how lifestyle factors, such as diet and nutrition, alter cellular phenotypes. Finally, the target tissues of globally prevalent diseases must be characterized to provide essential references for diagnosis and treatment.

*Given limited financial, technical, and human resources, how can the HCA ensure representative sampling?* One path forward might be to profile a minimum of five populations from each of the broader geographic regions defined by the 1000 Genomes Project. However, such a top-down approach could undermine the overall aim of equity. A different strategy might be to empower distributed generation of the HCA, allowing regional partners, working with the HCA's OC, to establish local priorities and performance goals. Together, they would determine the right technical and organizational controls — including the establishment of local centers of excellence to enable rapid on-site processing using standardized protocols, know-how, best practices, and quality-control standards developed by the HCA — as well as engage local support structures (government, hospitals, philanthropy, etc.) to ease sample collection. Although such decentralization might make it more challenging to maintain the level of uniformity and rigor needed for HCA sample processing, it will ensure scientific equity, enhance standards, and empower scientists and communities globally. The HCA should establish the infrastructure (physical, experimental, and computational) and SWAT teams necessary to support such a distributed approach. Moreover, HCA scientists and engineers from the Technical Forum should work with field deployment in mind as they innovate, modify, and optimize techniques for diverse settings, expanding their reach in the scientific community. This would ultimately improve data quality, all while developing and leveraging local scientific talent, establishing essential resources, and promoting the universal relevance of the HCA.

To make local profiling possible, all engaged communities must establish experimental infrastructure; develop controls that take into account the potential computational challenges of distributed sample collection (a topic of current focus for the AWG; **Section 4**); design and test methods for processing; and carry out hands-on training programs and workshops to educate engaged participants in preferred methods. To kick-start such an effort, and design pilots to inform how to link broad profiling to more targeted efforts, an initial planning meeting should be convened by the OC in the coming year, similar to those conducted for other aspects (technologies, computation, CCF. and model organisms).

**Organizational considerations**. Equity should not only be evident in the samples analyzed by the HCA but also in the scientists carrying out the research. This means including, from the inception of the project, individuals from many countries, disciplines, educational backgrounds, and stages of training. These individuals must also represent the world's countries, ethnicities, environments, and cultural traditions. The OC already has scientists from four continents — North America, Europe, Asia, and Australia — and the HCA community has scientists from all continents. We are working to better engage more scientists from Latin America and Africa.

A key starting point for engagement — as has been the HCA approach for every community — is the first planning meeting. This would include participants across the communities that the HCA is meant to engage. To judiciously consider the local complexities associated with a global atlas, each geographical location should have representatives with backgrounds in science, law, government, business, and social sciences, and from diverse career stages. While broad inclusion could lead to a prohibitively large membership, a limited hierarchy could enable collective decision-making for several of the major outstanding questions, such as where and when to train those involved, how to perform sample collection, how to prioritize processing, where it should be performed, and more. Properly addressing these issues will require substantial outreach to educate, and be educated by, the global community on the ultimate needs and goals of the HCA. In the collection phase, the OC and its working groups will partner with each community to develop and deploy best practices in experimental design and protocols and in computational analysis, and to disseminate them through portals, educational workshops and SWAT teams. The DCP and related portals will be important for sharing data and helping facilitate global collaborations.

**Educational considerations**. To establish equity, global buy-in to the HCA must be ensured. This will entail developing training programs, workshops, conferences, and outreach initiatives that captivate and inspire while also guiding overall thought processes. To facilitate global performance, training programs should empower local generation of high-quality data that can be incorporated into the HCA, including in regions and situations where sample transport is not feasible for social, legal, or safety reasons. Similarly, to inspire relevant technological innovations across professions and stages — essential for global performance, especially in resource-poor settings — workshops will be convened to identify and tackle existing bottlenecks and difficulties. To keep a critical eye on progress, conferences to present progress should occur at locations around the globe and have financial aid available to ensure diverse participation. Finally, to build public trust and ease the substantial legal and ethical complications associated with equitable engagement and atlas composition, the HCA will need outreach initiatives that effectively educate the public using all available media.

Disseminating and refining the wants and hopes of the HCA requires the formulation and effective sharing of materials to engender support across sectors. Hence, comprehensive educational plans should be established. These should involve creating globally accessible outreach materials that speak to people from all countries and backgrounds, perhaps by relying on the existing know-how of online education platforms. In parallel, the technologists and computational biologists involved in the HCA should help design training programs, workshops, jamborees, and conferences. Hosting responsibilities for these meetings should also rotate through the regional bodies to ensure equity, with additional centralized support.

## ETHICS, LEGAL, AND REGULATORY PRINCIPLES

**Ethics and regulatory principles**. Rapid, international and open data sharing is a key tenet of the HCA. As with other large-scale community resources that promise broad utility, the atlas will require a data-sharing framework. Previous approaches include the Bermuda, Fort Lauderdale, and Toronto Statement Principles (2009)[176] that ensure explicit, global cooperation between funding agencies, data producers, and data users. In these frameworks, data is rapidly made available and usable, proper incentives for data production are maintained, and the data's generation and use respect and protect participants. As noted in the relevant section of the HCA Governance (**Appendix I**): "*All HCA Projects will commit to ensuring that data developed by or for the Project will be (i) deposited in the DCP through regular data release, and (ii) made available in an open access manner to the maximal extent allowed by law and ethics (e.g., access to some metadata may be restricted). The DCP will tag the data to make it citable. Users will be free to use the data as they wish, apart from the obligation to cite the generators of the data and to not attempt to identify or contact individual participants who contributed samples. The OC will work with journals to define opportunities for key data collection and analysis papers.*"

A guide for achieving these goals can be found in the *Framework for Responsible Sharing of Genomic and Health Related Data* (http://genomicsandhealth.org/about-the-global-alliance/key-documents/framework-responsible-sharing-genomic-and-health-related-data) established by the Global Alliance for Genomics and Health (GA4GH), an international coalition with a mission to promote data sharing and to improve human health. The framework emphasizes the importance of maximizing the availability and re-use of data so future patients can benefit from scientific progress, while minimizing risks to participant privacy and recognizing the contributions of researchers.

Ideally, any and all data in the atlas will be immediately and publically accessible to the international community, with the only limit on use being the creativity of biologists, data scientists, and engaged members of the public. To achieve this, the HCA OC will negotiate varying regulatory constraints governing the collection, use, and international sharing of samples and data, especially those that can be linked to identifiable individuals[177]. Tissue acquisition efforts are already focused on appropriately consented samples that maximize the availability of open-access data. Nevertheless, specific constraints can arise, both in the context of living research participants and in diverse jurisdictions.

In cases when fully open consent cannot be achieved, steps will be taken to remove identifiers. Anonymization, however, is not straightforward for high-throughput molecular data, and it may be desirable to maintain a link with living research participants over time to collect more data (we do not generally anticipate returning results to research participants). A set of guidelines for deceased donors should be considered and can learn from experience learned from the GTEx project. Where necessary, sensitive samples or data will be sequestered with appropriate access-controlled portions of the DCP, or locally (if required), while enabling open sharing of less sensitive data[178]. Where safeguards are needed, streamlined access processes will be in place to ensure qualified, trustworthy researchers can rapidly access data for legitimate research uses without jeopardizing data security[179].

In certain jurisdictions, researchers contributing to the atlas may need to seek either the permission of individuals for international sharing or an ethics waiver (https://genomicsandhealth.org/consent-policy-read-online). When privacy risks are negligible, researchers should ideally seek open consent from participants to enable unrestricted access and reuse by the international community. Where privacy risks are slight, researchers should instead seek broad consent — which is increasingly recognized in international statements and national regulatory frameworks — to future sharing and use, subject to ongoing and transparent governance mechanisms (https://cioms.ch/wp-content/uploads/2017/01/WEB-CIOMS-EthicalGuidelines.pdf)[180]. Specific consent requirements limiting the domains of use may still pose a barrier to data sharing in some countries[177].

Because assembling the HCA will rely on emerging technologies such as scRNA-Seq and multiplex spatial approaches, unprecedented amounts and new varieties of molecular data about individuals — supported by metadata to clarify data quality and provenance — will need to be shared. It will be necessary to continue to assess privacy risks to ensure that data sharing will not expose participants to undue re-identification risks and harmful disclosure or misuse of sensitive health information. Security safeguards should be proportionate to the risk of data breach or misuse (https://genomicsandhealth.org/privacy-and-security-policy-read-online). Existing community monitoring and compliance infrastructure should be leveraged, and explicit responses prepared for intentional or malicious misuse (e.g., revocation of access or reporting to host institution) (https://genomicsandhealth.org/working-groups/our-work/accountability-policy). Communicating to research ethics committees and to participants the purposes, benefits, and risks of sharing these data will also be key. The approaches applied in many large-scale genomics projects show that solutions can be put in place to address each of these items.

Just as different sample preparation techniques or analysis platforms undermine data interoperability, use of different ethical and legal tools can create confusing or contradictory thickets of access and use conditions, leading to a lack of "legal" interoperability (http://www.codata.org/uploads/Legal%20Interoperability%20Principles%20and%20Implementation%20Guidelines_Final2.pdf). Harmonizing consent, privacy, and security practices, access policies, and terms of use as much as possible across the HCA will help ensure that data sharing proceeds in an effective and responsible manner.

# PUBLIC ENGAGEMENT

Public engagement throughout the course of the research will be essential to achieve the goals of the HCA. The research will better thrive with public support and involvement, and patients and the public should be engaged in all aspects of the HCA in a sustained manner. Thus, the HCA community must empower an ongoing dialogue between researchers, funders, patients, and the public.

Because the HCA will be built over several years it will require a public engagement strategy that evolves with the project, taking stakeholders and future beneficiaries on a journey of discovery and debate as the research progresses. As a global collaboration, the HCA will also require public engagement activities that are designed to cater to diverse audiences and are sensitive to cultural differences (for example, science popularization approaches that focus on debate are more common in the U.S. and U.K. than in other countries likely to be involved in the project). Importantly, the public engagement strategy will target a wide range of constituents, including generally interested members of the public, citizen-scientists, schoolchildren, and possible research participants.

Public engagement activities will take a wide range of forms and methods — from traditional didactic formats to immersive citizen-science approaches contributing to the research itself. In addition to traditional outreach strategies (such as press campaigns) and digital resources, such as the HCA website (http://www.humancellatlas.org) and YouTube channel (https://www.youtube.com/channel/UCK7wBjw53JdpLYBCvzHosWA) and streaming of HCA meetings), or yourgenome (http://www.yourgenome.org), ideas might include:

- involving the public in the research through citizen-science initiatives — for example, working with partners such as Institute for Research in Schools (U.K.) (http://www.researchinschools.org);
- using the open access, open source data platform and its ancillary portals to "gamify" analysis tasks (e.g., image segmentation and interpretation) or to "crowd-source" software solutions through open challenges;
- generating a data portal for the general public to facilitate exploration, including with emerging Virtual Reality applications;
- featuring the HCA at festivals, such as the Cambridge Science Festival in the U.K. (http://www.sciencefestival.cam.ac.uk) and the Cambridge Science Festival in the USA (http://www.cambridgesciencefestival.org), and in exhibition and event programs in partner localities.
- establishing artist residencies and similar partnerships with members of the creative industries; and
- developing science expos and projects to engage high school students with the HCA, especially in data analysis.

For any project of the magnitude and ambition of the HCA, the general public must be considered a target stakeholder community. An important aspect is making the fundamental principles and motivations of the project as accessible as possible, both via major media outlets and through social media. The name of the HCA itself should be helpful: The term *atlas* creates an instant image of the scale and objectives of the project; in the way that a reader can look with detail beyond continents and countries into individual regions and towns, the HCA will enable scientists to visualize and characterize individual cell types within different organs and tissues of the body. A portal to HCA data for the general public that uses a "Google Maps" approach to free exploration would help forge this connection.

That said, the science and approaches of the HCA are complex and sometimes may evoke emotive responses, so describing it and its impact clearly in a way that is compelling to nonscientists will be an important challenge. To succeed at this, the project's researchers will need to be comfortable engaging with the public. An important part of the HCA strategy should, therefore, be to support and empower researchers to build confidence and communication skills.

Of course, there will be many other, more specific target audiences for HCA engagement efforts. These include:

- *Patient advocacy groups.* Although the HCA will be made up of basic research, the data will ultimately translate into new ways of diagnosing and treating disease. For that reason, we must ensure that the findings are communicated to these groups correctly, without hype, and that public expectations are appropriately managed. Ongoing dialogue and discussion through a suite of public engagement activities will enable us to gauge potential areas of concern or misconception.
- *Teachers and students.* Groundbreaking research from the HCA will take our understanding of cell biology to a new level. It will also present an opportunity to implement new approaches to teaching the basic concepts of cell biology and shaping curricula to reflect this, from K–12 to well beyond. Teachers can be further engaged to communicate to others about HCA.
- *Potential research participants and donors.* The engagement team should be involved in developing strategies for how to sympathetically and effectively approach dialogue with potential tissue donors and their families, whose fully informed consent will be essential for key components of the HCA. Working with patient advocacy groups, such relationships can leverage new approaches for direct-to-patient interactions, especially in specific diseases, as is increasingly occurring in cancer (https://www.mbcproject.org/) and rare disease (https://atfamilies.org/?locale=en).

# FUNDING

The HCA will require financial support from many partners. Worldwide, this will involve collaborations with governmental agencies; partnerships with philanthropic foundations, organizations, and individuals; and collaborations with technology and pharmaceutical companies that can contribute equipment, expertise, or financial resources in support of HCA data collection, analysis, storage, and distribution (**Box 3**).

---

**Box 3. Opportunities to fund the HCA.**

Each of the four pillars of the Human Cell Atlas, and its overall organizational and training activities, requires financial support to meet its respective goals in areas such as:

- data generation in each organ, system, and tissue, including support of a full flagship project for each tissue;
- tissue acquisition pipelines from healthy research participants, organ donors, and postmortem;development of the Data Coordination Platform, from developing, revising, and maintaining the software through supporting data portals, data storage, and computational resources;
- technology development, including invention of new techniques, rigorous comparison, and benchmarking;
- technology dissemination and training, from necessary infrastructure to training on site by experienced SWAT teams of scientists (such support will be particularly critical for tissue handling and for spatial methods);
- computational methods development, from invention of new methods through their comparison and benchmarking across labs and in jamborees;
- computational methods dissemination and training, both through development of hardened and scaled software and portals and by hands-on and online training;
- development and testing of the Common Coordinates Framework;
- scientific planning and organization, including planning workshops and jamborees to develop and test aspects of the atlas, tracking progress across our many research sites and projects, and connecting and onboarding members of the community;
- convening HCA meetings, to bring together, educate, update, inspire, and elicit expertise from the HCA community, and to draft and refine designs and plans; and
- engaging the broader scientific community, the public, and the media.

---

**Funding partners**

The HCA will be internationally supported by science funders across government agencies, foundations, philanthropies, and the pharmaceutical, biotechnology, and technology industry.

*Government support.* National science funding agencies in countries where HCA researchers work have funding mechanisms that can be accessed to support various stages of the HCA effort.

In the U.S., for example, these include, among others, the NIH Common Fund, the NIH BRAIN Initiative, the National Cancer Institute, including the Cancer Moonshot, and institutes related to individual organs (e.g., the National Institute of Allergy and Infectious Diseases for the immune system and the National Institute of Diabetes and Digestive and Kidney Diseases for the kidney). Importantly, early grants have already been awarded by the NIH through pre-existing mechanisms for research in several tissues, in both human (e.g., colon) and mouse (e.g., through the BRAIN initiative); additional applications are pending. The NIH HubMAP program and the NCI Cancer Moonshot (through the Tumor Cell Atlas Network program) are important and engaged partners in the HCA. Other national and international agencies also have mechanisms to support research leading to the construction of the HCA, and several have expressed interest. These include the European Union through its H2020 funding mechanism and the German Federal Ministry of Education and Research, as well as the Japanese, Australian, and Indian governments.

*Philanthropy*. The HCA has benefitted from generous support of a number of philanthropic foundations that support science. The planning process — especially the scientific meetings and workshops that kick-started HCA — has been funded by both individual philanthropists and by foundations including (in chronological order) by an anonymous donor (U.S.), the Wellcome Trust (U.K.), the Chan Zuckerberg Initiative (U.S.), the Kavli Foundation (U.S.), and the Helmsley Charitable Trust (U.S.). The initial development of the DCP is funded by the Chan Zuckerberg Initiative, which has recently funded 38 grants for technical method development (https://chanzuckerberg.com/human-cell-atlas/collaborators) and another RFA has closed for computational method development and has supported the first computational jamboree.

Members of the OC and other members of the HCA Consortium continue to engage with additional foundations and medical research institutes with the goal of developing partnerships. As HCA grows, we will emphasize the importance of disease foundations as partners in both charting relevant organs and tissues and expanding the scope of HCA in disease areas.

*Partnering with technology and pharmaceutical companies.* The HCA provides a unique opportunity for tech companies: they can make important contributions to the collection of data through gifts of equipment or materials, but they can also benefit from the widespread adoption, standardization, rapid technology development, and new software tools built to use these companies' techniques. For example, the creation of the atlas helps establish and broaden the use of single-cell genomics techniques. And the technology and software we develop under the auspices of the HCA will have wider application to technology companies as it is adopted by more and more researchers around the world. While we strive to negotiate terms for use of technologies including early access programs and reduce costs, we remain open to all technologies. Given the HCA's principles of equity, transparency, openness. and nimbleness, we can make no exclusive commitments and promises to commercial partners.

In addition, pharmaceutical companies can benefit tremendously from the data produced which will likely shed light on new actionable targets for medical diagnostics and therapeutics. For example, information from the atlas could lay the groundwork for new diagnostic approaches, help determine inadvertent toxicities, and help identify new disease mechanisms. In this context, a precompetitive consortium of pharmaceutical companies could support some of the HCA activities, including open data generation. We will continue to explore this and additional possibilities.

# SECTION 8. APPENDICES

## APPENDIX I: GOVERNANCE

**Organizing Committee**

The HCA is steered and governed by an **Organizing Committee (OC)**, which is the decision-making body of the HCA.

**OC responsibilities.** The responsibilities of the OC include convening the community through regular meetings, workshops and jamborees; coordinating and authoring key documents; defining scientific values and ethical principles; defining and upholding processes including quality-control standards and analytic standards; governing the Data Coordination Platform (DCP) and Common Coordinate Framework (CCF); coordinating HCA work products; communicating on behalf of HCA; representing and negotiating on behalf of the HCA with other entities and organizations; and polling the HCA community at regular intervals for input on issues, including performance of the OC. The OC does not generate data and is not a direct grantee or grantor for such purposes.

**OC membership.** The OC will consist of up to ~35 scientists. Considerations for new OC members include expertise, geographical representation, and diversity. Additional members are added to the OC by majority vote of the OC. The OC will periodically seek input from the HCA community on the scientific scope of its members, performance, and potential new members. The current (founding) OC consists of 27 scientists from 10 countries and diverse areas of expertise (**Table 2**).

**Terms**. All members will have five-year terms, which can be renewed once by a majority vote of the OC.

**Co-chairs.** The OC is led by two co-chairs, who are members of the OC. The co-chairs have five-year terms, which can be renewed once by a majority vote of the OC.

**Executive Committee**. The OC has an Executive Committee that is responsible for performing routine tasks between OC meetings, preparing meeting agendas, and providing guidance to the executive offices. The EC includes the two co-chairs and five additional OC members, with two-year terms, which can be renewed once by a majority vote of the OC.

**Executive Offices**. The HCA is coordinated by Executive Offices (EOs), which staff the OC in performance of its duties. The OC established four Eos, which are located in the U.K. (Sanger), U.S. (Broad Institute), European Union (Karolinska Institute), and Asia (RIKEN). The EOs' responsibilities include meeting organization; community coordination; coordination of writing of community outputs (e.g., reviews, commentaries, white papers); coordination of interactions with companies; supporting interaction with funders; triage of press inquiries; triage of community inquiries; registry and tracking of projects; and registry and tracking of members. Each EO will also take the lead on some general duties, as well as regional activities.

**Quorum**. A quorum for decision making by the OC will constitute (**1**) a majority of OC members at an in-person meeting that has been announced to the OC at least one month in advance, or (**2**) at least 75% of OC members responding by email to a proposed action that has been circulated to the OC.

| OC Member | Affiliation |
|---|---|
| Ido Amit | Weizmann Institute of Science, Israel |
| Gary Bader | University of Toronto, Canada |
| Peter Campbell | Wellcome Trust Sanger Institute, U.K. |
| Piero Carninci | Riken, Japan |
| Hans Clevers | Hubrecht Institute, Netherlands |
| Roland Eils | German Cancer Research Center; University of Heidelberg, Germany |
| Nir Hacohen | Broad Institute, MGH, USA |
| Arnold Kriegstein | University of California, San Francisco, USA |
| Eric Lander | Broad Institute, USA |
| Sten Linnarsson | Karolinska Institutet, Sweden |
| Partha Majumdar | National Institute of Biomedical Genomics, India |
| Miriam Merad | Mount Sinai, USA |
| Shalin Naik | Walter + Eliza Hall Institute of Medical Research, Australia |
| Garry Nolan | Stanford University, USA |
| Dana Pe'er | Memorial Sloan Kettering Cancer Institute, USA |
| Chris Ponting | Edinburgh University, U.K. |
| Steve Quake | Stanford University / Chan Zuckerberg Biohub, USA |
| Nikolaus Rajewsky | Helmholtz Association, Germany |
| Aviv Regev, Co-Chair | Broad Institute, MIT, HHMI, USA |
| Ehud Shapiro | Weizmann Institute of Science, Israel |
| Jay Shin | Riken, Japan |
| Michael Stratton | Wellcome Trust Sanger Institute, U.K. |
| Henk Stunnenberg | Radboud University, Netherlands |
| Sarah Teichmann, Co-Chair | Wellcome Trust Sanger Institute, U.K. |
| Alexander Van Oudenaarden | Hubrecht Institute, Netherlands |
| Jonathan Weissman | University of California, San Francisco, USA |
| Barbara Wold | California Institute of Technology, USA |

**Table 2. HCA Organizing Committee members.** Current co-chairs are noted. (EC members have not yet been chosen by the OC).

**Working Groups**

The OC establishes **Working Groups** and mandates them to take on specific key areas. At the moment, these include the Analysis Working Group (AWG), the Meta Data Working Group (MDWG), the Common Coordinate Framework Working Group (CCFWG), the Standards and Technology Working Group (STWG), and the Ethics Working Group (EWG). Each Working Group has two co-chairs; one co-chair is a member of the OC and the other co-chair is external to the OC. The Working Group co-chairs together select the other members of the Working Group. Working Group members will have three-year terms, which can be renewed once by a majority vote of the OC.

**Data Coordination Platform**

The OC governs the Data Coordination Platform (DCP), which includes making all policy decisions concerning the DCP, approving the overall plan for the DCP, and ensuring the plan's successful execution by the major developers of the DCP. The OC will establish and appoint a **DCP Governance Group (DCPGG)**, which will report to the OC, to oversee the implementation of these policies, by providing guidance and making decisions concerning certain key topics, including definition of data manifest; official analysis pipelines; required metadata to reflect data collection standards; common coordinates framework; and any formal "release portal." The DCPGG will be led by two OC members; will include at least one member from each of the AWG, MDWG, and CCFWG; and will include at least three additional experts from the community.

The OC will convene, on a quarterly basis, a DCP Coordination Meeting — involving the DCPGG, the major developers of the DCP, and others as appropriate — to review progress and assist the OC in developing policy.

**HCA Meetings**

The HCA OC convenes and advertises meetings, workshops, and jamborees.

**HCA Members**

Any individual may become an HCA Member by registering at the HCA Member Registry and agreeing to abide by the principles of HCA, as stated in the HCA White Paper, especially including its ethical standards. HCA members are invited to attend HCA community-wide meetings and discussions; join the HCA Slack channel; be on mailing lists; and so on.

In addition, an HCA member who is a participant in an HCA project (defined below), or a member of an OC-designated HCA group will be designated as an HCA Collaborating Member. Certain types of scientific meetings, opportunities, and activities will specifically engage HCA Collaborating Members.

**HCA Projects**

Any scientific project related to systematic biological characterization at single-cell resolution may become an HCA Project by registering in the HCA Project Registry. The registry will include a description of the project, its strategy, and its investigators. By registering a project, its investigators affirm their commitment to abide by HCA standards, including the Information Release Policy (below). Projects will fall into three categories.

- **HCA Participating Projects.** Any project (including those focused on data generation, experimental method development, or computational methods development) may be an HCA Participating Project, simply by registering.
- **HCA Network Projects**. An HCA Network Project is an HCA Participating Project that commits to liaise with other network participants.
- **HCA Flagship Project.** An HCA Flagship Project is an HCA Network Project that is aimed a delivering a component of the HCA Draft Atlas Plan 1.0; adheres to the overall framework of Plan (e.g., being comprehensive, adhering to technology standards, engaging domain expertise, having available preliminary data, and having substantial impact); and involves funding of at least 20M Euros over its duration.

**Information Release Policy**

All HCA Projects will commit to ensuring that data developed by or for the project will be deposited in the DCP through regular data release and made available in an open access manner to the maximal extent allowed by ethics (e.g., some metadata may be restricted). The DCP will tag the data to make it citable. Users will be free to use the data as they wish, apart from the obligation to cite the generators of the data and to not attempt to identify or contact individual participants who contribute samples. The OC will work with journals to define opportunities for key data collection and analysis papers.

In addition, all HCA Projects will commit to publicly releasing all experimental and computational methods used for data generation and/or analysis and source code for software developed by or for the project.

# APPENDIX II: HCA DATA COORDINATION PLATFORM OVERVIEW

*Last modified 3/27/2017, current as of 3/27/2017*

*This document is attributed to all attendees of the meetings on Feburary 9th and 22nd, 2017, held jointly by Broad, EBI/Sanger UCSC, and the Chan Zuckerberg Initiative (CZI). It evolved from an initial proposal from EBI/Sanger and Broad, through the two meetings at CZI, and several follow up discussions.*

**Overview diagram**

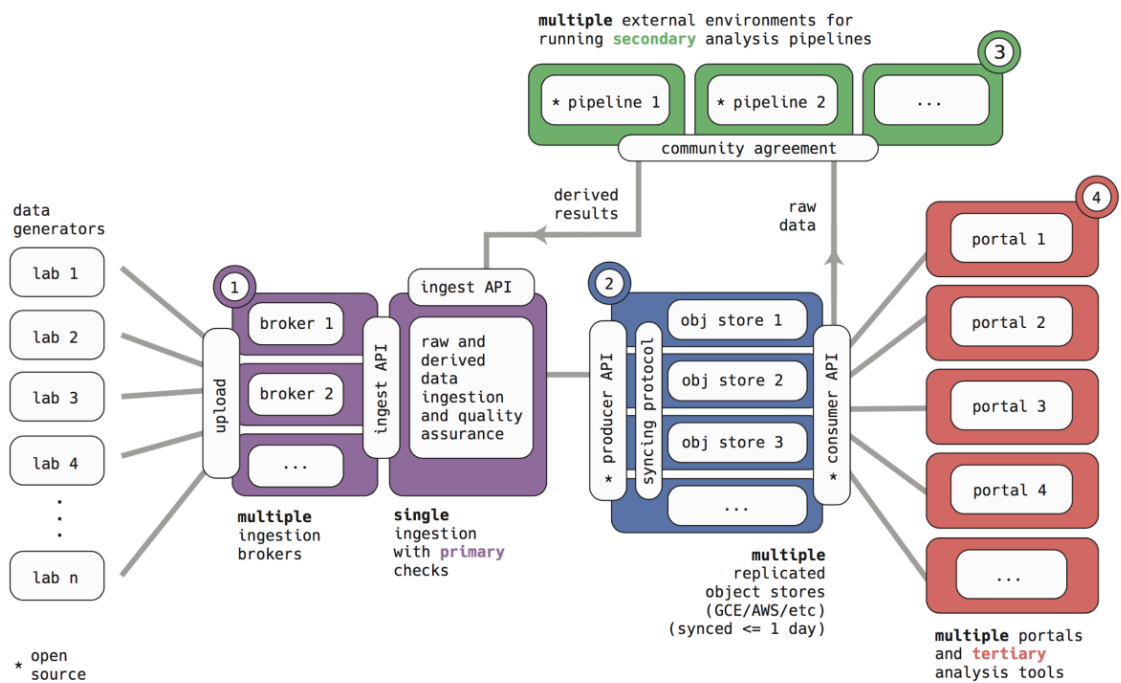

**Preface**

This document describes a dataflow architecture, and a set of principles, for coordinating data across the international Human Cell Atlas project.

The four key components, described in detail below, are the ingestion service [1], the synchronized data store [2], the secondary analysis pipelines [3], and the tertiary portals [4]. The overall data flow sequence is ingestion of raw data and metadata into the ingestion service, depositing into the data store, processing via the secondary analysis pipeline, ingestion of derived analysis results through the ingestion service, depositing those derived results into the data store, and then access from portals via the data store consumer API.

*Key principles informing design*

- Simple, open, and direct access to data.
- Modular components interact via standards and protocols rather than monolithic systems, which encourages good separation of concerns.
- Maximal opportunity for downstream innovation in how the data is analyzed and used.
- Extensible system encouraging a diversity of layered third-party components (e.g., pipelines, data replicas, portals, etc.).

- MIT or similar open source license for any code powering key project components.
- Apache-style governance for any software components requiring community agreement.
- Governance around three key areas of the project: the data ingestion system (for defining data and metadata standards), the synchronized data store (for development and operation), and the official secondary analysis pipelines (for validating and maintaining core analysis pipelines).

**Table of contents**

[1] Data ingestion

[2] Synchronized data store

[3] Secondary analysis pipelines

[4] Tertiary portals

[5] Quality checking

[6] Training

[7] Governance

**[1] Data ingestion**

The HCA Ingestion Service [1] is the single point of entry for all HCA data, including both raw data and metadata for projects, experiments, and samples, as well as a subset of approved derived analyses and quality metrics that result from running vetted analysis pipelines (see [3]). The ingestion service will both ingest data and perform basic quality assurance (see also [5]). Multiple data brokers will act as alternative user-facing entry points, providing any domain- or lab-specific handling or formatting — e.g., data from image-based transcriptomics may require different handling than single-cell RNA sequencing — and delivering data and metadata to the single Ingest API. A primary broker (dark purple) will be "branded" as HCA, but additional brokers will be developed and potentially branded as HCA over time, as determined by the ingestion governance (see [7]).

*Design principles*

- Standards will be defined for data and metadata for different aspects, including samples, experiments, and analyses, and will evolve with technological and biological understanding.
- Standards and validation processes will be defined and developed transparently, e.g., in open Github repositories like the GA4GH file formats group (https://github.com/samtools/hts-specs).
- Biomedical ontology terms will be used when possible, and HCA scientists will consult to help extend the ontologies where needed.
- The metadata validation ruleset will be exposed as an API to enable submitters to debug and validate before submission. The validation will eventually be published in containerized form to empower data brokers — in coordination with data producers and the ingestion service — to prevalidate submissions prior to upload.
- Forms of acceptable derived data from the secondary analysis pipelines will be determined by its governance group (see [3] and [7]).
- Users should get rapid feedback on the basic validity of their submissions.

- Users will be able to submit data in a manner that is most compatible with their workflows and level of expertise (e.g., Web UIs, spreadsheet templates, or bulk upload via APIs).
- Finalizing a data submission is an optionally interactive process, in which the submitter declares responsibility for accuracy and can review the results and metadata summary.
- For all data modalities, it will be possible to preregister sample metadata or any other study metadata before experimental data is collected, and updating preregistered metadata will be supported through versioning (see [3]).
- To obtain best network performance and availability, submitters should be able to upload bulk data either to their geographically proximal cloud provider, or directly to the ingestion service.

*Functional components*

The **Data Brokers** are user-facing sites or services for data upload that handle lab-specific or data-type specific formatting (e.g., a setting in which spatial information about cell locations within tissue is available). Brokers should consider only complete data sets, i.e., not samples isolated from a project. Brokers may be staffed by "data wranglers" and other support staff who can liaise with and establish relationships with labs providing data, ideally co-located with submitters for rapid iteration. This additional human interface, coupled to engineering, will allow a broker to adapt to the needs of different labs, and to communicate data requirements back to the core group running the ingestion service.

The **Ingest API** provides submission and validation functionality and will underlie all HCA broker operations and ingestion of data, metadata, and derived analyses. Validation will be applied at the metadata and file format level only, and check for data duplication across brokers, but will not involve quality control or data quality checks that require subsequent analysis (e.g., alignment). Users will be tracked upon account creation and at each data submission or data update, and users should be able to track their existing submissions.

The **Ingest CDN** is an upload acceleration network, operated using an independent replica of a subset of the synchronized data store [2] infrastructure, designed to allow data producers to upload data to their most performant, geographically proximal endpoint. From that endpoint, the Ingest CDN forwards data to the single ingestion point for subsequent processing. From the submitter's point of view, the Ingest CDN is write only, with the same authentication semantics as the Ingest API itself.

An **email help desk** will be available to support HCA submitters with follow-up phone conversations or in-person meetings where appropriate. Those providing help must understand the requirements, train users to structure information appropriately, and interact with domain experts to ensure standards are appropriate. Help desk activities should use a shared infrastructure (e.g., a ticketing system) to present a unified service for submitters and ensure responses are synchronized across brokers.

## [2] Synchronized data store

The HCA Data Store [2] is a multisite replicated storage system containing all raw data and metadata and certain forms of derived data, materialized and stored as flat files. Its goals are to ensure simple and direct data access for downstream consumers, to notify users of repository changes, and to synchronize data across multiple sites.

*Design principles*

- Data deposited in the data store will flow from the HCA ingestion service [1], which is solely responsible for data and metadata schemas, naming, accessioning, and validation.
- All access between the ingestion service and the data store is via well-defined APIs, ensuring modularity and separation of concerns.
- The data store will manage data replication across official HCA object stores. There will be at least three public cloud stores, and additional public or on-premise stores may be added over time.
- At the lowest level, all data and metadata will be stored as a set of identical objects, invariant across cloud stores, and synchronization across cloud providers will thus depend on synchronization of this set of objects.
- A common object layer across clouds will provide homogeneity and aid portability of analysis and access between different clouds, and allow cloud-independent data indexes to be built that link to each of the data copies on each cloud.
- The public reference atlas will not require authenticated access, but access control and authentication may be required for some subsets of data.
- It must be possible for any third party to add its own external pipeline or portal (as a consumer), to add a data replica, and to receive timely updates about changes to the data store.
- Data change notification and replication must be rapid (e.g., significantly less than a day).
- Tracking of data access via analytics will help determine data usage patterns to minimize costs (e.g., which data needs to be highly available or placed in cold storage).

*Functional components*

The **Producer API** is primarily responsible for receiving data directly from the authenticated ingestion service [1]. In addition, it will communicate with internal systems to trigger object synchronization across the replicas, and drive change notification Webhooks. All data and metadata will be explicitly versioned; the data store is effectively append-only, except in emergency circumstances. Updates to data will generally require new submissions, and updates to metadata will manifest as a series of files or explicit file versions.

The **Consumer API** provides read-only access to data, via cloud-native APIs. Access is subject to authentication and access control for subsets of data where required. The Consumer API also provides Webhook change notification (e.g., GitHub-style), allowing downstream computation to be triggered and ensuring that all secondary and tertiary analysis tools can be reliably and rapidly updated as changes occur.

The **Object synchronization system** and the **Pub/Sub bus and replication logs systems** are private subsystems that provide synchronization and change notification infrastructure across distributed object stores, and they communicate with both consumer and producer APIs.

The **Object stores** are the multiple S3-like object stores on public and/or private clouds.

**[3] Secondary analysis pipelines**

The HCA Secondary Analysis Pipelines are defined as analyses with results that may, subject to approval, be deposited back into the data store, in contrast to the tertiary analysis portals (see [4]). We consider it critical to have robust pipelines that run continuously on new data because most data types for the HCA will require some processing to support the majority of downstream use cases (e.g., alignment and demultiplexing for single-cell RNA sequencing, detection and segmentation for image-based transcriptomics). The analysis governance group (see [7]) will determine which analysis pipelines should be "approved" to deposit results back into the data store and which ones should additionally be designated as "official" for continuous deployment.

*Design principles*

- Encourage a diversity of analysis approaches, while ensuring that only vetted pipelines can deposit data back into the system and that one or more of those vetted pipelines are run continuously on all data from project onset.
- Any approved pipelines must be open source and under a MIT or similar license.
- Any approved pipelines must be packaged to ensure convenient reproducibility (e.g., via containerization and workflow description languages).
- It must be possible to run any approved pipeline against local data in local test environments.
- In support of the HCA, there will be continued development of two workflow standards (Workflow Description Language and Common Workflow Language) and two engines for running these workflows (FireCloud and Toil); both will be usable for analyzing HCA data across the three major cloud providers (AWS, GCE, Azure).
- Test and example datasets should be provided for core tasks and can be used by the governance group for the development of benchmarks and objective assessments of pipelines.
- Pipelines that need to aggregate multiple data sets, as opposed to running on individual samples, will run as close to continuous as possible, though delays may be inevitable; these analyses can reference appropriate metadata to ensure they only run on complete data sets.

*Functional components*

An extensible **pipelining service** for use by the HCA consortium. This service will have the ability for users to easily string together tools into pipelines and deploy them at scale and in a cost-effective manner across multiple computational backends. It should be container-based to support portability and ease-of-use by the community.

A **methods repository** for storing and sharing pipelines within the HCA community.

Approved **analysis pipelines** that consume raw data from the data store's consumer API and generate derived results (e.g., gene cell tables or quality-control metrics) to be deposited back into the ingestion service, running either once or multiple times depending on resources.

At least **one official HCA analysis pipeline per data type** [dark green], where data type could be, for example, single-cell RNA sequencing or image-based transcriptomics. These pipelines should reflect current best practices as determined by the analysis governance group (see [7]). They will be run continuously on all data of that type and be maintained by the HCA consortium. If multiple official pipelines are available for a single data type, they must provide information about their utility, features, and comparisons to other pipelines.

### [4] Tertiary portals

The HCA Tertiary Portals support a wide range of downstream user-facing analyses, visualizations, and forms of access for working with HCA data. We imagine a diversity of portal types to bloom, targeting different levels of users, including Web-based interfaces, analysis results, custom APIs for performing rich and structured queries, and other novel interfaces.

*Design principles*

- A rich ecosystem of portals covering a wide variety of use cases should be supported.
- Coordination on similar efforts should be encouraged within the community, where appropriate.
- Multiple portals can become branded with an "HCA" badge by following certain guidelines.

*Functional components*

At least **one developer-oriented portal** providing a platform (e.g. FireCloud or Toil) in which developers can bring containerized environments to perform analyses on the data

At least **one user-oriented portal** providing interactive interfaces to the data; for example:

- Quantifying the expression of a given gene (e.g., marker genes specified by user) across cell types, shown in several popular modalities (e.g., low-d plots, heatmaps, violin plots);
- Showing clustering of individual cells from an experiment based on expression profiles;
- Painting cell clusters (ordinations) by metadata (technical and experimental) to identify batch effects and visualize biological groupings (depending on the type of metadata);
- Visualizing gene signatures by several modalities, including heatmaps and dot plots of average expression by cell group; and
- Cross-correlating gene expression with epigenetic markers.

Multiple **query-oriented portals** with APIs targeting custom access patterns, for example:

- Querying all gene expression tables generated with a particular analysis;
- Querying all cells for those that match the expression pattern of a target cell and return the metadata for the matching cells; and
- Querying all raw data for a specific tissue type, ranked based on a custom combination of quality-control metrics.

### [5] Quality checking

Below is a framework for ensuring the quality of Human Cell Atlas data, including the types of quality checks that will be performed throughout the lifecycle of HCA data, how these quality checks are coupled to the flow of HCA data through the architecture, and how this framework will evolve over time as data grows and pipelines harden.

*Tiers of quality checking*

**Quality assurance (QA)** is focused on the structural soundness of the data, with greater emphasis on checking consistency of metadata than the experimental quality of the data itself. Key elements include checking that a file is not empty at point of ingest, that data and metadata formats are correct, that syntax is well formed (e.g., correct UUIDs), and that there is intra-sample consistency (e.g., if one cell has many files, all must be present). In general, failing this tier should result in rejection of data submission, unless there is an immediate and obvious fix, and passing this tier results in data ingestion (although it may be later blacklisted if it fails subsequent checks).

**Quality control (QC)** is focused on the experimental quality of the data itself. Defining these metrics will be governed by the analysis governance [7] of HCA, who will also determine the approved pipelines for secondary analyses [3]. In the example of single-cell RNA sequencing, checks will include mapping quality, potential contamination, and quality of cells. In collaboration with the analysis governance, criteria will be determined for a single sample, or entire batch, to pass this step, causing it to be whitelisted (official HCA data) or blacklisted (data has been withdrawn). These metrics will be made available as tables computed during secondary analyses by the official pipelines.

**Inspection** is focused on exceptional circumstances where a sample fails either QA or QC and the cause is not immediately obvious. In such situations, a human analyst must inspect the data to determine the cause of failure. The emphasis at this stage is ensuring that the human analyst has the data sets and analytical tools needed for rapid debugging. Data visualization will be key, and example steps, likely developed as part of the tertiary portals [4], and, in the example of single-cell RNA sequencing, might include plots focusing on potential technical batch effects or comparing mapping quality of cells within a study. Following this step, a sample or set of samples will either be repaired or not, resulting in whitelisting or blacklisting, and provenance of all changes will be stored.

*Relationship to data flow*

**Quality assurance** will primarily happen during ingestion [1] and be coupled to the date of submission receipt. Formats will evolve during the project, so it must be possible to repeat backward-compatible checks on old data. Data that pass this step will become part of the primary HCA data set and deposited into the data store [2].

**Quality control** results and metrics will be performed as part of the secondary analysis pipelines [3] that run on top of the data from the data store. Although in principle these analyses can run against any of the replicated cloud stores, in practice it will be most efficient if the continuously running pipelines run against one cloud store and distribute results.

*Evolution of quality checks*

**Quality assurance** during the first year will focus on format, compression, essential metadata, file integrity, and consistency with metadata ontologies. After the first year, checks can begin to incorporate the correct number and composition of files per sample.

**Quality control** during the first year will be closely coupled to ongoing developments in analysis approaches, and the focus will be maximizing experimentation and diversity of techniques. This rapid prototyping could benefit from running quality checks in a single cloud compute environment, alongside ad-hoc external deployments on other clouds. After the first year, emphasis can shift to standardizing, scaling, and optimizing the cost of continuously running analyses, and deployment strategies can tailor to this need.

## [6] Training

All components of the system will require training. The project will therefore have dedicated coordination of training activities. Training needs are broken down by project components but will ideally conform to common templates to provide an integrated user experience.

*Components*
- A training entry point will provide links to training materials and documentation.
- Training for use of the Ingest API will be needed for external users and brokers; example data files should form part of the materials as well as documentation of the API.
- Training for the secondary analysis portals will include webinars for analysis methods and hackathons to encourage analysis development and teach existing tools to new members.

- Training for developers of the tertiary portals will ensure that they understand how data updates occur and how to use the Webhook notifications from the data store's Consumer API.

**[7] Governance**

An oversight group with representatives from international funding bodies and the HCA consortium will ensure that all data storage and access protocols satisfy all relevant regulatory, legal, and ethical guidelines.

In addition, we imagine three broad areas of governance — ingestion, data management, and analysis — falling under the umbrella of the HCA consortium.

*Ingestion*

- Define standards around data and metadata formats.
- Ensure that standard definitions are developed transparently and are publicly available.
- Ensure that all quality assurance checks can be performed locally prior to data upload.
- Ensure harmonization and coordination of data broker activity and user-facing support.

*Data management*

- Define cost tiers for data storage based on usage.
- Define file organization, naming practices, and versioning within the data store.
- Define any foreign key indexing.
- Define protocols for handling authenticated data.

*Analysis*

- Responsible for shepherding pipeline development and determining approved project pipelines (which are allowed to deposit results back into the data store) and official pipelines (which, in addition, run continuously on all new data).
- Ensure that any approved analysis pipeline can be reproduced and deployed in multiple cloud environments, as well as run locally (e.g., via containerization).
- Decisions on approved and official pipelines should be handled via Apache-style voting.
- There should be one subgroup for each broad data type (e.g., single cell RNA sequencing, image-based transcriptomics), all operating under an analysis governance umbrella.
- Define requirements for portals becoming HCA branded.